\documentclass[11pt]{article}
\usepackage{acronym}
\usepackage{epsf,amsmath,amssymb,graphicx,scalefnt,rotating,enumitem,fancyhdr}
\usepackage{comment}
\usepackage[english]{babel}
\usepackage[final]{showkeys}
\usepackage{slashed}
\usepackage{dsfont}
\usepackage[acronym]{glossaries}
\glsdisablehyper
\let\oldglsentryshort\glsentryshort
\renewcommand{\glsentryshort}[1]{\abbrev{\oldglsentryshort{#1}}}
\usepackage{listings}
\usepackage{bbm}
\usepackage{mathtools}
\usepackage{subcaption}
\usepackage{multirow}
\usepackage{array}
\usepackage{cite}
\usepackage{tikz}
\usepackage{filemod}
\usepackage{tablefootnote}
\usepackage[pdftex, colorlinks=true, linkcolor=black, filecolor=black,
  citecolor=black, urlcolor=black, draft=false, bookmarks,
  bookmarksnumbered=true, plainpages=false,
  linktocpage={true}]{hyperref}
\usepackage[affil-it]{authblk}
\usepackage[capitalize]{cleveref}
\AddToHook{cmd/appendix/before}{\crefalias{section}{appendix}}
\usepackage[table]{xcolor}

\numberwithin{equation}{section}
\allowdisplaybreaks
\definecolor{codegreen}{rgb}{0,0.6,0}
\definecolor{codegray}{rgb}{0.5,0.5,0.5}
\definecolor{codepurple}{rgb}{0.58,0,0.82}
\definecolor{backcolour}{rgb}{0.95,0.95,0.92}
\lstdefinestyle{mystyle}{
    backgroundcolor=\color{backcolour},
    commentstyle=\color{codegreen},
    keywordstyle=\color{magenta},
    numberstyle=\tiny\color{codegray},
    stringstyle=\color{codepurple},
    basicstyle=\ttfamily\footnotesize,
    breakatwhitespace=false,
    breaklines=true,
    captionpos=b,
    keepspaces=true,
    numbers=left,
    numbersep=5pt,
    showspaces=false,
    showstringspaces=false,
    showtabs=false,
    tabsize=2
}
\newcounter{notecount}

\newcommand{\bsym}{}

\newcommand{\cp}{\abbrev{CP}}

\newcommand{\one}{one}
\newcommand{\two}{two}
\newcommand{\three}{three}
\newcommand{\four}{four}
\newcommand{\five}{five}

\newcommand{\rhs}{r.h.s.}
\newcommand{\wrt}{w.r.t.}

\newcommand{\eqmark}[1]{\rlap{\hspace{-0.1em}\raisebox{-1ex}{$\scriptscriptstyle \mathrm{#1}$}}}

\newcommand{\citere}[1]{Ref.~\cite{#1}}
\newcommand{\citeres}[1]{Refs.~\cite{#1}}

\newcommand{\abbrev}[1]{{\scalefont{.9}#1}}
\newcommand{\EulerGamma}{\gamma_\text{E}}

\newcommand{\bare}{\text{\abbrev{B}}}

\newcommand{\lmut}{L_{\mu t}}
\newcommand{\ep}{\epsilon}

\newcommand{\dd}{\mathrm{d}}
\newcommand{\deriv}[3]{\frac{\partial\ifthenelse{\equal{#1}{}}{}{^{#1}}%
    #2}{\partial #3\ifthenelse{\equal{#1}{}}{}{^{#1}}}}
\newcommand{\dderiv}[3]{\frac{\dd\ifthenelse{\equal{#1}{}}{}{^{#1}}%
    #2}{\dd #3\ifthenelse{\equal{#1}{}}{}{^{#1}}}}
\newcommand{\order}[1]{{\cal O}(#1)}

\newcommand{\msbar}{\ensuremath{\overline{\mbox{\abbrev{MS}}}}}

\newcommand{\ccf}{C_\text{F}}
\newcommand{\cca}{C_\text{A}}
\newcommand{\ctr}{T_\text{R}}
\newcommand{\nf}{n_\text{f}}
\newcommand{\nh}{n_\text{h}}
\newcommand{\nl}{n_\text{l}}
\newcommand{\na}{n_\text{A}}

\newcommand{\nc}{n_\text{c}}
\newcommand{\apis}{a_\text{s}}

\newcommand{\dsquared}{D^2}

\newacronym{NDR}{NDR}{naive dimensional regularization}
\newcommand{\ndr}{\gls{NDR}}
\newacronym{OPE}{OPE}{operator-product expansion}
\newcommand{\ope}{\gls{OPE}}
\newacronym{LEFT}{LEFT}{low-energy effective field theory}
\newcommand{\LEFT}{\gls{LEFT}}
\newacronym{SFTX}{SFTX}{short-flow-time expansion}
\newcommand{\sftx}{\gls{SFTX}}
\newcommand{\gf}{\abbrev{GF}}
\newcommand{\gfs}{\abbrev{GF} scheme}
\newacronym{CMDM}{CMDM}{chromomagnetic dipole}
\newcommand{\cmdm}{\gls{CMDM}}
\newacronym{CEDM}{CEDM}{chromoelectric dipole}
\newcommand{\cedm}{\gls{CEDM}}
\newacronym[\glslongpluralkey={degrees of freedom},%
  shortplural=d.o.f.]{dof}{d.o.f.}{degree of freedom}

\newacronym[\glslongpluralkey={renormalization-group equations},%
  shortplural=\abbrev{RGE}s]{RGE}{RGE}{renormalization-group equation}

\newacronym{LSZ}{LSZ}{Lehmann-Symanzik-Zimmermann}

\newacronym[\glslongpluralkey={equation-of-motions},%
  shortplural=\abbrev{EOM}s]{EOM}{EOM}{equation-of-motion}
\newcommand{\eom}{\gls{EOM}}

\newacronym[\glslongpluralkey={Effective Field Theories},%
  shortplural=\abbrev{EFT}s]{EFT}{EFT}{Effective Field Theory}

\newacronym{QED}{QED}{quantum elecrodynamics}

\newacronym{SSB}{SSB}{spontaneous symmetry breaking}

\newacronym[\glslongpluralkey={vacuum expectation values},%
  shortplural=\abbrev{VEV}s]{VEV}{VEV}{vacuum expectation value}
\newcommand{\vev}{\gls{VEV}}
\newcommand{\vevs}{\glspl{VEV}}
\newacronym{QM}{QM}{quantum mechanics}

\newacronym[\glslongpluralkey={quantum field theories},%
  shortplural=\abbrev{QFT}s]{QFT}{QFT}{quantum field theory}

\newacronym{UV}{UV}{ultraviolet}
\newcommand{\uv}{\gls{UV}}
\newacronym{QCD}{QCD}{quantum chromodynamics}
\newcommand{\qcd}{\gls{QCD}}
\newacronym{LHC}{LHC}{Large Hadron Collider}

\newacronym{PMNS}{PMNS}{Pontecorvo–Maki–Nakagawa–Sakata}

\newacronym{CKM}{CKM}{Cabbibo-Kobayashi-Maskawa}

\newacronym{IbP}{IbP}{integration-by-parts}

\newacronym{LO}{LO}{leading order}
\newcommand{\lo}{\gls{LO}}
\newacronym{NLO}{NLO}{next-to-leading order}
\newcommand{\nlo}{\gls{NLO}}
\newacronym{NNLO}{NNLO}{next-to-next-to-leading order}
\newcommand{\nnlo}{\gls{NNLO}}
\newacronym{LL}{LL}{leading logarithmic}
\newacronym{NLL}{NLL}{next-to-leading logarithmic}

\newacronym{NNLL}{NNLL}{next-to-next-to-leading logarithmic}

\newacronym[\glslongpluralkey={parton density functions},%
  shortplural=\abbrev{PDF}s]{PDF}{PDF}{parton density function}
\newcommand{\pdf}{\gls{PDF}}
\newcommand{\pdfs}{\glspl{PDF}}
\newacronym{SM}{SM}{Standard Model}

\newacronym{SMEFT}{SMEFT}{Standard Model Effective Field Theory}

\newacronym{BSM}{BSM}{beyond-the-\gls{SM}}

\newacronym{MSSM}{MSSM}{Minimal Supersymmetric \gls{SM}}

\newacronym{SUSY}{SUSY}{Supersymmetry}

\newacronym{DREG}{DREG}{Dimensional Regularization}

\newacronym{DRED}{DRED}{Dimensional Reduction}

\newacronym{EMT}{EMT}{energy-momentum tensor}

\newcommand{\RHheaderline}{\textsf{TTK-26-42, P3H-26-094, November 2026}}
\fancypagestyle{firstpage}
{
  
  \fancyhead[R]{\RHheaderline}
}
\title{Quark dipole operators in the gradient-flow scheme\\
at \abbrev{NNLO QCD}}
\author[a]{J.~Borgulat}
\author[a]{L.~Georg}
\author[a]{R.V.~Harlander}
\author[a,b]{J.T.~Kohnen}
\author[a,c,d]{A.~Shindler}
\affil[a]{TTK, RWTH~Aachen University, Sommerfeldstr.~16, 52074~Aachen,
  Germany}
\affil[b]{Institute for Theoretical Physics, ELTE Eötvös
  Lor\'and University, P\'azm\'any P\'eter 1/A, H–1117 Budapest, Hungary}
\affil[c]{Department of Physics, University of California,
  Berkeley, California 94720, USA}
\affil[d]{Nuclear Science Division,
  Lawrence Berkeley National Laboratory, Berkeley, California 94720, USA}
\date{}
\begin{document}
\maketitle
\thispagestyle{firstpage}
\begin{abstract}
Conversion factors for the dimension-\five\ \abbrev{QCD} operator basis
from the \msbar\ to the gradient-flow scheme are presented through \nnlo\ in
the strong coupling. In the \cp-even sector, the results are provided for
general quark masses and arbitrary flavor content of the operators.
Specifically, this concerns the quark \cmdm\ operator and its mixing with
lower-dimensional \abbrev{QCD} operators.  In the massless case, our results
immediately apply to the quark chromoelectric dipole operator. As a
by-product, we also determine missing entries in the \msbar\ anomalous
dimension matrix of the \cmdm\ operator to \nnlo\ \abbrev{QCD}.
\end{abstract}
\glsresetall

\parskip.0cm
\tableofcontents
\parskip.2cm


\section{Introduction}\label{sec:Introduction}

Over the past few years, the benefits of the gradient flow as a tool to deal
with the renormalization of composite operators have been demonstrated through
various phenomenological applications, among them an evaluation of the strange
and charm quark masses~\cite{Black:2025gft}, the determination of heavy-meson
bag parameters~\cite{Black:2026qfv,Black:2026dzp}, and the calculation of
higher Mellin moments of \pdfs~\cite{Shindler:2023xpd,Francis:2025pgf,Francis:2025rya}.

The origin of this success is twofold: The gradient flow not only provides a
gauge-invariant \uv\ regularization of composite operators which preserves the
O(4) symmetry of Euclidean space; it can also be implemented in a perturbative
formulation as well as on the lattice.  At fixed positive flow time, matrix
elements of composite operators built from renormalized flowed fields admit a
continuum limit without additional operator-mixing subtractions.  Conversion
to the \msbar\ scheme can be done via perturbative matching coefficients. The
operator mixing is thus kept fully within the perturbative part of the
calculation, where the \uv\ divergences can be canceled analytically. This
eliminates the necessity of taking into account Lorentz-symmetry violating
operators which arise from the breaking of the O(4)-symmetry on the lattice,
for example. Early applications of the gradient flow established methods to
define and renormalize the energy-momentum tensor on the
lattice~\cite{Suzuki:2013gza,Makino:2014taa,DelDebbio:2013zaa}.
\citere{Shindler:2023xpd} developed a method for extracting \pdf\ moments of
arbitrary order, using continuum rotational symmetry to control the mixing of
higher-spin operators. This development enabled the calculations of higher
\pdf\ moments mentioned above~\cite{Francis:2025pgf,Francis:2025rya}.

From a general point of view, the gradient flow allows for a formulation of
the operator-product expansion which is better suited to practical
calculations than the usual $\msbar$
scheme~\cite{Harlander:2026lte,Monahan:2015lha,Harlander:2020duo,
  Harlander:2022tgk,Beneke:2025hlg}. It provides a common renormalization
scheme for the Wilson coefficients and the operator matrix elements which, at
the same time, allows a consistent separation of perturbative coefficients and
flowed matrix elements, removing the infrared renormalon ambiguities
associated with the power corrections included in the
\ope~\cite{Beneke:2025hlg}.

Crucial ingredients for these applications of the gradient flow are the
process-independent matching coefficients which describe the conversion of
either the operators or the Wilson coefficients between the \msbar\ and the
gradient-flow (\abbrev{GF}) scheme.  It is important to note that the
conversion is understood at the level of a complete operator basis which
involves all operators of the same symmetries up to a certain mass dimension.
Over the past few years, the matching coefficients for a number of operator
bases have been evaluated through \nnlo\ in \qcd. They include the
quark-bilinear operators~\cite{Borgulat:2023xml}, the \cp-even
dimension-\four\ operator basis of Lorentz rank zero~\cite{Harlander:2020duo}
and two~\cite{Harlander:2018zpi}, a subset of \four-quark
operators~\cite{Harlander:2021jcq,Harlander:2022tgk,Black:2026dzp}, and flavor
nonsinglet twist-\two\ operators up to Lorentz rank
six~\cite{Harlander:2025qsa}.  The integrated topological charge is
independent of flow time in the continuum, provided boundary terms
vanish~\cite{Luscher:2010iy,Shindler:2015aqa,Ce:2015qha}; the local
topological charge density, however, changes with flow time by a total
derivative; the corresponding coefficient is known through
\nnlo\ \qcd~\cite{Borgulat:2026diss}.  Corresponding \nlo\ results for the
matching coefficients of these operators were obtained in
\citeres{Suzuki:2013gza,Makino:2014taa,Hieda:2016lly,
  Suzuki:2020zue,Shindler:2023xpd}.  In fact, at this perturbative order, the
\qcd\ matching of the complete baryon- and lepton-number-conserving
\LEFT\ basis through mass dimension six to the \gfs\ is
available~\cite{Crosas:2026ofx}. Likewise, the matching coefficients in the
\cp-odd\ sector are known at \nlo\ for operators up to mass dimension six
\cite{Rizik:2020naq,Mereghetti:2021nkt,Buhler:2023gsg,Crosas:2023anw}. In the
case of ``scalar \qcd'', the matching coefficients for the scalar bilinear and
the scalar Noether current are known through
\nnlo\ \qcd~\cite{Borgulat:2025gys}.

In this paper, we will predominantly consider the \cp-even Lorentz-scalar
\qcd\ operator basis of mass dimension five. In the massless case, it consists
of a single operator which we will refer to as the \cmdm\ operator in this
paper. Preliminary results for its conversion to the \gfs\ through \nnlo\ have
been presented in \citere{Harlander:2022vgf}. Here, we will generalize these
results to an arbitrary quark flavor configuration of the operator, and
arbitrary quark masses. For consistency, this also requires the matching of
the flowed scalar density to the dimension-five operators which appear at
order $t$. For the case of massless quarks, we argue that the matching
coefficients for the \cmdm\ operator are the same as for the \cp-odd case of
the \cedm\ operator due to chiral symmetry, which means that our results (and
those of \citere{Harlander:2022vgf}) also apply in this case. The
\cedm\ mixing with the pseudoscalar density has also been determined
nonperturbatively using the gradient flow~\cite{Kim:2021qae}.

The work of the present paper can be viewed as part of a coherent program to
evaluate the set of all \qcd\ operators up to mass dimension six in the
\gfs\ through \nnlo\ \qcd. It should provide useful input for phenomenological
studies on the lattice, in particular in the context of quarkonium
physics~\cite{Brambilla:2010cs} or the study of the origin of \cp-violation.


\section{Theoretical framework}


\subsection{Lagrangian}

\label{sec:lagrangian}
We work in $D$-dimensional Euclidean space-time with $D=4-2\ep$.  Furthermore,
we suppress space-time arguments and spinor indices unless they are required
for clarity. Our gauge group (color) conventions are those of
\citeres{Luscher:2010iy, Luscher:2011bx, Harlander:2020duo}.  The $\na$
anti-Hermitian generators $T^a$ are normalized according to
$\mathrm{Tr}(T^aT^b)=-\ctr\delta^{ab}$, and the structure constants $f^{abc}$
are defined by $\left[T^a,T^b\right]=f^{abc}T^c$.  For $\mathrm{SU}(\nc)$, the
relevant group factors are given by
\begin{gather}
\begin{aligned}
    \na &= \nc^2-1\,, & \ccf &= \ctr\frac{\nc^2-1}{\nc}\,, & \cca &= 2\,\ctr
    \nc\,,
\end{aligned}
\end{gather}
where $\ccf$ and $\cca$ denote the quadratic Casimir eigenvalues of the
fundamental and adjoint representations, respectively.

Our calculations are based on the Lagrangian
\begin{equation}\label{eq:basi:brix}
  \begin{aligned}
    \mathcal{L} &= \frac{1}{4}G_{\mu\nu}^aG_{\mu\nu}^a +
    \bar{\bsym{\psi}}_L\slashed{D}\bsym{\psi}_L
    +\bar{\bsym{\psi}}_R\slashed{D}\bsym{\psi}_R
    +\bar{\bsym{\psi}}_L
    M_\bare \bsym{\psi}_R+\bar{\bsym{\psi}}_R M_\bare^\dagger
    \bsym{\psi}_L + \mathcal{L}_\text{gf} + \mathcal{L}_\text{gh}\,,
  \end{aligned}
\end{equation}
where $\mathcal{L}_\text{gf}$ and $\mathcal{L}_\text{gh}$ are the usual
gauge-fixing and Faddeev-Popov ghost terms,
\begin{equation}\label{eq:SpinorVec}
    \bsym{\psi}
    =
    \left(
        \psi_1,\ldots,\psi_{\nf}
    \right)^T
\end{equation}
is a vector of $\nf$ quark flavors, and $\bsym{\psi}_{L/R} =
(1\mp\gamma_5)\bsym{\psi}/2$ are the left/right-handed components of the quark
spinors.\footnote{For the \cp-even quantities treated in this paper, $\psi_L$
and $\psi_R$ will always combine to $\psi$, and $\gamma_5$ never appears
explicitly in our calculation, making any discussion about its $D$-dimensional
treatment irrelevant. Concerning the \cp-odd case, see \cref{sec:cedm}.}

The field-strength tensor $G_{\mu\nu}^a$ is given by
\begin{equation}
\begin{gathered}
    \label{eq:unfConv1}
    G_{\mu\nu}^a
    =
    \partial_\mu A_\nu^a
    -
    \partial_\nu A_\mu^a
    +
    g_\bare f^{abc}A_\mu^b A_\nu^c\,,
\end{gathered}
\end{equation}
with the gauge fields $A_\mu^a$ (gluons), and the covariant derivative is
\begin{equation}\label{eq:basi:kerf}
  \begin{aligned}
    D_\mu=\partial_\mu+g_\bare A_\mu^aT^a\,,
  \end{aligned}
\end{equation}
where $g_\bare$ is the bare gauge coupling, related to the renormalized
coupling $g(\mu)$ as
\begin{equation}\label{eq:basi:ebon}
  \begin{aligned}
    g_\bare = \left(\frac{\mu^2 e^{\EulerGamma}}{4\pi}\right)^{\ep/2}
    \sqrt{Z_\text{s}}\,g(\mu)\,,
  \end{aligned}
\end{equation}
with $\mu$ the renormalization scale, $\EulerGamma$ the Euler-Mascheroni
constant, and $Z_\text{s}$ is given in \cref{eq:Zs}.  We express our results
as an expansion in the rescaled strong coupling
\begin{equation}\label{eq:basi:esme}
  \begin{aligned}
    \apis\equiv\frac{\alpha_{\textrm{s}}}{\pi}=\frac{g^2}{4\pi^2}\,.
  \end{aligned}
\end{equation}

For the discussion of chiral flavor symmetry, we have introduced a general bare
mass matrix $M_\bare$ in \cref{eq:basi:brix}. Since \qcd\ interactions are
flavor blind, it is proportional to the renormalized mass matrix $M(\mu)$,
\begin{equation}\label{eq:basi:krym}
  \begin{aligned}
    M_\bare = Z_m\,M(\mu)\,.
  \end{aligned}
\end{equation}
Both for the mass and the coupling renormalization constants $Z_m$ and
$Z_\text{s}$, we adopt the \msbar\ scheme. The corresponding expressions are
given in \cref{app:rg-functions} to the required order.

The Lagrangian in \cref{eq:basi:brix} can be made formally invariant under
chiral flavor transformations
\begin{equation}\label{eq:basi:boat}
  \begin{aligned}
    \begin{pmatrix}
      \bsym{\psi}_{L}\\
      \bsym{\psi}_{R}
    \end{pmatrix} \to 
    \begin{pmatrix}
      U_L & 0 \\
      0 & U_R
    \end{pmatrix}
    \begin{pmatrix}
      \bsym{\psi}_{L}\\
      \bsym{\psi}_{R}
    \end{pmatrix}
 \end{aligned}
\end{equation}
by interpreting $M_\bare$ as a spurion which transforms as 
\begin{equation}\label{eq:basi:jato}
  \begin{aligned}
    M_\bare &\to U_LM_\bare U_R^\dagger\,,
  \end{aligned}
\end{equation}
where $U_{L/R}\in \text{U}(\nf)$.


\subsection{Operator basis}\label{sec:basis}

Following \citere{Bhattacharya:2015rsa}, we introduce an $\nf\times\nf$ flavor
matrix $\Lambda$ which transforms as
\begin{equation}\label{eq:basi:aunt}
  \begin{aligned}
    \Lambda\to U_R\Lambda U_L^\dagger
  \end{aligned}
\end{equation}
under the chiral flavor transformation of \cref{eq:basi:boat}.  The operator
basis is then constructed by requiring chiral flavor invariance while
restricting the operators to be linear in $\Lambda$. This excludes operators
that contain structures such as $\mathrm{Tr}(M_\bare)$,
$\mathrm{Tr}(\Lambda)$, $\mathrm{Tr}(\Lambda (M_\bare^\dagger M_\bare)^{n})$
($n\in\mathbbm{N}$), etc. Such terms would require an additional source of
chiral symmetry breaking, whereas in our setup the quark-mass matrix is taken
to be the only chiral-breaking spurion. After constructing all chirally
covariant structures, we select only operators which are \cp-even by assuming
a real diagonal mass matrix and a real symmetric flavor matrix; explicit
choices of $\Lambda$ will be discussed in \cref{sec:special}. In
\cref{sec:cedm}, it will be shown to what extent our results also apply to the
\cp-odd case.

It is necessary to distinguish \textit{physical} and \textit{unphysical}
operators, characterized by whether they have non-vanishing matrix elements
between physical states or not, respectively.
\footnote{In the classic paper by
Kluberg-Stern and Zuber~\cite{Kluberg-Stern:1975ebk}, they are referred to as
\textit{class-I} and \textit{class-II} operators, respectively.} Let us first
consider the physical operators of mass dimension five and group them
according to powers of the mass matrix.  The \cmdm\ operator is the only one
without a factor of $M_\bare$:
\begin{equation}\label{eq:CMDMbare}
  \mathcal{O}_\text{\abbrev{CM}}^\bare[\Lambda]\equiv
  \mathcal{O}_1^\bare[\Lambda]
    =
    i g_\bare\,
    \bar{\bsym{\psi}}\Lambda\sigma_{\mu\nu}T^a\bsym{\psi}\,
    G_{\mu\nu}^a\,,
\end{equation}
where $\sigma_{\mu\nu}\equiv\frac{i}{2}\left[\gamma_\mu,\gamma_\nu\right]$,
and we have introduced a factor $ig_\bare$ for convenience.  Similarly, there
is only a single operator with one insertion of the mass matrix:
\begin{align}\label{eq:mGG}
  \mathcal{O}_2^\bare[\Lambda]
    &=
    \mathrm{Tr}(\Lambda M_\bare)\,G_{\mu\nu}^aG_{\mu\nu}^a\,.
\end{align}
The physical dimension-five operators with two insertions of the mass matrix
are given by
\begin{gather}
  \begin{aligned}\label{eq:m2FF}
    \mathcal{O}_{3}^\bare[\Lambda]
    &=
    \frac{1}{2}\bar{\bsym{\psi}}
    \left\{M_\bare^2,\Lambda\right\}\bsym{\psi}\,,
    &
    \mathcal{O}_{4}^\bare[\Lambda]
    &=
    \bar{\bsym{\psi}}M_\bare\Lambda M_\bare\bsym{\psi}\,,
    \\
    \mathcal{O}_{5}^\bare[\Lambda]
    &=
    \mathrm{Tr}(M_\bare^2)\,
    \bar{\bsym{\psi}}\Lambda\bsym{\psi}\,,
    &
    \mathcal{O}_{6}^\bare[\Lambda]
    &=
    \mathrm{Tr}(\Lambda M_\bare)\,
    \bar{\bsym{\psi}}M_\bare\bsym{\psi}\,,
\end{aligned}
\end{gather}
while the operators with five powers of $M_\bare$ are all proportional to the
identity,
\begin{gather}
\begin{aligned}\label{eq:m5}
    \mathcal{O}_{7}^\bare[\Lambda]
    &=
    \mathrm{Tr}(\Lambda M_\bare^5)\,\mathbbm{1}\,,
    &
    \mathcal{O}_{8}^\bare[\Lambda]
    &=
    \mathrm{Tr}(\Lambda M_\bare^3)\mathrm{Tr}(M_\bare^2)\,\mathbbm{1}\,,
    \\
    \mathcal{O}_{9}^\bare[\Lambda]
    &=
    \mathrm{Tr}(\Lambda M_\bare)\mathrm{Tr}(M_\bare^4)\,\mathbbm{1}\,,
    &
    \mathcal{O}_{10}^\bare[\Lambda]
    &=
    \mathrm{Tr}(\Lambda M_\bare)\mathrm{Tr}(M_\bare^2)^2\,\mathbbm{1}\,.
\end{aligned}
\end{gather}
For the purpose of this paper, we also need the operators of mass dimension
three and one. The physical set is given by
\begin{gather}
\begin{aligned}
\mathcal{O}_{11}^\bare[\Lambda]\label{eq:OpsDim31}
&=\bar{\bsym{\psi}}\Lambda\bsym{\psi}\equiv
\mathcal{O}_\text{\abbrev{S}}^\bare[\Lambda]\,,
    \\
    \mathcal{O}_{12}^\bare[\Lambda]
   &=
    \mathrm{Tr}(\Lambda M_\bare^3)\,\mathbbm{1}\,,
\\ \mathcal{O}_{13}^\bare[\Lambda]
    &=
    \mathrm{Tr}(\Lambda M_\bare)\mathrm{Tr}(M_\bare^2)\,\mathbbm{1}\,,\\
    \mathcal{O}_{14}^\bare[\Lambda]
    &=
    \mathrm{Tr}(\Lambda M_\bare)\,\mathbbm{1}\,,
\end{aligned}
\end{gather}
where, for later purposes, we introduced the notation
$\mathcal{O}_\text{\abbrev{S}}^\bare[\Lambda]$ for the scalar bilinear operator.

Green's functions containing insertions of these local composite operators
generally exhibit \uv\ divergences which are not removed by the
renormalization of the \qcd\ fields and parameters. For on-shell Green's
functions, these divergences can be absorbed by defining renormalized
operators as suitable linear combinations of the bare operators,
\begin{equation}\label{eq:RenOpsDef}
    \mathcal{O} [\Lambda]= Z\,\mathcal{O}^\bare[\Lambda]\,,
\end{equation}
where $\mathcal{O}[\Lambda]$ is a vector containing the renormalized
operators,
\begin{equation}\label{eq:VecNotation}
    \mathcal{O}[\Lambda]
    \equiv
    \left(
        \mathcal{O}_1[\Lambda],\ldots,\mathcal{O}_{14}[\Lambda]
    \right)^{T}\,,
\end{equation}
and analogously for the bare operators $\mathcal{O}^\bare[\Lambda]$.  For the
renormalization matrix $Z$, we adopt the \msbar\ scheme. Operators of
different mass dimensions\footnote{We include powers of explicit mass factors
in the mass dimension of an operator throughout this paper.} do not mix in
this case, meaning that the renormalization matrix is block diagonal with
respect to the operator dimension,
\begin{equation}\label{eq:RenMatrixStructure}
\renewcommand*{\arraystretch}{1.5}
Z=\begin{pmatrix}Z_{1,1}
& Z_{1\times9}& & \\
    &Z_{9\times9}& & \\
    & & Z_{3\times3} & \\
    & & & Z_{14,14}
    \end{pmatrix}\,,
\end{equation}
where entries that are not shown are understood to vanish. Because of
\cref{eq:basi:krym}, it is $Z_{14,14}=Z_m^{-1}$. The submatrix $Z_{9\times 9}$
renormalizes the mass-dressed dimension-five operators of
\cref{eq:mGG,eq:m2FF,eq:m5}, whereas $Z_{3\times 3}$ renormalizes the
dimension-three operators of \cref{eq:OpsDim31}. Both submatrices can be
expressed to all orders in terms of the \qcd\ $\beta$ function and the
anomalous dimensions of the quark mass and the vacuum
energy~\cite{Spiridonov:1984br,Spiridonov:1988md,Chetyrkin:1994ex}.

The entry $Z_{1,1}$ can be extracted from \citere{Gorbahn:2005sa} through
\nnlo. In \citere{Naterop:2024cfx}, \qcd\ with a single quark flavor was
considered in the pure-\msbar\ calculation. Its parameter counterterms allow
one to reconstruct $Z_{1,2}$, i.e.\ the first element of $Z_{1\times9}$, as
well as a linear combination of $Z_{1,3},\ldots,Z_{1,6}$.

The entry $Z_{1,7}$, describing the mixing of the \cmdm\ operator into the
dimension-five identity operator $\operatorname{Tr}(\Lambda M^5)\mathbbm{1}$,
is of order $\apis$; it arises from two-loop vacuum diagrams and can be
inferred from \citere{Mereghetti:2021nkt}. The three-loop contribution,
corresponding to $\mathcal{O}(\apis^2)$, as well as the mixing into the
remaining independent dimension-five identity operators, $Z_{1,8}$, $Z_{1,9}$,
and $Z_{1,10}$ is---to the best of our knowledge---not yet available in the
literature. These contributions are determined in the present work, along with
all the known results listed above; their reproduction serves as a welcome
check of our calculation.  For completeness, we provide the full
renormalization matrix $Z$, generalized to arbitrary $\nf$ and expressed in
terms of the \msbar\ anomalous dimension matrix, in \cref{sec:gamma}.

The considerations up to this point are sufficient when dealing with physical
matrix elements only. However, since the calculation described in
\cref{sec:zeta} involves off-shell matrix elements, we also need to take into
account unphysical operators. With the shorthand notation
\begin{equation}
  \bar{\bsym{\psi}}_{\text{\abbrev{E}}}\equiv\bar{\bsym{\psi}}
  \Big(-\overleftarrow{\slashed{D}}+M_\bare\Big)\qquad \textrm{and}\qquad
  \bsym{\psi}_{\text{\abbrev{E}}}\equiv
  \Big(\slashed{D}+M_\bare\Big)\bsym{\psi}
\end{equation}
for the equations of motion, the relevant unphysical operators in our case
read
\begin{gather}\label{eq:unphysOps}
\begin{aligned}
    \mathcal{N}_{+}^\bare[\Lambda]
    &=\bar{\bsym{\psi}}\Big\{M_\bare,\Lambda\Big\}\bsym{\psi}_{\text{\abbrev{E}}}+\bar{\bsym{\psi}}_{\text{\abbrev{E}}}\Big\{M_\bare,\Lambda\Big\}\bsym{\psi}\,,
    \\
    \mathcal{N}_{-}^\bare[\Lambda]
    &=\bar{\bsym{\psi}}\Big[M_\bare,\Lambda\Big]\bsym{\psi}_{\text{\abbrev{E}}}-\bar{\bsym{\psi}}_{\text{\abbrev{E}}}\Big[M_\bare,\Lambda\Big]\bsym{\psi}\,,
    \\
    \mathcal{N}_1^\bare[\Lambda]&=\mathrm{Tr}(\Lambda M_\bare)\Big[\bar{\bsym{\psi}}\bsym{\psi}_{\text{\abbrev{E}}}+\bar{\bsym{\psi}}_{\text{\abbrev{E}}}\bsym{\psi}\Big]\,,\\
    \mathcal{N}_{2}^\bare[\Lambda]
    &=\bar{\bsym{\psi}}\Lambda\Big(\slashed{D}-M_\bare\Big)\bsym{\psi}_{\text{\abbrev{E}}}-\bar{\bsym{\psi}}_{\text{\abbrev{E}}}\left(\overleftarrow{\slashed{D}}+M_\bare\right)\Lambda\bsym{\psi}\,,
    \\
    \mathcal{N}_{3}^\bare[\Lambda]
    &=\bar{\bsym{\psi}}M_\bare\Lambda\bsym{\psi}_{\text{\abbrev{E}}}+\bar{\bsym{\psi}}_{\text{\abbrev{E}}}\Lambda M_\bare\bsym{\psi}-\bar{\bsym{\psi}}_{\text{\abbrev{E}}}\Lambda \bsym{\psi}_{\text{\abbrev{E}}}\,,\\
    \mathcal{N}_{4}^\bare[\Lambda]
    &=g_\bare\Big[\bar{\bsym{\psi}}\Lambda \slashed{A}\bsym{\psi}_{\text{\abbrev{E}}}+\bar{\bsym{\psi}}_{\text{\abbrev{E}}}\slashed{A}\Lambda \bsym{\psi}\Big]\,,
    \\
    \mathcal{N}_5^\bare[\Lambda]&
    =\mathrm{Tr}(\Lambda M_\bare)A_\mu^a \frac{\delta S}{\delta A_\mu^a}
     = -\mathrm{Tr}(\Lambda M_\bare)A_\mu^a\left[D_\nu^{ab}G^{b}_{\nu\mu}-g_\bare\bar{\bsym{\psi}}\gamma_\mu T^a\bsym{\psi}\right]+\ldots\,,
\end{aligned}
\end{gather}
where $S=\int\dd^D x\,\mathcal{L}$ is the \qcd\ action, and the ellipsis
denotes ghost and gauge-fixing terms which are irrelevant for the purpose of
this paper. The left-acting derivative is defined as
\begin{equation}\label{eq:basi:gale}
  \begin{aligned}
 \overleftarrow{D}_\mu=\overleftarrow
               {\partial}_\mu-g_\bare A_\mu^aT^a
  \end{aligned}
\end{equation}
and we used $D_\mu$ from \cref{eq:basi:kerf}. All operators of
\cref{eq:unphysOps} vanish between physical states by virtue of the equations
of motion. In addition, we include the following total-derivative operator
into our list of unphysical operators, since eventually we are interested in
zero-momentum insertions of the physical operators only:
\begin{equation}\label{eq:lagr:icao}
  \begin{aligned}
        \mathcal{N}_{6}^\bare[\Lambda]
    &=
    \partial^2
    \left(
        \bar{\bsym{\psi}}\Lambda\bsym{\psi}
    \right)\,.
  \end{aligned}
\end{equation}
Note that our basis is a rotated
version of the operator basis of~\citere{Bhattacharya:2015rsa} (see their
Table~5), supplemented by operators which do not arise for the off-diagonal
choice of the flavor matrix considered there.

According to the Joglekar-Lee theorem~\cite{Joglekar:1975nu,Joglekar:1976eb}
(see also \citere{Collins:1974bg}), the renormalization matrix of the
off-shell operator basis is triangular and reads \begingroup
\renewcommand*{\arraystretch}{1.2}
\begin{equation}\label{eq:Joglekar-Lee}
    \begin{pmatrix}
        \mathcal{O}[\Lambda] \\
        \mathcal{N}[\Lambda]
    \end{pmatrix}
    =
    \begin{pmatrix}
        Z_{\mathcal{O}\mathcal{O}} & Z_{\mathcal{O}\mathcal{N}} \\
        0 & Z_{\mathcal{N}\mathcal{N}}
    \end{pmatrix}
    \begin{pmatrix}
        \mathcal{O}^\bare[\Lambda]\\
        \mathcal{N}^\bare[\Lambda]
    \end{pmatrix}\,,
\end{equation}
\endgroup where $Z_{\mathcal{O}\mathcal{O}}=Z$ is the renormalization matrix
of the physical operators of \cref{eq:RenMatrixStructure}, and we collected
all unphysical operators in the vectors $\mathcal{N}$ and $\mathcal{N}^\bare$.


\subsection{Gradient flow and matching coefficients}

\label{sec:gf}
In the gradient-flow formalism, the usual \qcd\ fields are promoted to flowed
fields by introducing an additional parameter, the flow time $t$.  The
evolution of the flowed fields with respect to the flow time is governed by
the corresponding flow
equations~\cite{Narayanan:2006rf,Luscher:2009eq,Luscher:2010iy,
  Luscher:2011bx,Luscher:2013cpa}
\begin{gather}\begin{aligned}\label{eq:FlowEq}
    0 &= \partial_t B_\mu^a(t) - \mathcal{D}_\nu^{ab}(t)
    \mathcal{G}_{\nu\mu}^b(t) - \kappa\, \mathcal{D}_\mu^{ab}(t) \partial_\nu
    B_\nu^b(t)\,, \\0&= \partial_t \bsym {\chi}(t) - \Delta(t)
    \bsym{\chi}(t) + \kappa g_\bare\, \partial_\mu B_\mu^a(t) T^a
    \bsym{\chi}(t)\,,\\ 0&= \partial_t \bar{\bsym {\chi}}(t) -
    \bar{\bsym{\chi}}(t)\overleftarrow{\Delta}(t) - \kappa
    g_\bare\,\bar{\bsym{\chi}}(t)\partial_\mu B_\mu^a(t) T^a \,.
\end{aligned} \end{gather}
Here, $\kappa$ denotes a gauge parameter~\cite{Zwanziger:1981kg}, which we set
to $\kappa=1$ for simplicity. Physical observables, however, are independent
of this choice.  $B_\mu^a(t)\equiv B_\mu^a(t,x)$ denotes the flowed gauge
field, while $\bsym{\chi}(t)\equiv\bsym{\chi}(t,x)$ collects the
flowed fermion fields,
\begin{equation}\label{eq:SpinorVec2}
  \bsym{\chi}(t)
  =
  \left(
  \chi_1(t),\ldots,\chi_{\nf}(t)
  \right)^T\,.
\end{equation}
Furthermore,
\begin{equation}\label{eq:ConvUnf}
    \mathcal{D}^{ab}_\mu(t)=\delta^{ab}\partial_\mu-g_\bare f^{abc} B_\mu^c(t)
\end{equation}
is the flowed covariant derivative in the adjoint representation.  In the same
way, $\mathcal{G}_{\mu\nu}^a(t)$ and $\mathcal{D}_\mu(t)$ denote the flowed
field-strength tensor and the flowed covariant derivative in the fundamental
representation, respectively.  They are obtained from their unflowed
counterparts, given in \cref{eq:unfConv1,eq:basi:kerf}, by replacing
$A_\mu^a\rightarrow B_\mu^a(t)$. The flowed Laplace operator is given by
$\Delta(t)=\mathcal{D}_\mu(t)\mathcal{D}_\mu(t)$. A key requirement of the
gradient flow is that the flowed fields reduce to the ordinary \qcd\ fields at
vanishing flow time,
\begin{align}\label{eq:BC}
    B_\mu^a(t=0)\equiv A_\mu^a\qquad \textrm{and} \qquad \bsym{\chi}(t=0)\equiv\bsym{\psi}\,.
\end{align}
The perturbative solution of the flow equation leads to a new set of Feynman
rules.  They contain exponential damping factors for the quark and gluon
propagators, as well as so-called flow lines, which govern the evolution of
the fields in the flow time.  The flow lines couple to the fields through flow
vertices. For a more detailed discussion of the basic properties and technical
aspects of the gradient flow, we refer to \citeres{Luscher:2010iy,
  Luscher:2011bx, Artz:2019bpr}.

Using the flowed fields, we can define flowed bare counterparts
$\tilde{\mathcal{O}}_i^\bare[t;\Lambda]$ of the operators defined in
\cref{sec:basis} by replacing $\bsym{\psi}\to\bsym{\chi}(t)$ and
$A_\mu^a\to B_\mu^{a}(t)$ in $\mathcal{O}_i^\bare[\Lambda]$.  An
important property of the gradient flow is that composite operators built from
flowed fields do not mix under
renormalization~\cite{Luscher:2011bx,Luscher:2013cpa,Hieda:2016xpq}.
Consequently, we can write the renormalized flowed operators as
\begin{equation}
  \label{eq:flowedOpsRen}
    \tilde{\mathcal{O}}[t;\Lambda] =
    \mathcal{Z}(t)\tilde{\mathcal{O}}^{\bare}[t;\Lambda]\,,
\end{equation}
with a diagonal matrix $\mathcal{Z}(t)$; its entries are determined entirely
by \qcd\ parameter- and flowed field-renormalization constants.  When employing
the \msbar\ scheme for the flowed fields, $\mathcal{Z}$ is independent of $t$.
In the following, however, we employ the ringed scheme for the flowed fermion
fields~\cite{Makino:2014taa}, where their renormalization is given by
\begin{gather}
  \begin{aligned}
    \label{eq:ParRen}
    \mathring{\bsym{\chi}}(t)
    &=
    \mathring{Z}_\chi^{1/2}(t)\bsym{\chi}(t)
    =
    \zeta_\chi^{1/2}(t)Z_\chi^{1/2}\bsym{\chi}(t)\,.
  \end{aligned}
\end{gather}
The \msbar\ flowed-fermion renormalization constant $Z_\chi$ as well as the
conversion factor to the ringed scheme $\zeta_\chi$ are given in
\cref{app:rg-functions}. In our normalization of the flowed gluon field
$B_\mu^a$, its renormalization factor is identical to the strong coupling
renormalization $Z_\text{s}$~\cite{Luscher:2011bx}. The \uv-finite
renormalized flowed operators are then obtained from \cref{eq:flowedOpsRen},
with
\begin{equation}\label{eq:CursZdef}
    \mathcal{Z}(t)
    =
    \mathrm{diag}\!\left(
        \mathring{Z}_\chi(t),\,
        Z_\text{s} Z_m^{-1},\,
        \left(\mathring{Z}_\chi(t) Z_m^{-2}\right)^{\times 4},\,
        \left(Z_m^{-5}\right)^{\times 4},\,
        \mathring{Z}_\chi(t),\,
        \left(Z_m^{-3}\right)^{\times 2},\,
        Z_m^{-1}
    \right)\,,
\end{equation}
where $(X)^{\times n}$ denotes $n$ consecutive occurrences of the diagonal
entry $X$.

The flowed operators can now be expressed in terms of the unflowed operators
by means of the so-called \sftx,
\begin{equation}\label{eq:SFTX}
  \tilde{\mathcal{O}}[t;\Lambda]=\zeta^\bare(t)\mathcal{O}^\bare[\Lambda]
  \equiv \zeta(t)\mathcal{O}[\Lambda]\,,
\end{equation}
where we have restricted ourselves to the physical sector of the operators and
used the vector notation of \cref{eq:VecNotation} also for the flowed
operators.  The quantities $\zeta^\bare(t)$ and $\zeta(t)$ are the bare and
renormalized matching matrices, respectively, related through
\begin{equation}\label{eq:RenMixingMatrix}
    \zeta(t)=\zeta^\bare(t)Z^{-1}\,,
\end{equation}
with the renormalization matrix $Z$ defined in
\cref{eq:RenMatrixStructure}. Note that the triangular form of the matrix in
\cref{eq:Joglekar-Lee} prevents contributions from the unphysical sector on
the \rhs\ of this relation.

The elements of the matching matrix are also referred to as matching
coefficients in this paper. For a consistent conversion between the flowed and
the unflowed operator basis, their $t$-dependence is bound to be
\begin{equation}\label{eq:lagr:alec}
  \begin{aligned}
    \zeta_{ij}(t) &\sim t^{(d_j-d_i)/2}\,,
  \end{aligned}
\end{equation}
modulo logarithmic terms, where $d_i$ and $d_j$ are the mass dimensions of the
corresponding operators. For example, the $\msbar$/gradient-flow conversion of
the \cmdm\ operator involves $1/t^2$ and $1/t$ terms due to the contributions
of the lower-dimensional unflowed operators.  On the other hand, the only
relevant terms of order $t$ arise between the flowed dimension-\three\ quark
bilinear operator
$\tilde{\mathcal{O}}_{11}=\tilde{\mathcal{O}}_\text{\abbrev{S}}$ and the
dimension-\five\ operators.  In the context of a fixed operator basis as
considered here, the term \textit{short-flow-time expansion} is thus actually
misleading, since \cref{eq:SFTX}, together with \cref{eq:lagr:alec}, is exact
in $t$~\cite{Harlander:2026lte}.

Note that, due to the linearity in $\Lambda$, for any two flavor matrices
$\Lambda_1,\Lambda_2$, we have
\begin{gather}
\begin{aligned}
    \tilde{\mathcal{O}}[t;\Lambda_1+\Lambda_2]
    &=
    \zeta(t;\Lambda_1+\Lambda_2)\mathcal{O}[\Lambda_1]
    +
    \zeta(t;\Lambda_1+\Lambda_2)\mathcal{O}[\Lambda_2]\,,\\
    \tilde{\mathcal{O}}[t;\Lambda_1+\Lambda_2]
    &=
    \tilde{\mathcal{O}}[t;\Lambda_1]
    +
    \tilde{\mathcal{O}}[t;\Lambda_2]
    =
    \zeta(t;\Lambda_1)\mathcal{O}[\Lambda_1]
    +
    \zeta(t;\Lambda_2)\mathcal{O}[\Lambda_2]\,.
\end{aligned}
\end{gather}
Matching the independent flavor and mass structures of the complete operator
basis then gives
\begin{equation}
    \zeta(t;\Lambda_1+\Lambda_2)
    =
    \zeta(t;\Lambda_1)
    =
    \zeta(t;\Lambda_2) \equiv \zeta(t)\,,
\end{equation}
demonstrating that the matching matrix is independent of the flavor spurion
$\Lambda$.  It can be calculated by choosing any fixed flavor matrix
$\Lambda_\star$ as long as all operators remain linearly independent with this
choice. For example, while
$\Lambda_{ij}=\delta_{ip}\delta_{jq}+\delta_{iq}\delta_{jp}$ for $p\neq q$
might seem a particularly convenient choice, it implies that all operators
proportional to $\mathrm{Tr}(\Lambda M_\bare)$ vanish; similarly, the singlet
choice $\Lambda_{ij}=\delta_{ij}$ introduces the relation
$\mathcal{O}^\bare_3=\mathcal{O}^\bare_4$.

A convenient choice that retains access to all matching coefficients is
\begin{equation}\label{eq:LamStar}
(\Lambda_\star)_{ij}
  =
  \delta_{ip}\delta_{jp}
  +
  \delta_{ip}\delta_{jq}+\delta_{iq}\delta_{jp}\,,
\end{equation}
with generic but fixed $p,q\in\{1,\ldots,\nf\}$ and $p\neq q$.  With this
choice, the full matching matrix of the general-$\Lambda$ expansion can be
determined uniquely from a single matching calculation.


\begin{table}[!ht]
    \centering
    \caption{Structure and non-vanishing entries of the matching
      matrix. Left$|$right markers indicate the
      $\abbrev{\textrm{NLO}}|\abbrev{\textrm{NNLO}}$ coefficients. A
      $\checkmark$ denotes known results (see \cref{tab:OverviewOfCalcs}),
      while $\bsym{\times}$ marks new results obtained in this work. Empty
      positions indicate vanishing contributions at the respective order,
      entirely empty cells indicate coefficients vanishing to all orders, and
      a vertical bar alone indicates contributions that vanish through
      NNLO. The field content of the unflowed operators is shown above the
      columns. Identical uppercase letters or colors indicate identical
      entries.}
    \label{tab:sftx_structure}
    \setlength{\arrayrulewidth}{0.8pt}
    \renewcommand{\arraystretch}{1.5}
    \setlength{\tabcolsep}{4.5pt}
    \resizebox{\textwidth}{!}{%
    \begin{tabular}{
        |c||
        c|c|c|c|c|c|c|c|c|c||
        c|c|c||
        c|
    }
        \hline
        & \multicolumn{1}{c|}{$\bsym{\psi}^2 G$}
& \multicolumn{1}{c|}{$M_\bare G^2$}
& \multicolumn{4}{c|}{$M^2_\bare\bsym{\psi}^2$}
& \multicolumn{4}{c||}{$M_\bare^5 \mathbbm 1$}
& \multicolumn{1}{c|}{$\bsym{\psi}^2$}
& \multicolumn{2}{c||}{$M_\bare^3 \mathbbm 1$}
        & \multicolumn{1}{c|}{$M_\bare\mathbbm 1$}
\\
\cline{2-15}
        &
        $\mathcal O_\text{\abbrev{CM}}$
        & $\mathcal O_2$
        & $\mathcal O_3$
        & $\mathcal O_4$
        & $\mathcal O_5$
        & $\mathcal O_6$
        & $\mathcal O_7$
        & $\mathcal O_8$
        & $\mathcal O_9$
        & $\mathcal O_{10}$
        & $\mathcal O_\text{\abbrev{S}}$
        & $\mathcal O_{12}$
        & $\mathcal O_{13}$
        & $\mathcal O_{14}$
        \\
        \hline\hline

        $\tilde{\mathcal O}_\text{\abbrev{CM}}$
        & $\checkmark|\checkmark$ & $\checkmark|\bsym{\times}$ &$\checkmark|\bsym{\times}$ & $\textcolor{white}{\checkmark}|\bsym{\times}$ & $\textcolor{white}{\checkmark}|\bsym{\times}$ & $\textcolor{white}{\checkmark}|\bsym{\times}$
        & $\checkmark|\bsym{\times}$ & $\textcolor{white}{\checkmark}|\bsym{\times}$ & $\textcolor{white}{\checkmark}|\bsym{\times}$ & $\textcolor{white}{\checkmark}|\textcolor{white}{\bsym{\times}}$
        & $\checkmark|\checkmark$ & $\checkmark|\bsym{\times}$ & $\textcolor{white}{\checkmark}|\bsym{\times}$ & $\checkmark|\bsym{\times}$
        \\
        \hline

        $\tilde{\mathcal O}_2$
        & & $\checkmark|\checkmark$ & & & & $\checkmark|\checkmark$
        & & & $\checkmark|\checkmark$ & $\textcolor{white}{\checkmark}|\checkmark$
        & & & $\checkmark|\checkmark$ & $\checkmark|\checkmark$
        \\
        \hline

        $\tilde{\mathcal O}_3$
        & & & $\cellcolor{green!20}\checkmark|\checkmark\eqmark{A}$ & & &
        & $\cellcolor{orange!20}\checkmark|\checkmark\eqmark{B}$ &$\cellcolor{violet!20}\textcolor{violet!20}{\checkmark}|\checkmark\eqmark{C}$ & &
        & & $\cellcolor{blue!20}\checkmark|\checkmark\eqmark{D}$& &
        \\
        \hline

        $\tilde{\mathcal O}_4$
        & & & & $\cellcolor{green!20}\checkmark|\checkmark\eqmark{A}$ &
        & &$\cellcolor{orange!20} \checkmark|\checkmark\eqmark{B}$ & $\cellcolor{violet!20}\textcolor{violet!20}{\checkmark}|\checkmark\eqmark{C}$ &
        & & & $\cellcolor{blue!20}\checkmark|\checkmark\eqmark{D}$ & & 
        \\
        \hline

        $\tilde{\mathcal O}_5$
        & & & & & $\cellcolor{green!20}\checkmark|\checkmark\eqmark{A}$
        & & & $\cellcolor{orange!20}\checkmark|\checkmark\eqmark{B}$ & &$\cellcolor{violet!20}\textcolor{violet!20}{\checkmark}|\checkmark\eqmark{C}$
        & & & $\cellcolor{blue!20}\checkmark|\checkmark\eqmark{D}$ &
        \\
        \hline

        $\tilde{\mathcal O}_6$
        & & & & & & $\cellcolor{green!20}\checkmark|\checkmark\eqmark{A}$
        & & & $\cellcolor{orange!20}\checkmark|\checkmark\eqmark{B}$ & $\cellcolor{violet!20}\textcolor{violet!20}{\checkmark}|\checkmark\eqmark{C}$
        & & & $\cellcolor{blue!20}\checkmark|\checkmark\eqmark{D}$ &
        \\
        \hline

        $\tilde{\mathcal O}_7$
        & & & & & &
        & $1$ & & &
        & & & &
        \\
        \hline

        $\tilde{\mathcal O}_8$
        & & & & & &
        & & $1$ &
        & & & & & 
        \\
        \hline

        $\tilde{\mathcal O}_9$
        & & & & & &
        & & & $1$
        & & & & &
        \\
        \hline

        $\tilde{\mathcal O}_{10}$
        & & & & & &
        & & & & $1$
        & & & &
        \\
        \hline\hline

        $\tilde{\mathcal O}_\text{\abbrev{S}}$
       & $\bsym{\times}|\bsym{\times}$ &$\bsym{\times}|\bsym{\times}$ & $\bsym{\times}|\bsym{\times}$&$\bsym{\times}|\bsym{\times}$ & $\textcolor{white}{\bsym{\times}}|\bsym{\times}$
        &$\textcolor{white}{\bsym{\times}}|\bsym{\times}$ & $\bsym{\times}|\bsym{\times}$ & $\textcolor{white}{\bsym{\times}}|\bsym{\times}$ & $\textcolor{white}{\bsym{\times}}|\bsym{\times}$ & $|$ & $\cellcolor{green!20}\checkmark|\checkmark\eqmark{A}$ & $\cellcolor{orange!20}\checkmark|\checkmark\eqmark{B}$ & $\cellcolor{violet!20}\textcolor{violet!20}{\checkmark}|\checkmark\eqmark{C}$ & $\cellcolor{blue!20}\checkmark|\checkmark\eqmark{D}$
        \\
        \hline

        $\tilde{\mathcal O}_{12}$
        & & & & & &
        & & & &
        & & $1$ & &
        \\
        \hline

        $\tilde{\mathcal O}_{13}$
        & & & & & &
        & & & &
        & & & $1$ &
        \\
        \hline\hline

        $\tilde{\mathcal O}_{14}$
        & & & & & &
        & & & &
        & & & & $1$
        \\
        \hline
    \end{tabular}%
    }
\end{table}


An overview of the required matching coefficients is given in
\cref{tab:sftx_structure,tab:OverviewOfCalcs}. It reflects the fact that the
conversion of lower-dimensional operators has already been studied
extensively in the literature, as mentioned in the introduction of this paper.
Despite the fact that these earlier works were mostly carried out for specific
choices of the flavor matrix, the corresponding matching coefficients can be
generalized to an arbitrary flavor matrix by using the arguments
above.\footnote{For example, \citere{Borgulat:2023xml} considers $\nh$
degenerate quarks with mass $m$ and $\nl=\nf-\nh$ massless quarks, and
computes the \msbar-\gf\ conversion of the singlet bilinear, in which the
operator $m^3$ appears. The matching coefficients of the general operators
$\mathrm{Tr}(\Lambda M^3)$ and $\mathrm{Tr}(\Lambda M)\mathrm{Tr}(M^2)$ can
then be disentangled by separating the contributions $\sim\nh$ from those
$\sim\nh^2$.}  The goal of the present work is to determine the missing
entries of the matching matrix through \nnlo.  These include, in particular,
the contributions of the mass-dependent operators to the \sftx\ of the
\cmdm\ operator, as well as the contributions of the
dimension-\five\ operators to the \sftx\ of the scalar quark bilinear.

To refer to the \sftx\ of an individual flowed operator, we introduce the
vectors
\begin{equation}\label{eq:ZetaVec}
    \zeta_i^T(t)=e_i^T\zeta(t)\,,
    \qquad
    \textrm{such that}
    \qquad
    \tilde{\mathcal{O}}_i[t;\Lambda]=\zeta_i^T(t)\mathcal{O}[\Lambda]\,,
\end{equation}
where $e_i$ denotes the $i$th unit vector and $i\in\{1,\ldots,14\}$.  
As mentioned above, the \sftx\ of the \cmdm\ operator and the scalar bilinear are of particular interest.  
We therefore introduce the more suggestive notation
\begin{equation}\label{eq:AbbrevSFTX}
    \zeta_{\text{\abbrev{CM}}}(t)\equiv\zeta_1(t)\,,
    \qquad
    \textrm{and}
    \qquad
    \zeta_{\text{\abbrev{S}}}(t)\equiv\zeta_{11}(t)\,.
\end{equation}
Since the matching coefficients are independent of $\Lambda$, the \sftx\ of
the bilinears dressed with two mass matrices, i.e.~$\tilde{\mathcal{O}}_k$
with $k\in\{3,4,5,6\}$, can be derived directly from the \sftx\ of the
dimension-three bilinear $\tilde{\mathcal{O}}_{\text{\abbrev{S}}}$.  This
follows from the relations
\begin{gather}
  \begin{aligned}\label{eq:3-6_11}
    \tilde{\mathcal{O}}_{3}[t;\Lambda] &=
    \frac{1}{2}\tilde{\mathcal{O}}_{\text{\abbrev{S}}}[t;\{M^2,\Lambda\}]\,, &
    \tilde{\mathcal{O}}_{4}[t;\Lambda] &=
    \tilde{\mathcal{O}}_{\text{\abbrev{S}}}[t;M\Lambda M]\,,
    \\ \tilde{\mathcal{O}}_{5}[t;\Lambda] &=
    \mathrm{Tr}\left(M^2\right)
    \tilde{\mathcal{O}}_{\text{\abbrev{S}}}[t;\Lambda]\,, &
    \tilde{\mathcal{O}}_{6}[t;\Lambda] &= \mathrm{Tr}\left(\Lambda
    M\right)\tilde{\mathcal{O}}_{\text{\abbrev{S}}}[t;M]\,.
\end{aligned}
\end{gather}
These identities imply the following relations between the non-vanishing
entries of $\zeta_{\text{\abbrev{S}}}$ and those of $\zeta_3$, $\zeta_4$, $\zeta_5$, and
$\zeta_6$:
\begin{equation}\label{eq:lagr:huff}
  \begin{alignedat}{4}
    (\zeta_3)_3 &= 
    (\zeta_4)_4 &&= 
    (\zeta_5)_5 &&= 
    (\zeta_6)_6 &&= 
    (\zeta_\text{\abbrev{S}})_{11}\,,\\
    (\zeta_3)_7 &= 
    (\zeta_4)_7 &&= 
    (\zeta_5)_8 &&= 
    (\zeta_6)_9 &&= 
    (\zeta_\text{\abbrev{S}})_{12}\,,\\
    (\zeta_3)_8 &= 
    (\zeta_4)_8 &&= 
    (\zeta_5)_{10} &&= 
    (\zeta_6)_{10} &&= 
    (\zeta_\text{\abbrev{S}})_{13}\,,\\
    (\zeta_3)_{12} &= 
    (\zeta_4)_{12} &&= 
    (\zeta_5)_{13} &&= 
    (\zeta_6)_{13} &&= 
    (\zeta_\text{\abbrev{S}})_{14}\,.
  \end{alignedat}
\end{equation}
This is also indicated in \cref{tab:sftx_structure}. The relevant coefficients
of $\zeta_\text{\abbrev{S}}$ are given in
\cref{eq:zetaS11,eq:zetaS12,eq:zetaS13,eq:zetaS14}.


\begin{table}[h!]
    \centering
    \caption{Perturbative status of the matching coefficients through
      $\mathcal{O}(\apis^2)$ from the literature, labeled according to \cref{eq:ZetaVec,eq:AbbrevSFTX}. All other non-vanishing entries shown in
      \cref{tab:sftx_structure} are, to the best of our knowledge,
      calculated here for the first time.}
    \label{tab:OverviewOfCalcs}
    \renewcommand{\arraystretch}{1.5}
    \begin{tabular}{||l|c|c||}
    \hline
         entry & $\mathcal{O}\left(\apis\right)$ & $\mathcal{O}\left(\apis^2\right)$  \\\hline
         \multicolumn{3}{|c|}{\itshape \sftx\ of the \cmdm\ operator (row 1)}\\\hline
        $\left(\zeta_{\text{\abbrev{CM}}}\right)_1$ & \citere{Mereghetti:2021nkt} & \citere{Harlander:2022vgf}\\
        $\left(\zeta_{\text{\abbrev{CM}}}\right)_{2,3,7,12,14}$ & \citere{Mereghetti:2021nkt} & ---\\
        $\left(\zeta_{\text{\abbrev{CM}}}\right)_{11}$ & \citere{Mereghetti:2021nkt} & \citere{Harlander:2022vgf}\\\hline
        \multicolumn{3}{|c|}{\itshape \sftx\ of the mass-dressed action density (row 2)}\\\hline
        $(\zeta_{2})_{2,6}$ & \citeres{Makino:2014taa,Suzuki:2013gza} & \citeres{Harlander:2018zpi, Harlander:2020duo}\\
        $(\zeta_{2})_{9,10,13}$ & \citeres{Harlander:2016vzb,Harlander:2020duo} & \citere{Harlander:2020duo}\\
        $(\zeta_{2})_{14}$ & \citere{Luscher:2010iy} & \citeres{Harlander:2016vzb,Artz:2019bpr}\\\hline
        \multicolumn{3}{|c|}{\itshape \sftx\ of the scalar bilinear (row 11)}\\\hline
        $(\zeta_{\text{\abbrev{S}}})_{11}$ & \citere{Makino:2014taa} & \citeres{Harlander:2022vgf,Borgulat:2023xml}\\
         $(\zeta_{\text{\abbrev{S}}})_{12,13}$ & \citere{Borgulat:2023xml} & \citere{Borgulat:2023xml}\\
        $(\zeta_{\text{\abbrev{S}}})_{14}$ & \citere{Makino:2014taa} & \citere{Artz:2019bpr}\\\hline
    \end{tabular}
\end{table}



\subsection{Calculating the matching matrix}
\label{sec:zeta}
The bare matching matrix $\zeta^{\bare}(t)$ can be determined using the method
of projectors~\cite{Gorishnii:1983su,Harlander:2018zpi}. Schematically, for
each operator $\mathcal{O}_i^\bare$ we define a projector\footnote{Here,
``amp'' refers to the amputation of the unflowed \qcd\ external
propagators. If an external leg with momentum $p$ is attached to a vertex or
operator at non-zero flow time $t$, the corresponding flowed exponential
factor $e^{-tp^2}$ therefore remains after amputation.}
\begin{equation}\label{eq:ProjStructure}
P_i[X]
\equiv\mathcal{P}_i(\partial_k,\partial_m)
\langle i \vert \Gamma_i X \vert 0 \rangle^{\mathrm{1PI,\,amp}}_{k,m_\bare=0}\,,
\end{equation}
where $\mathcal{P}_i(x,y)$ is a multivariate polynomial and $\Gamma_i$ is a
matrix in Dirac and color space. These projectors are constructed such that
\begin{equation}\label{eq:lagr:etta}
  \begin{aligned}
    P_i\left[\mathcal{O}_i^\bare[\Lambda_\star]\right]\Big|_\text{tree}
    \neq 0\quad \forall i\,.
  \end{aligned}
\end{equation}
Their definition is not unique; the specific choice used in our calculation is
given in \cref{sec:Projectors}. The masses and external momenta are set to
zero after differentiation, but before performing the loop integrations.

Note that at this point we do not require the projectors $P_i$ to be
orthogonal in the sense that $P_i[\mathcal{O}_j^\bare] \sim \delta_{ij}$.  In
fact, since \cref{eq:ProjStructure} generally involves an off-shell matrix
element with a non-vanishing momentum insertion at the operator, we may even
have $P_i[\mathcal{N}^\bare_j]\neq 0$, where $\mathcal{N}^\bare_j$ is one of
the unphysical operators of \cref{eq:unphysOps}. To remove these
contributions, we also construct projectors $P_{\mathcal{N}_i}$ onto the
unphysical operators; their explicit form is again given in
\cref{sec:Projectors}.  We then determine orthogonal linear combinations
$P_i^\perp$ of the $P_i$ and the $P_{\mathcal{N}_i}$ such that
\begin{equation}\label{eq:ProjDef}
  \begin{alignedat}{2}
  P_i^{\perp}\left[\mathcal{O}_j^\bare[\Lambda_\star]\right]
  &= \delta_{ij}\,,\qquad
  P_i^{\perp}\left[\mathcal{N}_j^\bare[\Lambda_\star]\right]
  = 0\,.
  \end{alignedat}
\end{equation}
We find
\begin{gather}
\begin{aligned}\label{eq:OrthProj}
    P_{1}^{\perp}
    &=P_{1}
-P_{\mathcal{N}_{2}}
+\frac{1}{2}P_{\mathcal{N}_{3}}
+P_{\mathcal{N}_{4}}\,,
    \\
    P_{3}^{\perp}
    &=P_{3}
-P_{4}
-P_{5}
-P_{6}
-4P_{\mathcal{N}_{+}}
+2P_{\mathcal{N}_{-}}
+2P_{\mathcal{N}_{1}}
-2P_{\mathcal{N}_{2}}
+P_{\mathcal{N}_{3}}
+2P_{\mathcal{N}_{4}}
     \,,
    \\
    P_{4}^{\perp}
    &=P_{4}
-2P_{\mathcal{N}_{-}}
-P_{\mathcal{N}_{3}}\,,\\
    P_{6}^{\perp}
    &=P_{6}
-2P_{\mathcal{N}_{1}}\,,\\
P_{7}^{\perp}
    &=P_{7}
-P_{8}
-P_{9}
+2P_{10}\,,\\
P_{8}^{\perp}
    &=P_{8}
-2P_{10}\,,\\
P_{9}^{\perp}
    &=P_{9}
-P_{10}\,,\\
    P_{12}^{\perp}
    &=P_{12}
-P_{13}\,,\\
P_{i}^{\perp}
    &=
    P_i\,,
    \qquad \text{otherwise}.
\end{aligned}
\end{gather}
All our projectors automatically satisfy
$P_i^\mathrm{\perp}[\mathcal{N}_5]=P_i^\mathrm{\perp}[\mathcal{N}_6]=0$ since
we include the antisymmetric Dirac structure $\sigma_{\mu\nu}$ in
$P_{\mathcal{N}_3}$ and the derivative \wrt\ both momenta in $P_{2}$; see
\cref{sec:Projectors}. Since the unphysical operators enter the physical
matching coefficients only through this orthogonalization procedure,
$\mathcal{N}_5$ and $\mathcal{N}_6$ can be omitted from the basis in our
specific case.

The entries of the renormalized matching matrix can now be calculated by
applying the orthogonal projectors to \cref{eq:SFTX} and subsequently
renormalizing the result according to \cref{eq:RenMixingMatrix},
\begin{equation}
  \zeta(t)=\left(\sum_jP_{j}^{\perp}
  \left[\tilde{\mathcal{O}}[t;\Lambda_\star]\right]e^{T}_j\right)Z^{-1}\,.
\end{equation}

We have calculated the complete matching matrix in the general-flavor case and
explicitly verified that, upon taking the corresponding flavor and mass
limits, all matching coefficients previously known in the literature (see
\cref{tab:OverviewOfCalcs}) are reproduced.

To perform these calculations, we use the tool chain described in \citeres{
  Artz:2019bpr,Borgulat:2025gys, Lange:2021vqg}.  It consists of
\texttt{qgraf} \cite{Nogueira:1991ex} for the generation of Feynman diagrams,
\texttt{tapir}/\texttt{exp} \cite{Gerlach:2022qnc,Harlander:1998cmq} for
identifying momentum topologies and factorizing the color and Lorentz
structures, and subsequently
\texttt{FORM}~\cite{Vermaseren:2000nd,Kuipers:2012rf} routines to carry out
algebraic operations such as applying derivatives, evaluating Dirac traces,
and simplifying tensor structures.

Using integration-by-parts identities adapted to the gradient-flow
framework~\cite{Harlander:2021esn}, we reduce the resulting integrals to
master integrals with the help of
\texttt{Kira}/\texttt{FireFly}~\cite{Maierhofer:2018gpa,
  Klappert:2020nbg,Klappert:2019emp}.  The corresponding \two-loop master
integrals were calculated analytically in
\citeres{Artz:2019bpr,Harlander:2018zpi}. At \three-loop level, we evaluate
the required master integrals numerically using the program
\texttt{ftint}~\cite{Harlander:2024vmn}, which employs
\texttt{pySecDec}~\cite{Borowka:2017idc}.
\newcolumntype{C}[1]{>{\centering\arraybackslash}m{#1}}
\begin{table}[h]
    \centering
    \caption{Total number of \nnlo\ diagrams for the projectors (before
      orthogonalization) acting on $\tilde{\mathcal{O}}_\text{\abbrev{CM}}$
      (third column) and $\tilde{\mathcal{O}}_\text{\abbrev{S}}$ (fourth
      column).  Diagrams without external legs are three-loop diagrams,
      whereas those with external legs are two-loop diagrams.  Our calculation
      uses three distinct flavors $p$, $q$, and $r$ explicitly appearing in
      the projectors and one additional spectator flavor. The results are
      subsequently generalized to arbitrary $\nf$; see \cref{sec:Projectors}
      for further details.}
    \label{tab:diagrams}
    \setlength{\tabcolsep}{8pt}
    \renewcommand{\arraystretch}{1.3}
    \begin{tabular}{||c|c|C{3cm}|C{3cm}||}
    \hline
    \multirow{2}{*}{\textbf{External legs}}
    & \multirow{2}{*}{\textbf{Projectors}}
    & \multicolumn{2}{c||}{%
        \shortstack{\textbf{\# Diagrams}}}
    \\
    \cline{3-4}
    &
    & $\tilde{\mathcal{O}}_\text{\abbrev{CM}}[t;\Lambda_\star]$
    & $\tilde{\mathcal{O}}_\text{\abbrev{S}}[t;\Lambda_\star]$
    \\
    \hline\hline

        $\varnothing$
        & $P_{7,\ldots,10}$, $P_{12,\ldots,14}$
        & $137+20\,\nf$
        & $218+40\,\nf$
        \\

        $\bar\psi_p$, $\psi_p$
        & $P_3$, $P_5$, $P_{\mathcal N_+}$,$P_{\mathcal N_3}$
        & $197+25\,\nf$
        & $325+50\,\nf$
        \\
        
        $\bar\psi_p$, $\psi_q$
        & $P_4$, $P_{11}$, $P_{\mathcal N_-}$
        & $189+25\,\nf$
        & $305+50\,\nf$
        \\

        $\bar\psi_r$, $\psi_r$
        & $P_6$, $P_{\mathcal{N}_1}$
        & $8$
        & $16$
        \\

        $A$, $A$
        & $P_2$
        & $497$
        & $772$
        \\

        $\bar\psi$, $\psi$, $A$
        & $P_1$, $P_{\mathcal N_4}$
        & $2718+371\,\nf$
        & $4156+672\,\nf$
        \\

        $\bar\psi$, $\psi$, $A$, $A$
        & $P_{\mathcal N_2}$
        & $43531+6335\,\nf$
        & $63064+10704\,\nf$\\
        
        \hline
    \end{tabular}
\end{table}



\section{Results}\label{sec:Results}


\subsection{Matching coefficients of the \abbrev{CMDM} operator}
\label{sec:sftx}
\begin{figure}[t]
  \centering
  \setlength{\tabcolsep}{6pt} 
  \begin{tabular}{@{}c c c@{}}
    \begin{subfigure}[t]{0.25\textwidth}
      \centering
      \includegraphics[width=\linewidth]{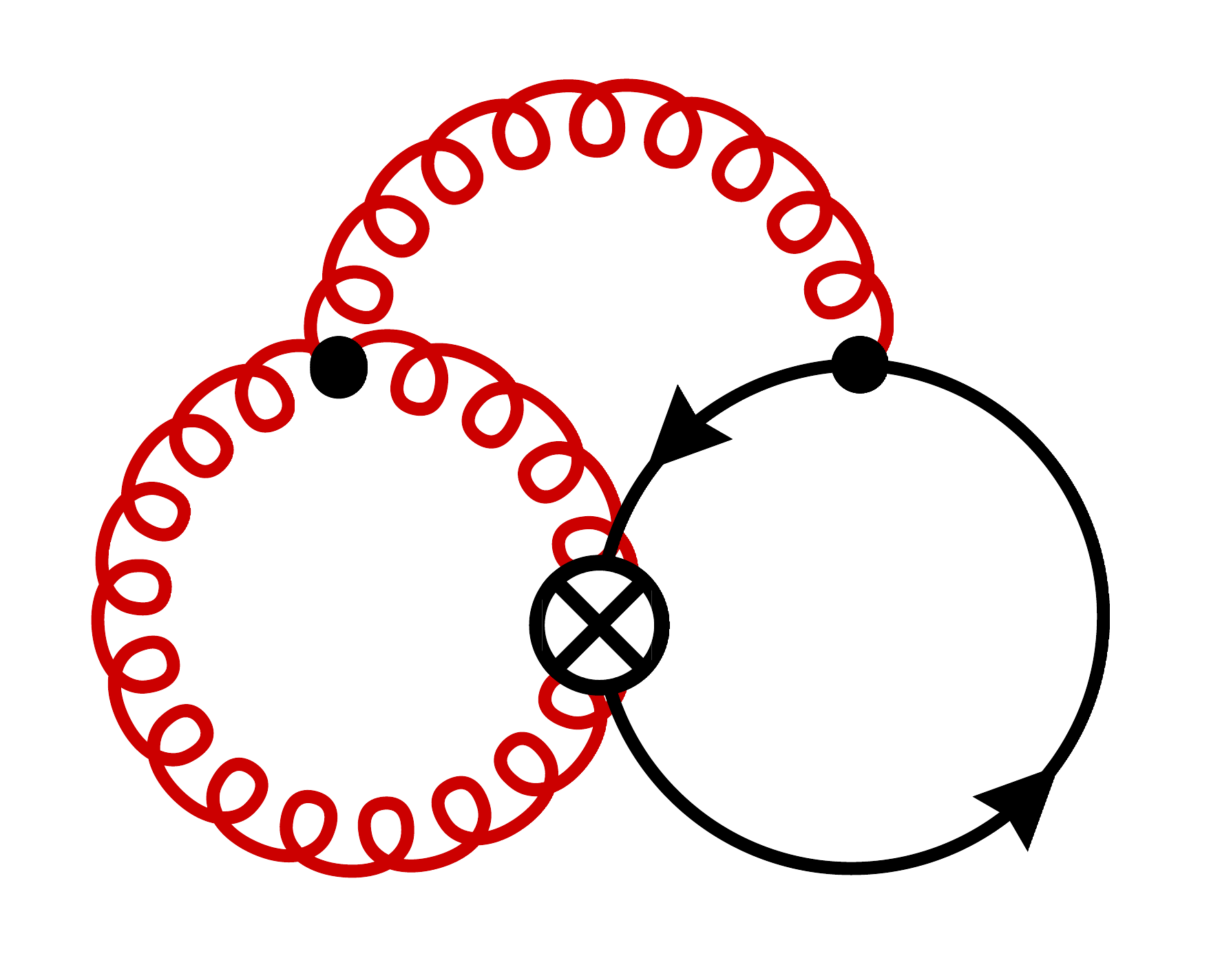}
    \end{subfigure} &
    \begin{subfigure}[t]{0.25\textwidth}
      \centering
      \includegraphics[width=\linewidth]{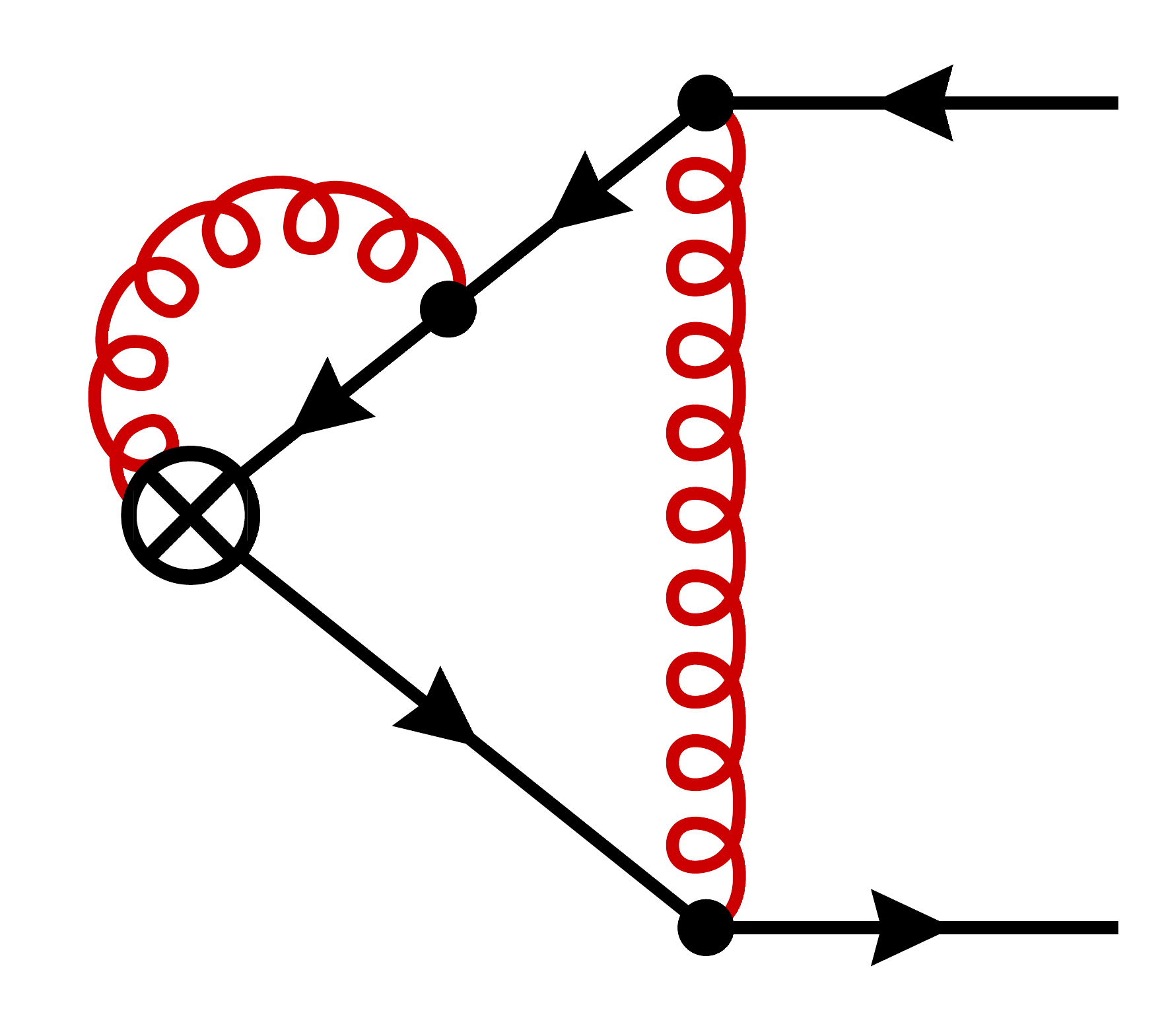}
    \end{subfigure} &
    \begin{subfigure}[t]{0.25\textwidth}
      \centering
      \includegraphics[width=\linewidth]{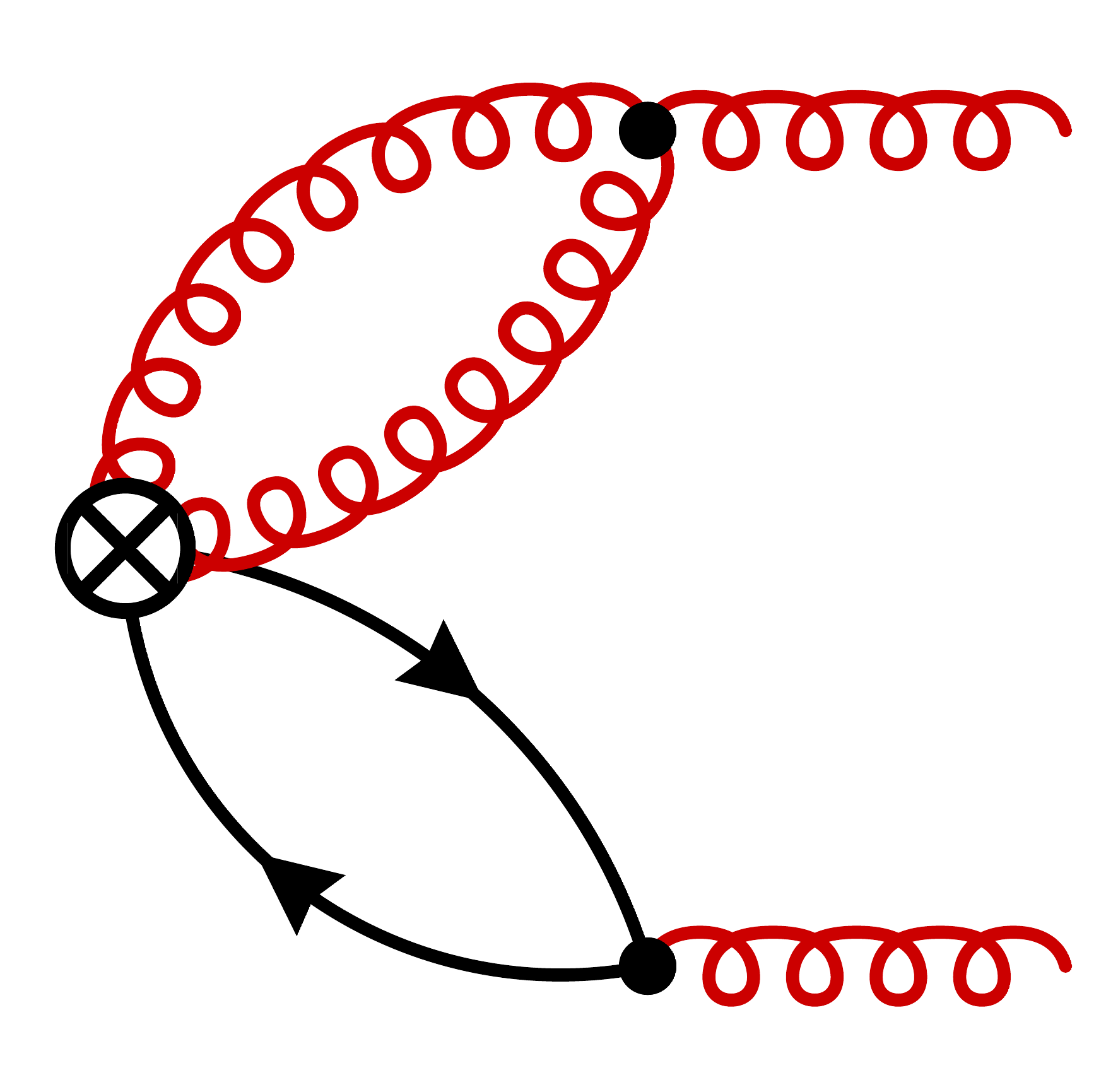}
    \end{subfigure} \\
\multicolumn{3}{@{}c@{}}{%
  \begin{subfigure}[t]{0.25\textwidth}
    \centering
    \includegraphics[width=\linewidth]
      {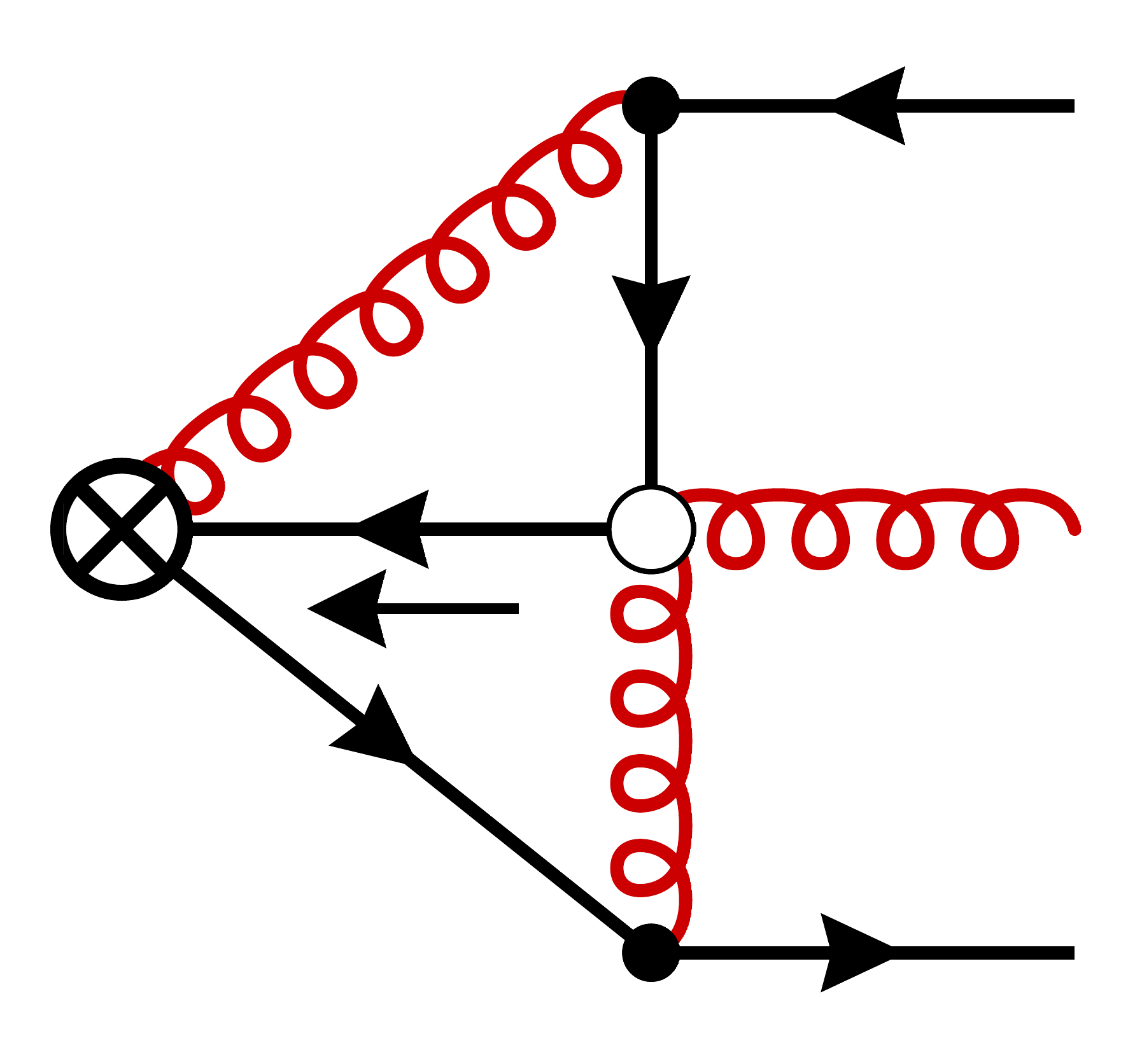}
  \end{subfigure}
  \hspace{2\tabcolsep}%
  \begin{subfigure}[t]{0.25\textwidth}
    \centering
    \includegraphics[width=\linewidth]
      {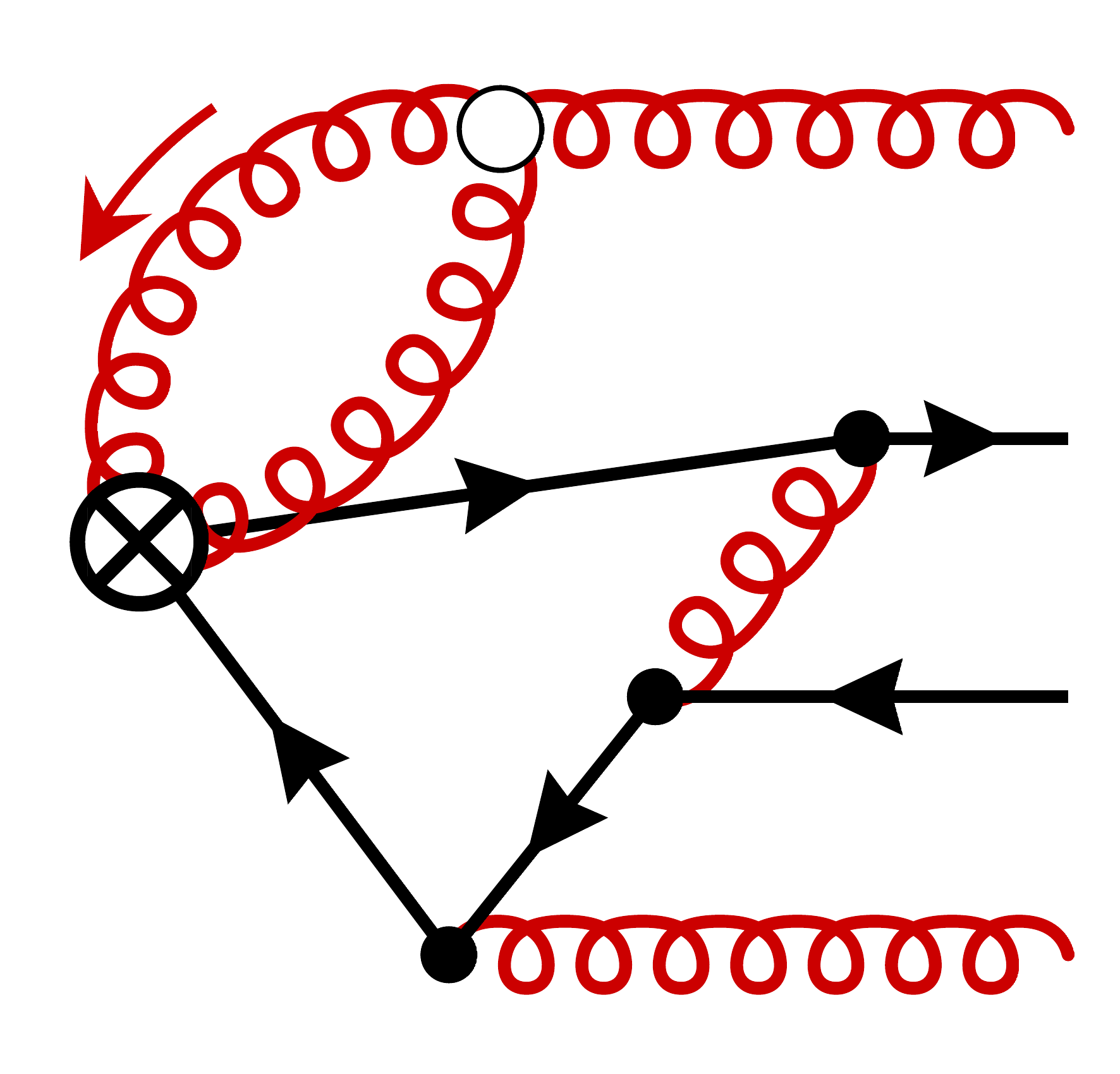}
  \end{subfigure}%
}
  \end{tabular}
  \caption{Sample diagrams contributing to the \sftx\ of the \cmdm\ operator
    at \nnlo. The operator insertion is denoted by $\otimes$. Solid lines
    represent fermions, curly lines represent gluons, and lines accompanied by
    an arrow denote fermion or gluon flow lines. We do not assign explicit
    flavors, since these diagrams occur for several flavor combinations in
    both the projectors and the operator insertion. All Feynman diagrams in
    this work are drawn with
    \textsc{FeynGame}~\cite{Harlander:2020cyh,Harlander:2024qbn,
      Bundgen:2025utt}.}
  \label{fig:gridO1}
\end{figure}

In the following, we present the results obtained for the \sftx\ of the
\cmdm\ operator.  Besides the explicit power dependence on $t$,
cf.~\cref{eq:lagr:alec}, the matching matrix depends logarithmically on the
flow time through the combination
\begin{equation}\label{eq:LogMut}
    \lmut
    \equiv
    \ln\left(\frac{\mu^2}{\mu_t^2}\right),
    \qquad
    \text{with}
    \qquad
    \mu_t^2
    =
    \frac{e^{-\gamma_{\abbrev{\mathrm{E}}}}}{2t}\,.
\end{equation}
We provide the results for general color factors and $\nf$, retaining the
ringed-scheme constant $c_\chi^{(2)}$ of \cref{eq:RingedScheme}.  The
coefficients $\left(\zeta_{\text{\abbrev{CM}}}\right)_1$ and
$\left(\zeta_{\text{\abbrev{CM}}}\right)_{11}$ were already presented in
\citere{Harlander:2022vgf} for $\nc=3$ and a single quark flavor in numerical
form; we fully confirm these results.  The $\mathcal{O}(\apis)$ contributions
agree with the results of \citere{Mereghetti:2021nkt}, while the
$\mathcal{O}(\apis^2)$ contributions of
$\left(\zeta_{\text{\abbrev{CM}}}\right)_i$ for $i\in\{2,\ldots,10,12,13,14\}$
are new results. Example diagrams for different projectors are shown in
\cref{fig:gridO1}, and the total number of diagrams for each external state is
given in \cref{tab:diagrams}.

Up to and including $\mathcal{O}\left(\apis^2\right)$, we find
\begin{align}\label{eq:ResultCMDM}
\left(\zeta_{\text{\abbrev{CM}}}\right)_1&=1
+\apis\bigg\{
    -\frac{1}{2}\cca
    -\ccf\left(
        \frac{3}{8}
        +\ln 2
        +\frac{3}{4}\ln 3
    \right)
    +\lmut\left[
        -\frac{1}{2}\cca
        +\frac{5}{4}\ccf
    \right]
\bigg\}
\notag\\
&+\apis^2\bigg\{
    \frac{c_\chi^{(2)}}{16}
    +\cca^2\left[
        -\frac{311}{108}
        -\frac{5}{24}\text{Li}_2\!\left(1\right)
        -\frac{251}{48}\ln 2
        +\frac{129}{32}\ln 3
        +\frac{7}{32}\text{Li}_2\!\left(\frac{1}{4}\right)
    \right]
\notag\\
&\quad
    +\cca\ccf\left[
        \frac{1601}{1728}
        +\frac{5}{16}\text{Li}_2\!\left(1\right)
        +\frac{425}{72}\ln 2
        +\frac{1}{4}\ln^2 2
        -\frac{61}{32}\ln 3
        -\frac{9}{16}\text{Li}_2\!\left(\frac{1}{4}\right)
    \right]
\notag\\
&\quad
    +\ccf^2\left[
        \frac{10}{9}
        -\frac{5}{12}\text{Li}_2\!\left(1\right)
        +\frac{265}{144}\ln 2
        +\frac{1}{4}\ln^2 2
        -\frac{13}{4}\ln 3
        +\frac{19}{12}\text{Li}_2\!\left(\frac{1}{4}\right)
    \right]
\notag\\
&\quad
    +\cca\nf\ctr\left[
        \frac{1021}{1728}
        +\frac{1}{12}\text{Li}_2\!\left(1\right)
    \right]
    +\ccf\nf\ctr\left[
        \frac{7}{54}
        -\frac{1}{12}\text{Li}_2\!\left(1\right)
    \right]
\notag\\
&\quad
    +\lmut\left[
        -\frac{103}{72}\cca^2
        +\cca\ccf\left(
            \frac{26}{9}
            -\frac{5}{12}\ln 2
            -\frac{5}{16}\ln 3
        \right)+\frac{67}{144}\cca\nf\ctr
\right.
\notag\\
&\quad\quad\left.
        -\ccf^2\left(
            \frac{33}{16}
            +\frac{5}{4}\ln 2
            +\frac{15}{16}\ln 3
        \right)
        +\ccf\nf\ctr\left(
            -\frac{4}{9}
            +\frac{1}{3}\ln 2
            +\frac{1}{4}\ln 3
        \right)
    \right]
\notag\\
&\quad
    +\lmut^2\left[
        -\frac{5}{48}\cca^2
        -\frac{5}{96}\cca\ccf
        +\frac{25}{32}\ccf^2
        +\frac{1}{12}\cca\nf\ctr
        -\frac{5}{24}\ccf\nf\ctr
    \right]
\bigg\}\,,\\
\left(\zeta_{\text{\abbrev{CM}}}\right)_2&=-\apis\ctr\bigg\{
    1+\lmut
    +\apis\bigg[
        \cca\left(
            \frac{35}{48}
            +\frac{1}{8}\text{Li}_2\!\left(1\right)
            +6\ln 2
            -\frac{13}{4}\ln 3
            +\text{Li}_2\!\left(\frac{1}{4}\right)
        \right)
\notag\\
&\quad\quad
        +\ccf\left(
            \frac{101}{24}
            +\frac{9}{4}\ln 2
            -\frac{39}{8}\ln 3
            -\frac{9}{4}\text{Li}_2\!\left(\frac{1}{4}\right)
        \right)
\notag\\
&\quad\quad
        +\lmut\left[
            \frac{3}{4}\cca
            +\ccf\left(
                \frac{15}{8}
                -\ln 2
                -\frac{3}{4}\ln 3
            \right)
        \right]
        +\lmut^2\left(
            -\frac{1}{4}\cca
            +\ccf
        \right)
    \bigg]
\bigg\}\,,\\
\left(\zeta_{\text{\abbrev{CM}}}\right)_3&=-\apis\,\ccf\bigg\{
    \frac{29}{8}
    +3\lmut-\apis\bigg[\nf\ctr\left(
            \frac{21}{8}
            +\frac{1}{2}\text{Li}_2\!\left(1\right)
        \right)
\notag\\
&\quad\quad
        +\cca\left(
            -\frac{2093}{160}
            +\frac{7}{16}\text{Li}_2\!\left(1\right)
            -\frac{6323}{240}\ln 2
            +\frac{901}{40}\ln 3
            +\frac{5}{16}\text{Li}_2\!\left(\frac{1}{4}\right)
        \right)
\notag\\
&\quad\quad
        +\ccf\left(
            -\frac{2311}{480}
            -\frac{9}{8}\text{Li}_2\!\left(1\right)
            +\frac{8243}{240}\ln 2
            -\frac{1077}{80}\ln 3
            +\frac{7}{4}\text{Li}_2\!\left(\frac{1}{4}\right)
        \right)\notag\\
&\quad\quad
        +\lmut\left[
            -\frac{851}{96}\cca
            +\ccf\left(
                \frac{45}{32}
                +3\ln 2
                +\frac{9}{4}\ln 3
            \right)
            +\frac{49}{24}\nf\ctr
        \right]
\notag\\
&\quad\quad
        +\lmut^2\left(
            -\frac{5}{8}\cca
            -3\ccf
            +\frac{1}{2}\nf\ctr
        \right)
    \bigg]
\bigg\}\,,\\
\left(\zeta_{\text{\abbrev{CM}}}\right)_4&=\apis^2\,\ccf\bigg\{
    \cca\left(
        \frac{23}{48}
        +\frac{33}{4}\ln 2
        -\frac{47}{8}\ln 3
        +\frac{3}{8}\text{Li}_2\!\left(\frac{1}{4}\right)
    \right)
\notag\\
&\quad
    +\ccf\left(
        -\frac{13}{48}
        -13\ln 2
        +\frac{15}{2}\ln 3
    \right)
    -\lmut\left(
        \frac{1}{4}\cca
        +\frac{1}{2}\ccf
    \right)
\bigg\}\,,\\
\left(\zeta_{\text{\abbrev{CM}}}\right)_5&=\apis^2\,\ccf\ctr\left(
    7
    +6\lmut
\right)\,,\\
\left(\zeta_{\text{\abbrev{CM}}}\right)_6&=\apis^2\,\ccf\ctr\left(
    \frac{15}{8}
    +18\ln 2
    -9\ln 3
    -\frac{3}{2}\text{Li}_2\!\left(\frac{1}{4}\right)
    +\frac{1}{2}\lmut
    -\frac{3}{2}\lmut^2
\right)\,,\\
\left(\zeta_{\text{\abbrev{CM}}}\right)_7&=\frac{3}{8}\frac{\apis\,\ccf\nc}{\pi^2}
\bigg\{
    -\frac{7}{8}
    +\frac{21}{2}\ln 2
    -\frac{21}{4}\ln 3
    +\text{Li}_2\!\left(\frac{1}{4}\right)
    +\frac{15}{4}\lmut
    +\lmut^2
\notag\\
&\quad
    +\apis\bigg[
        -1.6777\,\cca
        +4.8135\,\ccf
        -3.1037\,\nf\ctr
        +\lmut\Biggl[
            \frac{4}{3}\Gamma_7
\notag\\
&\quad\quad\quad
            +\cca\left(
                \frac{9983}{1440}
                -\frac{7}{24}\text{Li}_2\!\left(1\right)
                +\frac{976}{45}\ln 2
                -\frac{1273}{80}\ln 3
                +\frac{11}{24}\text{Li}_2\!\left(\frac{1}{4}\right)
            \right)
\notag\\
&\quad\quad\quad
            +\ccf\left(
                \frac{2917}{1440}
                +\frac{3}{4}\text{Li}_2\!\left(1\right)
                +\frac{2143}{90}\ln 2
                -\frac{1337}{80}\ln 3
                +\frac{31}{12}\text{Li}_2\!\left(\frac{1}{4}\right)
            \right)
\notag\\
&\quad\quad\quad
            -\nf\ctr\left(
                \frac{35}{24}
                +\frac{1}{3}\text{Li}_2\!\left(1\right)
                +\frac{7}{2}\ln 2
                -\frac{7}{4}\ln 3
                +\frac{1}{3}\text{Li}_2\!\left(\frac{1}{4}\right)
            \right)
        \Biggr]
\notag\\
&\quad\quad
        +\lmut^2\left[
            \frac{637}{144}\cca
            +\ccf\left(
                \frac{137}{16}
                -\ln 2
                -\frac{3}{4}\ln 3
            \right)
            -\frac{47}{36}\nf\ctr
        \right]
\notag\\
&\quad\quad
        +\lmut^3\left(
            \frac{4}{9}\cca
            +\frac{23}{12}\ccf
            -\frac{2}{9}\nf\ctr
        \right)
    \bigg]
\bigg\}\,,\label{eq:zetaCM7}\\
\left(\zeta_{\text{\abbrev{CM}}}\right)_8&=\frac{\apis^2\,\ccf\nc}{2\pi^2}
\left\{
    4.6877\,\ctr
    + \lmut\left(\Gamma_8 - \frac72 \ctr\right)
    -\frac{3}{2}\lmut^2\,\ctr
\right\}\,,\label{eq:zetaCM8}\\
\left(\zeta_{\text{\abbrev{CM}}}\right)_9&=\frac{\apis^2\,\ccf\nc}{2\pi^2}
\left\{
    -1.1388\,\ctr
    +\lmut\left[
        \Gamma_9
        -\ctr\left(
            \frac{23}{16}
            +9\ln 2
            -\frac{9}{2}\ln 3
            -\frac{3}{4}\text{Li}_2\!\left(\frac{1}{4}\right)
        \right)
    \right]\right.\notag\\
    &\qquad\qquad\qquad
    \left.-\frac{3}{8}\lmut^2\,\ctr
    +\frac{1}{4}\lmut^3\,\ctr
    \right\}\,,\label{eq:zetaCM9}\\
\left(\zeta_{\text{\abbrev{CM}}}\right)_{10}&=0\,,\label{eq:zetaCM10}\\
\left(\zeta_{\text{\abbrev{CM}}}\right)_{11}&=-\frac{3}{2}\frac{\apis\,\ccf}{t}
\left\{
    1
    +\apis\left[
        \cca\left(
            \frac{91}{36}
            +\frac{13}{6}\ln 2
            -\frac{3}{2}\ln 3
        \right) -\frac{5}{9}\nf\ctr
\right.\right.
\notag\\
&\quad\left.\left.
        +\ccf\left(
            -\frac{1}{2}
            -\frac{28}{3}\ln 2
            +\frac{15}{4}\ln 3
        \right)+\lmut\left(
            \frac{11}{12}\cca
            -\frac{3}{4}\ccf
            -\frac{1}{3}\nf\ctr
        \right)\right]\right\}\,,\label{eq:zetaCM11}\\
\left(\zeta_{\text{\abbrev{CM}}}\right)_{12}&=\frac{3}{8}\frac{\apis\,\ccf\nc}{\pi^2 t}
\bigg\{
    6\ln 2
    -3\ln 3
    +\lmut
    +\apis\bigg[
        1.9394\,\cca
        -1.3474\,\ccf
        -0.8701\,\nf\ctr
\notag\\
&\quad\quad\left.
        +\lmut\left[
            \cca\left(
                \frac{91}{36}
                +\frac{23}{3}\ln 2
                -\frac{17}{4}\ln 3
            \right)
            +\ccf\left(
                \frac{1}{2}
                +\frac{25}{6}\ln 2
                -3\ln 3
            \right)\right.
\right.
\notag\\
&\quad\quad\left.\left.\left.
            -\nf\ctr\left(
                \frac{5}{9}
                +2\ln 2
                -\ln 3
            \right)
        \right]
        +\lmut^2\left(
            \frac{11}{12}\cca
            +\frac{3}{4}\ccf
            -\frac{1}{3}\nf\ctr
        \right)
    \right]
\right\}\,,\\
\left(\zeta_{\text{\abbrev{CM}}}\right)_{13}&=\frac{\apis^2\,\ccf\nc\ctr}{2\pi^2 t}\,(1.2946)\,,\\
\left(\zeta_{\text{\abbrev{CM}}}\right)_{14}&=\frac{3}{8}\frac{\apis\,\ccf\nc}{\pi^2 t^2}
\bigg\{
    2\ln 2
    -\ln 3
    +\apis\bigg[
        0.5208\,\cca
        +0.2680\,\ccf
        -0.0879\,\nf\ctr
\notag\\
&\quad\quad
        +\lmut\left[
            \cca\left(
                \frac{11}{6}\ln 2
                -\frac{11}{12}\ln 3
            \right)
            +\ccf\left(
                \frac{3}{2}\ln 2
                -\frac{3}{4}\ln 3
            \right)
\right.
\notag\\
&\quad\quad\quad\left.\left.\left.
            +\nf\ctr\left(
                -\frac{2}{3}\ln 2
                +\frac{1}{3}\ln 3
            \right)
        \right]
    \right]
\right\}\,,
\end{align}
where $\operatorname{Li}_s(z)=\sum_{k=1}^\infty z^k/k^s$ is the polylogarithm
of order $s$ and argument $z$; specifically, we use
\begin{equation}\label{eq:Resu:eyry}
  \begin{aligned}
    \text{Li}_2(1)&=\frac{\pi^2}{6}=1.64493\ldots\,,\\
    \text{Li}_2\left(\frac{1}{4}\right)&=0.267653\ldots\,,\\
    \text{Li}_3(1)&=1.20206\ldots\,.
  \end{aligned}
\end{equation}
$c_\chi^{(2)}$ originates from the ringed scheme and is given in
\cref{eq:RingedScheme}.  All results presented in this work were calculated
using a general $R_\xi$ gauge, and we explicitly verified their independence
of the gauge parameter.

The entries $\left(\zeta_{\text{\abbrev{CM}}}\right)_{7},\ldots,
\left(\zeta_{\text{\abbrev{CM}}}\right)_{9}$ and
$\left(\zeta_{\text{\abbrev{CM}}}\right)_{12},\ldots,
\left(\zeta_{\text{\abbrev{CM}}}\right)_{14}$ contain three-loop
contributions, which we quote numerically to four decimal places in the main
text, and to six significant digits in the ancillary file. By varying the
target accuracy of \texttt{ftint}, we have checked that these digits are not
affected by the numerical uncertainty.

The logarithmic terms of these results can be derived analytically from the
\msbar\ anomalous dimension.  For
$\left(\zeta_{\text{\abbrev{CM}}}\right)_{3,\ldots,10}$, this requires the
newly evaluated entries $\gamma_{1,3},\ldots,\gamma_{1,10}$, see
\cref{eq:regAnoDimResult3,eq:GammasNum}. This is the origin of the
coefficients $\Gamma_7$, $\Gamma_8$, and $\Gamma_9$ appearing in
\cref{eq:zetaCM7,eq:zetaCM8,eq:zetaCM9}.  Further details, together with the
value obtained for these coefficients, are given in \cref{sec:gamma}.

\begin{figure}[h]
  \centering
  \setlength{\tabcolsep}{6pt} 
  \begin{tabular}{@{}c c c@{}}
    \begin{subfigure}[t]{0.30\textwidth}
      \centering
      \includegraphics[width=\linewidth]{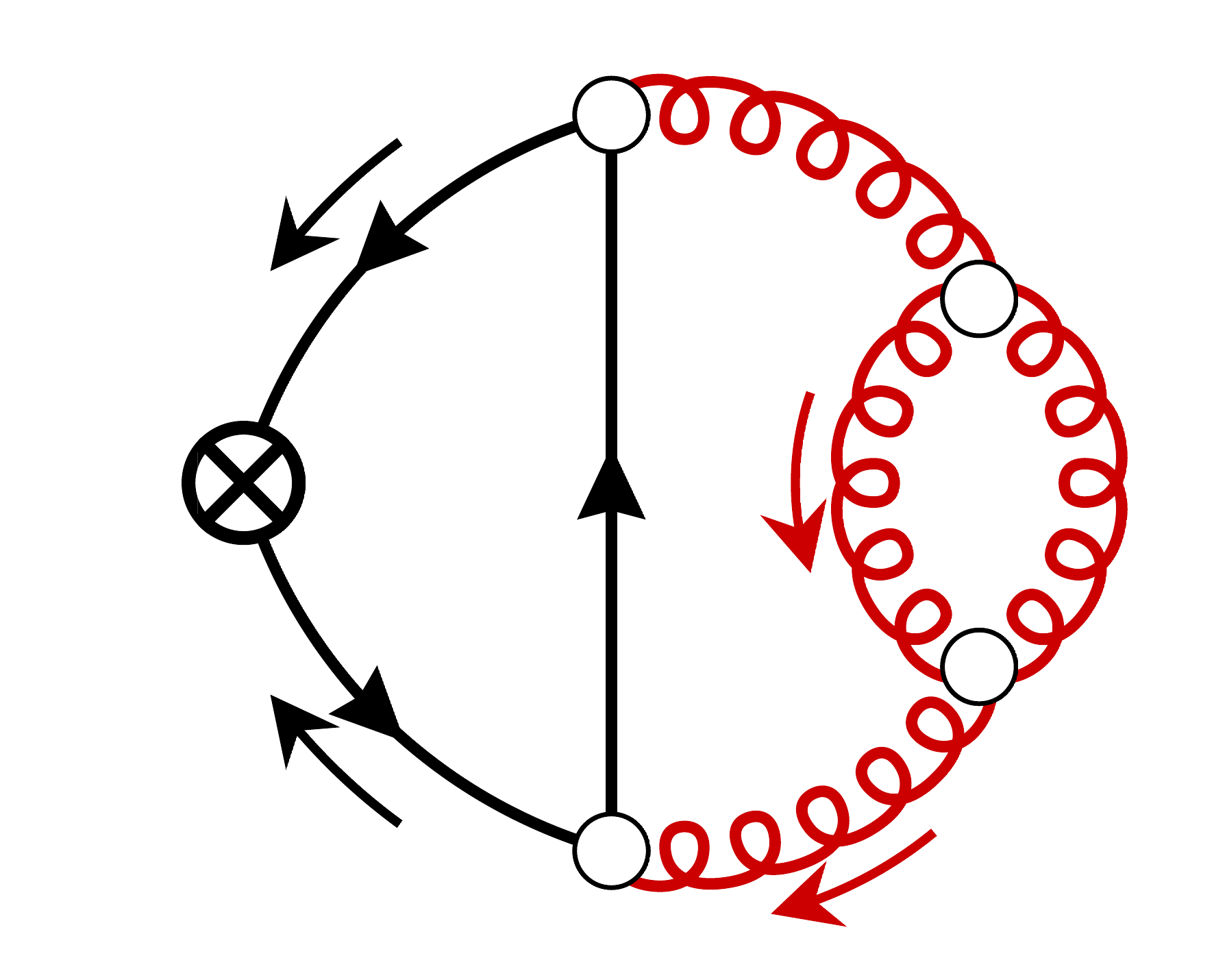}
    \end{subfigure} &
    \begin{subfigure}[t]{0.28\textwidth}
      \centering
      \includegraphics[width=\linewidth]{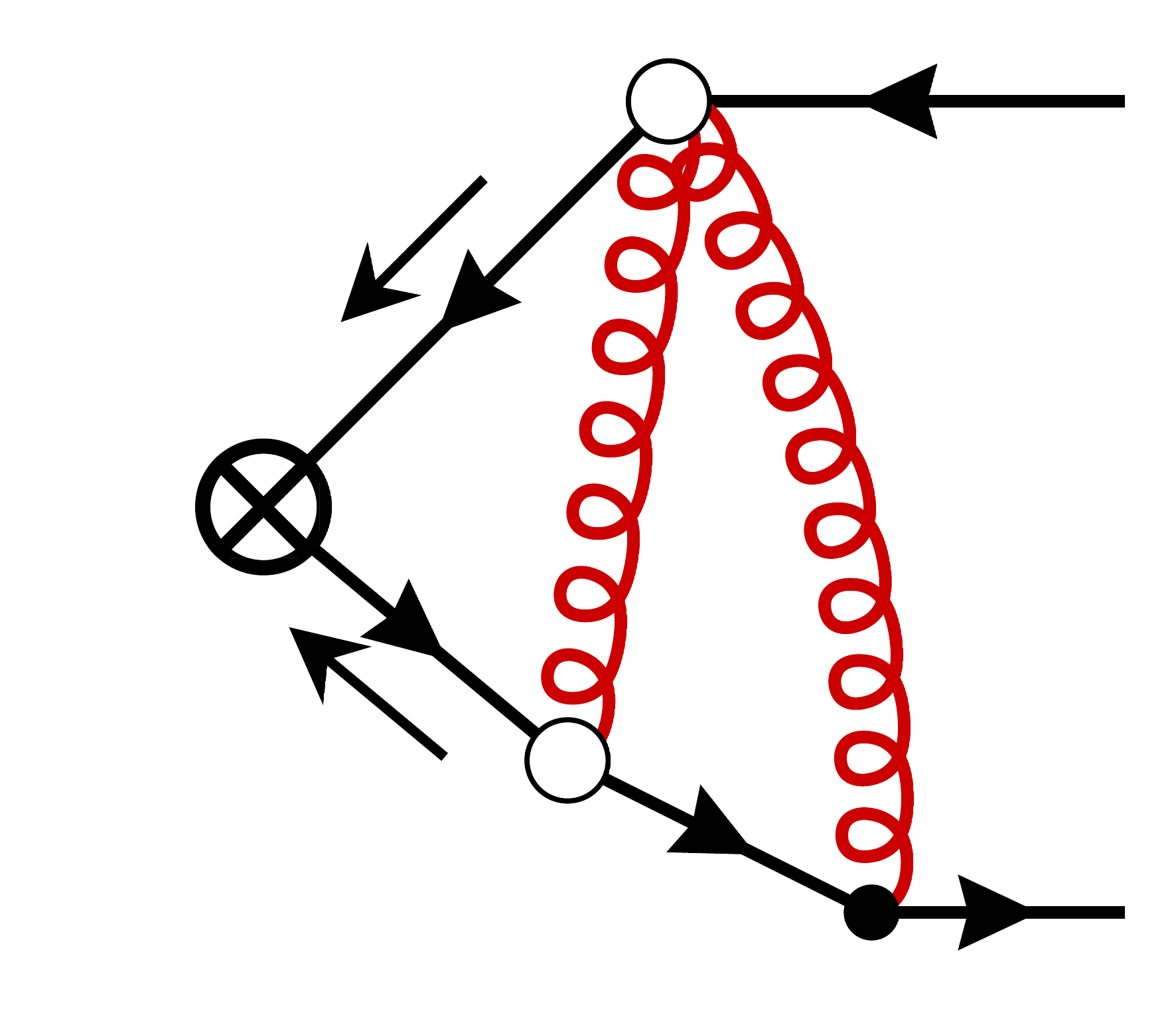}
    \end{subfigure} &
    \begin{subfigure}[t]{0.28\textwidth}
      \centering
      \includegraphics[width=\linewidth]{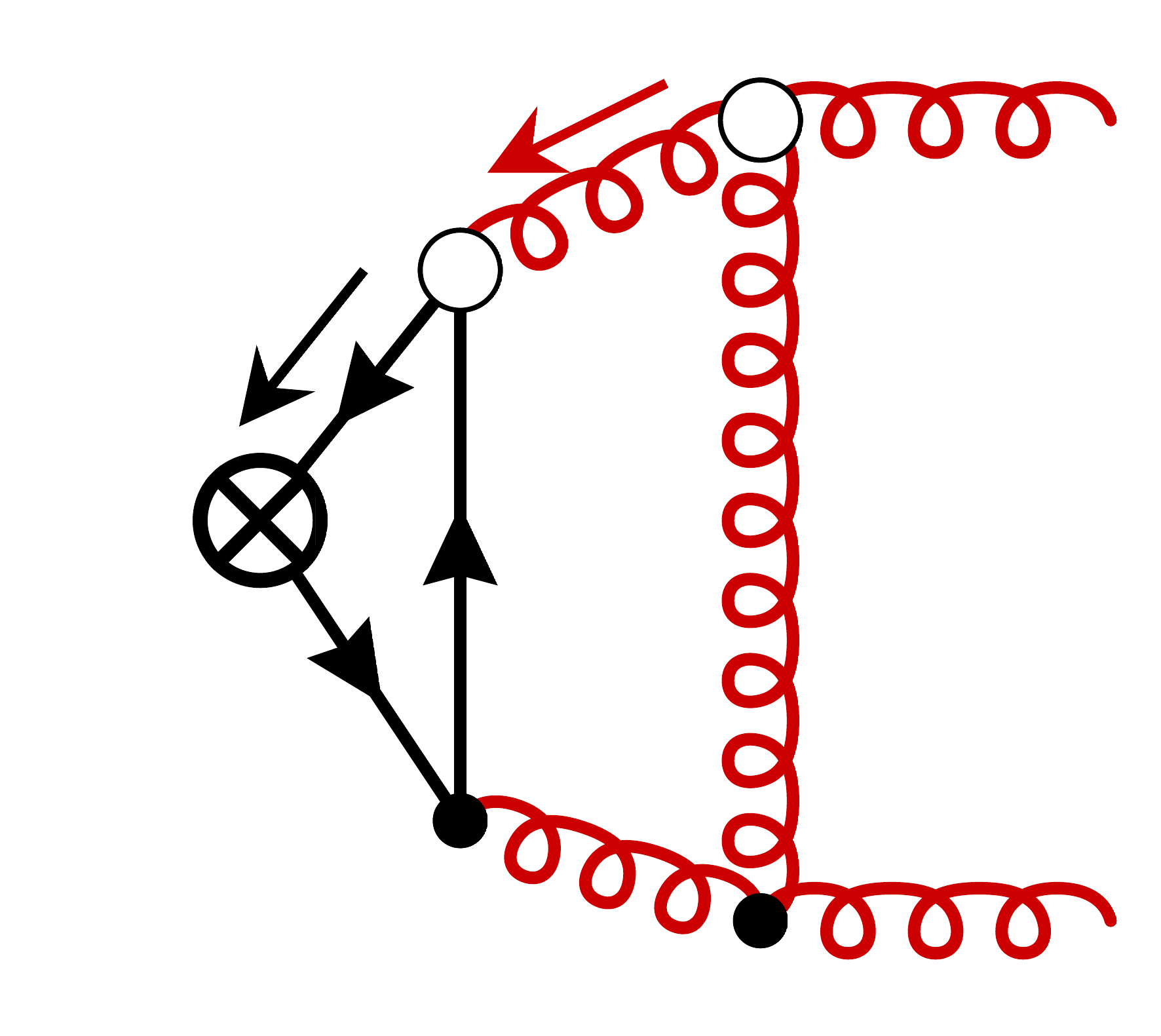}
    \end{subfigure} \\
\multicolumn{3}{@{}c@{}}{%
  \begin{subfigure}[t]{0.28\textwidth}
    \centering
    \includegraphics[width=\linewidth]
      {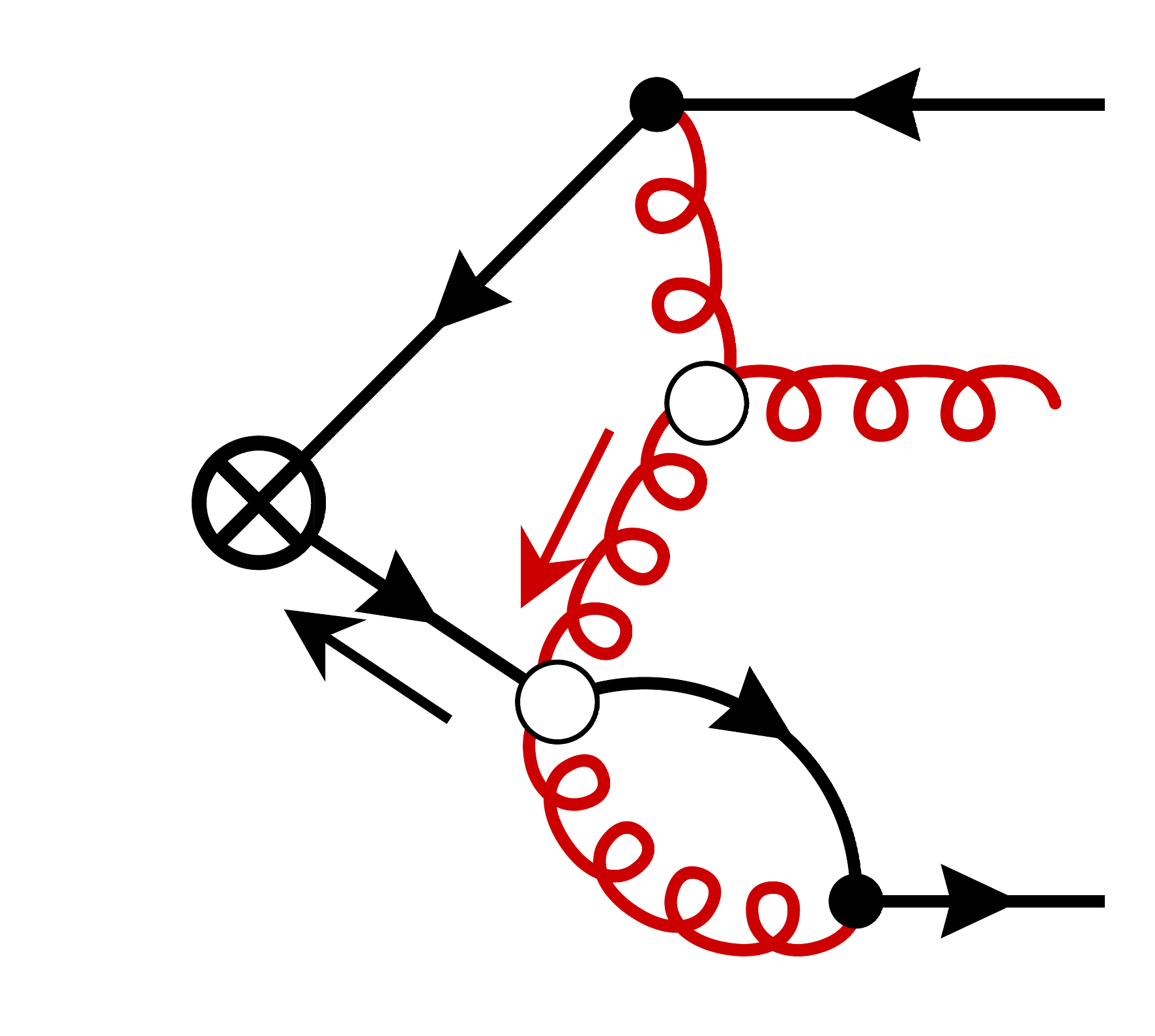}
  \end{subfigure}
  \hspace{2\tabcolsep}%
  \begin{subfigure}[t]{0.28\textwidth}
    \centering
    \includegraphics[width=\linewidth]
      {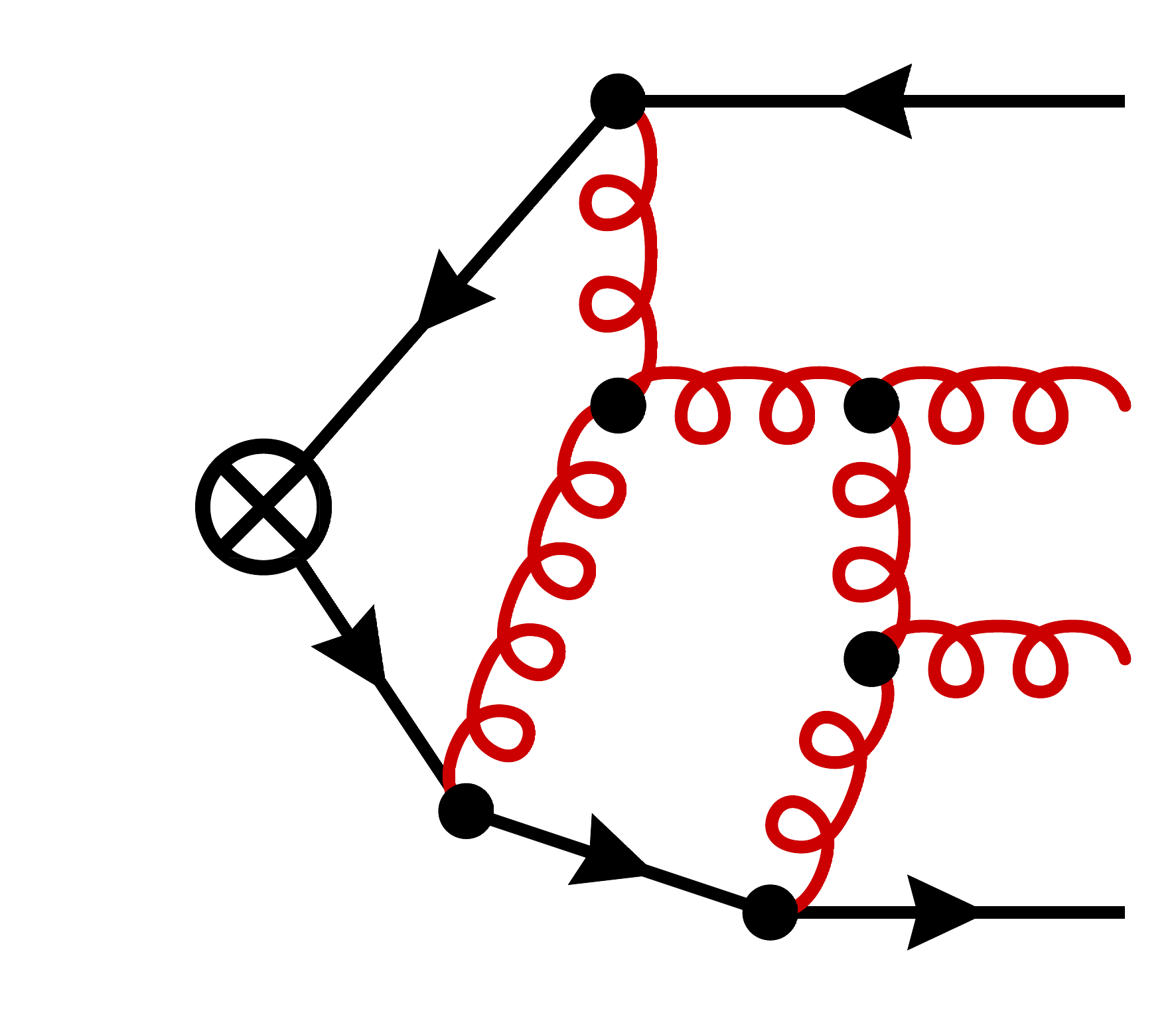}
  \end{subfigure}%
}
  \end{tabular}
  \caption{Example \nnlo\ diagrams contributing to the $\mathcal{O}(t)$ terms
    in the \sftx\ of the flowed scalar bilinear
    $\tilde{\mathcal{O}}_{\text{\abbrev{S}}}(t)$, which contribute at \nnlo.}
  \label{fig:gridO5}
\end{figure}


\subsection[$\mathcal{O}(t)$ corrections
  to the quark bilinear]{$\boldsymbol{\mathcal{O}(t)}$ corrections
  to the quark bilinear}\label{sec:ordert}

\begin{figure}[ht]
  \centering
    \begin{subfigure}[b]{0.4\textwidth}
    \centering
    \includegraphics[width=\linewidth]{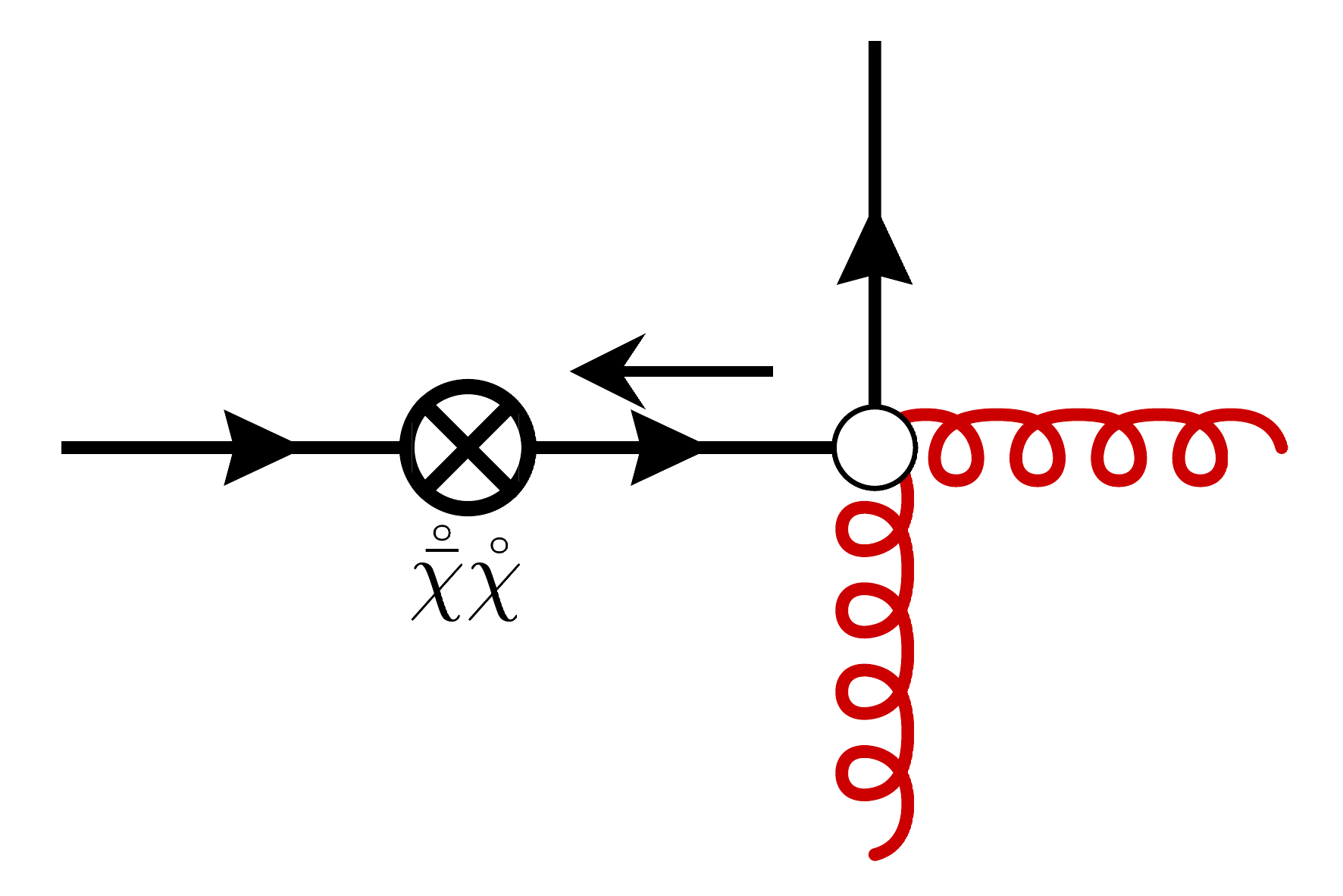}
    \caption{}
    \label{fig:twopics-a}
  \end{subfigure}
  \begin{subfigure}[b]{0.3\textwidth}
    \centering
    \includegraphics[width=\linewidth]{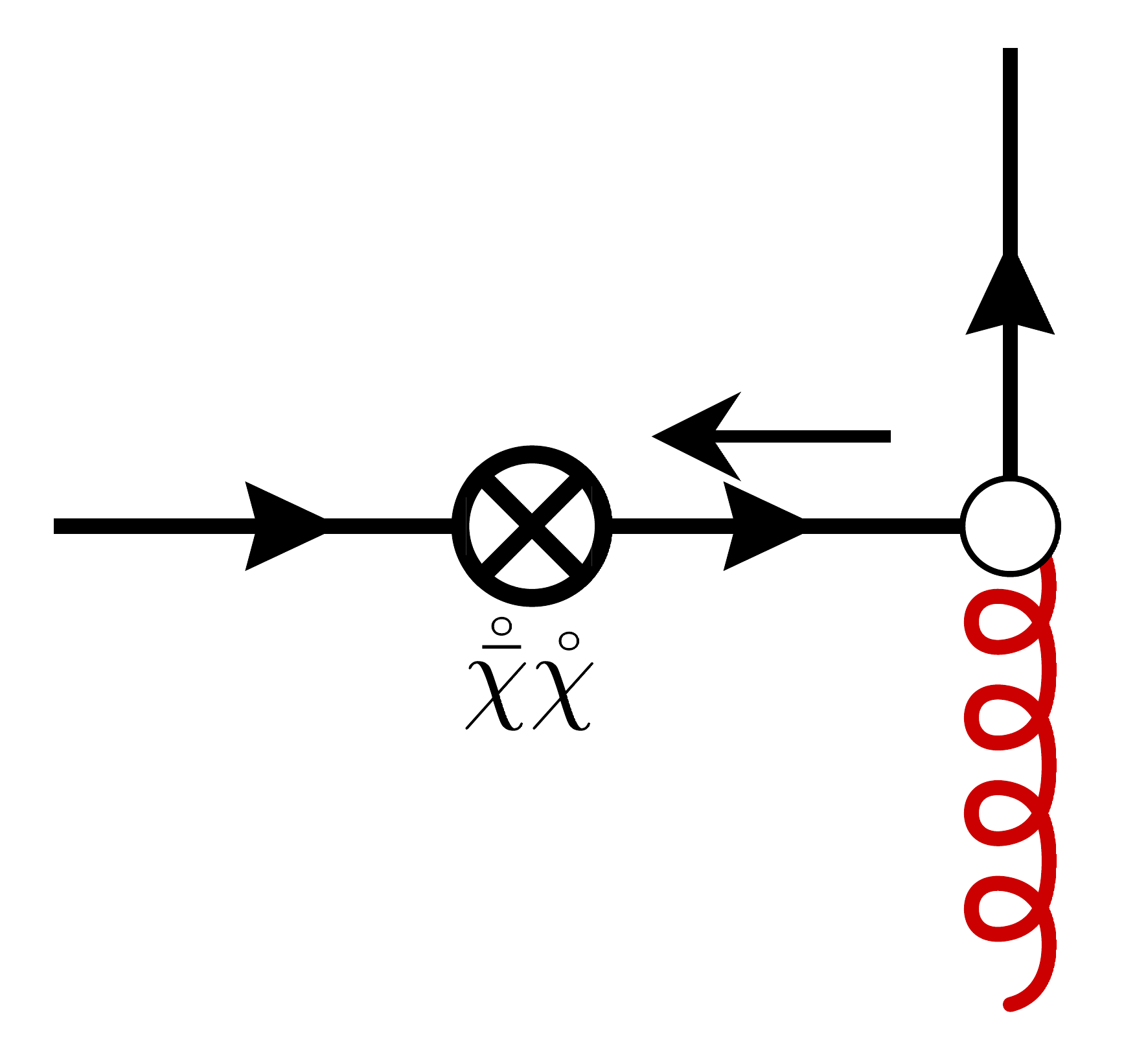}
    \caption{}
    \label{fig:twopics-b}
  \end{subfigure}
  \caption{Example diagrams contributing to the $\mathcal{O}(\apis^0t)$ terms
    in $(\zeta_{\text{\abbrev{S}}})_i$, originating from
    $P_{\mathcal{N}_2}[\tilde{\mathcal{O}}_{\text{\abbrev{S}}}]$ in (a) and
    $P_{\mathcal{N}_4}[\tilde{\mathcal{O}}_{\text{\abbrev{S}}}]$ in (b). According
    to \cref{eq:OrthProj}, both diagrams contribute to the \lo\ terms in
    $(\zeta_{\text{\abbrev{S}}})_1$ and $(\zeta_{\text{\abbrev{S}}})_3$.  }
  \label{fig:twopics}
\end{figure}

As discussed in \cref{sec:gf}, a consistent conversion between flowed and
unflowed operators requires us to retain terms of $\mathcal{O}(t)$ in the
\sftx\ of the lower-dimensional flowed operators.  In our case, the only
non-trivial such terms occur for the flowed scalar bilinear
$\tilde{\mathcal{O}}_\text{\abbrev{S}}= \tilde{\mathcal{O}}_{11}$.
Example diagrams contributing to the newly calculated entries
$(\zeta_{\mathrm{S}})_i$ are shown in \cref{fig:gridO5}.  The corresponding
external states used in the projectors $P_i$, together with the total number
of diagrams, are listed in \cref{tab:diagrams}.  We obtain
\begin{align}\label{eq:ResultBilinear}
   \frac{1}{t}\left(\zeta_{\text{\abbrev{S}}}\right)_1&=1
+\apis\bigg\{
    -\ccf\left(
        \frac{5}{8}
        +\ln 2
        +\frac{3}{4}\ln 3
    \right)
    +\lmut\left[
        -\frac{1}{2}\cca
        +\frac{5}{4}\ccf
    \right]
\bigg\}
\notag\\
&+\apis^2\bigg\{
    \frac{c_\chi^{(2)}}{16}
    +\cca^2\left[
        -\frac{179}{108}
        -\frac{1}{12}\text{Li}_2\!\left(1\right)
        -\frac{157}{48}\ln 2
        +\frac{87}{32}\ln 3
        -\frac{1}{32}\text{Li}_2\!\left(\frac{1}{4}\right)
    \right]
\notag\\
&\quad\quad
    +\cca\ccf\left[
        \frac{1703}{1728}
        -\frac{1}{6}\text{Li}_2\!\left(1\right)
        -\frac{143}{36}\ln 2
        +\frac{1}{4}\ln^2 2
        +\frac{47}{16}\ln 3
        +\frac{19}{48}\text{Li}_2\!\left(\frac{1}{4}\right)
    \right]
\notag\\
&\quad\quad
    +\ccf^2\left[
        \frac{55}{48}
        +\frac{7}{24}\text{Li}_2\!\left(1\right)
        +\frac{175}{144}\ln 2
        +\frac{1}{4}\ln^2 2
        -\frac{35}{16}\ln 3
        +\frac{1}{6}\text{Li}_2\!\left(\frac{1}{4}\right)
    \right]
\notag\\
&\quad\quad
    +\cca\nf\ctr\left[
        \frac{505}{1728}
        +\frac{1}{12}\text{Li}_2\!\left(1\right)
    \right]
    -\ccf\nf\ctr\left[
        \frac{5}{216}
        +\frac{1}{12}\text{Li}_2\!\left(1\right)
    \right]
\notag\\
&\quad
    +\lmut\Biggl[
        -\frac{11}{9}\cca^2
        +\cca\ccf\left(
            \frac{491}{144}
            -\frac{5}{12}\ln 2
            -\frac{5}{16}\ln 3
        \right)+\frac{43}{144}\cca\nf\ctr
\notag\\
&\quad\quad
        -\ccf^2\left(
            \frac{19}{8}
            +\frac{5}{4}\ln 2
            +\frac{15}{16}\ln 3
        \right)
        +\ccf\nf\ctr\left(
            -\frac{13}{36}
            +\frac{1}{3}\ln 2
            +\frac{1}{4}\ln 3
        \right)
    \Biggr]
\notag\\
&\quad
    +\lmut^2\left[
        -\frac{5}{48}\cca^2
        -\frac{5}{96}\cca\ccf
        +\frac{25}{32}\ccf^2
        +\frac{1}{12}\cca\nf\ctr
        -\frac{5}{24}\ccf\nf\ctr
    \right]
\bigg\}\,,\\
\frac{1}{t}\left(\zeta_{\text{\abbrev{S}}}\right)_2&=-\frac{1}{6}\apis\,\ctr
\bigg\{
    1
    +6\lmut
    +\apis\bigg[
        \cca\left(
            -\frac{13}{40}
            +\frac{111}{5}\ln 2
            -\frac{72}{5}\ln 3
            +\frac{15}{2}\text{Li}_2\!\left(\frac{1}{4}\right)
        \right)
\notag\\
&\quad\quad
        +\ccf\left(
            \frac{51}{2}
            +\frac{97}{2}\ln 2
            -48\ln 3
            -\frac{27}{2}\text{Li}_2\!\left(\frac{1}{4}\right)
        \right)+\lmut\left[
            \frac{15}{2}\cca\right.
\notag\\
&\quad\quad\quad
        \left.
            +\ccf\left(
                6
                -6\ln 2
                -\frac{9}{2}\ln 3
            \right)
        \right]
        +\lmut^2\left(
            -\frac{3}{2}\cca
            +6\ccf
        \right)
    \bigg]
\bigg\}\,,\\
\frac{1}{t}\left(\zeta_{\text{\abbrev{S}}}\right)_3&=2
+\apis\,\ccf\left(
    \frac{7}{8}
    -2\ln 2
    -\frac{3}{2}\ln 3
    -\frac{3}{2}\lmut
\right)+\apis^2\bigg\{
    \frac{c_\chi^{(2)}}{8}+\ccf\nf\ctr\left(
        \frac{37}{24}
        +\frac{1}{2}\text{Li}_2\!\left(1\right)
    \right)
\notag\\
&\quad\quad
    +\cca\ccf\left(
        -\frac{5597}{1440}
        -\frac{11}{16}\text{Li}_2\!\left(1\right)
        +\frac{1001}{720}\ln 2
        +\frac{1}{2}\ln^2 2
        +\frac{179}{160}\ln 3
        +\frac{7}{3}\text{Li}_2\!\left(\frac{1}{4}\right)
    \right)
\notag\\
&\quad\quad
    +\ccf^2\left(
        -\frac{1201}{180}
        +\frac{23}{48}\text{Li}_2\!\left(1\right)
        -\frac{21341}{720}\ln 2
        +\frac{1}{2}\ln^2 2
        +\frac{1713}{80}\ln 3
        +\frac{107}{16}\text{Li}_2\!\left(\frac{1}{4}\right)
    \right)
\notag\\
&\quad
    +\lmut\left[
        -\cca\ccf\left(
            \frac{135}{32}
            +\frac{11}{6}\ln 2
            +\frac{11}{8}\ln 3
        \right)
        +\ccf^2\left(
            \frac{183}{32}
            +\frac{3}{2}\ln 2
            +\frac{9}{8}\ln 3
        \right)
\right.
\notag\\
&\quad\quad\left.
        +\ccf\nf\ctr\left(
            \frac{1}{8}
            +\frac{2}{3}\ln 2
            +\frac{1}{2}\ln 3
        \right)
    \right]
    +\lmut^2\left(
        \frac{1}{16}\cca\ccf
        -\frac{39}{16}\ccf^2
        +\frac{1}{4}\ccf\nf\ctr
    \right)
\bigg\}\,,\\
\frac{1}{t}\left(\zeta_{\text{\abbrev{S}}}\right)_4&=\frac{3}{4}\apis\,\ccf
\left\{
    1
    +\apis\left[
        \cca\left(
            \frac{535}{216}
            +\frac{1}{4}\text{Li}_2\!\left(1\right)
            -\frac{125}{54}\ln 2
            -\frac{11}{12}\ln 3
            -\frac{37}{36}\text{Li}_2\!\left(\frac{1}{4}\right)
        \right)
\right.\right.
\notag\\
&\quad\quad\left.
        +\ccf\left(
            -\frac{85}{108}
            -\frac{17}{36}\text{Li}_2\!\left(1\right)
            +\frac{395}{54}\ln 2
            -3\ln 3
            +\frac{3}{4}\text{Li}_2\!\left(\frac{1}{4}\right)
        \right)
        -\frac{1}{9}\nf\ctr
\right.
\notag\\
&\quad\left.\left.
        +\lmut\left(
            \frac{7}{12}\cca
            +\frac{1}{12}\ccf
            -\frac{1}{3}\nf\ctr
        \right)
    \right]
\right\}\,,\\
\frac{1}{t}\left(\zeta_{\text{\abbrev{S}}}\right)_5&=\apis^2\,\ccf\ctr\left(
    1
    +6\lmut
\right)\,,\\
\frac{1}{t}\left(\zeta_{\text{\abbrev{S}}}\right)_6&=\apis^2\,\ccf\ctr\left(
    -\frac{31}{40}
    -\text{Li}_2\!\left(1\right)
    +\frac{41}{15}\ln 2
    -\frac{3}{10}\ln 3
    -\frac{1}{2}\text{Li}_2\!\left(\frac{1}{4}\right)
    +3\lmut
    -\frac{3}{2}\lmut^2
\right)\,,\\
\frac{1}{t}\left(\zeta_{\text{\abbrev{S}}}\right)_7&=-\frac{\nc}{2\pi^2}\bigg\{
    \lmut
    +\apis\,\ccf\bigg[
        \frac{125}{32}
        +\frac{71}{8}\ln 2
        -\frac{135}{16}\ln 3
        -\frac{9}{4}\text{Li}_2\!\left(\frac{1}{4}\right)
        +\lmut\left(
            \frac{13}{16}-\ln 2\right.
\notag\\
&\quad\quad\quad\left.
            -\frac{3}{4}\ln 3
        \right)
        +\frac{3}{2}\lmut^2
    \bigg]
    +\apis^2\bigg[
        0.7835\,\cca\ccf
        +3.2146\,\ccf^2
        -1.7140\,\ccf\nf\ctr
\notag\\
&\quad\quad
        +\lmut\Biggl[
            \frac{c_\chi^{(2)}}{16}
            -\ccf\Gamma_7+\cca\ccf\left(
                \frac{11467}{1920}
                -\frac{1}{4}\text{Li}_2\!\left(1\right)
                +\frac{637}{80}\ln 2
                +\frac{1}{4}\ln^2 2\right.
\notag\\
&\quad\quad\quad\left.
                -\frac{1203}{160}\ln 3-\frac{41}{32}\text{Li}_2\!\left(\frac{1}{4}\right)
                -\frac{3}{2}\text{Li}_3\!\left(1\right)
            \right)+\ccf^2\left(
                \frac{5351}{640}
                +\frac{1}{16}\text{Li}_2\!\left(1\right)
                \right.
\notag\\
&\quad\quad\quad
            \left.
                +\frac{5089}{240}\ln 2 +\frac{1}{4}\ln^2 2-\frac{7059}{320}\ln 3
                -\frac{77}{16}\text{Li}_2\!\left(\frac{1}{4}\right)
                +3\text{Li}_3\!\left(1\right)
            \right)
\notag\\
&\quad\quad\quad
+\ccf\nf\ctr\left(
                -\frac{115}{96}+\frac{1}{4}\text{Li}_2\!\left(1\right)-\frac{71}{24}\ln 2
                +\frac{45}{16}\ln 3
                +\frac{3}{4}\text{Li}_2\!\left(\frac{1}{4}\right)
            \right)
        \Biggr]
\notag\\
&\quad\quad
+\lmut^2\Biggl[
            \cca\ccf\left(
                \frac{141}{64}
                -\frac{11}{12}\ln 2
                -\frac{11}{16}\ln 3
            \right)
+\ccf^2\left(
                \frac{141}{64}
                -\frac{3}{2}\ln 2
                -\frac{9}{8}\ln 3
            \right)
\notag\\
&\quad\quad\quad
        +\ccf\nf\ctr\left(
                -\frac{11}{16}
                +\frac{1}{3}\ln 2
                +\frac{1}{4}\ln 3
            \right)
        \Biggr]\notag\\
&\quad\quad
        +\lmut^3\left(
            \frac{15}{32}\cca\ccf
            +\frac{47}{32}\ccf^2
            -\frac{1}{8}\ccf\nf\ctr
        \right)
    \bigg]
\bigg\}\,,\\
\frac{1}{t}\left(\zeta_{\text{\abbrev{S}}}\right)_8&=\frac{\apis^2\,\ccf\nc}{2\pi^2}
\left\{
    -0.5238\,\ctr
    +\lmut\left(
        \frac{5}{2}\ctr
        +\Gamma_8
    \right)
    -\frac{3}{2}\lmut^2\,\ctr
\right\}\,,\\
\frac{1}{t}\left(\zeta_{\text{\abbrev{S}}}\right)_9&=\frac{\apis^2\,\ccf\nc}{2\pi^2}
\left\{
    -2.1460\,\ctr
    +\lmut\left[
        \Gamma_9
        +\ctr\left(
            \frac{73}{240}
            +\frac{1}{2}\text{Li}_2\!\left(1\right)
            -\frac{41}{30}\ln 2
            +\frac{3}{20}\ln 3
            \right.\right.\right.
            \notag\\
&\quad\quad\quad \left.\left.\left.
            +\frac{1}{4}\text{Li}_2\!\left(\frac{1}{4}\right)
        \right)
    \right]
    -\lmut^2\,\ctr
    +\frac{1}{4}\lmut^3\,\ctr
    \right\}\,,\\
\frac{1}{t}\left(\zeta_{\text{\abbrev{S}}}\right)_{10}&=0\,.
\end{align}
All of these results are new.\footnote{The \nlo\ coefficient of
$(\zeta_\text{S})_1$ was independently obtained in \citere{Crosas:private}.}
Let us emphasize that projections of flowed operators onto higher-dimensional
unflowed operators can have non-vanishing contributions of order $t$ already
at \lo, even though the corresponding projectors are orthogonal \wrt\ the
unflowed operators.  Examples are the \lo\ contributions to
$(\zeta_\text{\abbrev{S}})_i$ for $i\in\{1,3,7\}$.  The origin of this effect
is illustrated by diagrams such as those shown in \cref{fig:twopics}, which
arise because we impose the 1\abbrev{PI} condition only with respect to the
unflowed fields. Note also that the \lo\ contribution to $(\zeta_\text{S})_7$
arises from a \one-loop vacuum diagram.

The remaining entries $(\zeta_\text{S})_{11}$ through
$(\zeta_\text{S})_{14}$ have been evaluated in \citere{Borgulat:2023xml} and
are collected for completeness in
\cref{eq:zetaS11,eq:zetaS12,eq:zetaS13,eq:zetaS14}.


\subsection{Double-derivative operator}\label{sec:FlowTimeRel}

The operators $\mathcal{O}^\bare[\Lambda]$ constructed in \cref{sec:basis}
represent a complete physical basis of \qcd\ operators at mass dimension
\five. The specific set of operators in this basis can be modified by adding
arbitrary linear combinations of the unphysical operators of
\cref{eq:unphysOps}. Inversion of \cref{eq:SFTX} therefore allows us to
express an arbitrary \cp-even \qcd\ operator of mass dimension five in terms
of gradient-flow operators. The opposite is not true though: with the results
obtained up to now, one can only convert the flowed versions of the
$\mathcal{O}[\Lambda]$ defined in \cref{sec:basis} to the \msbar\ scheme;
flowed versions of linear combinations with the unphysical operators require a
separate calculation of the \sftx\ in general. In other words, the conversion
between different bases of flowed operators $\tilde{\mathcal{O}}(t)$ and
$\tilde{\mathcal{O}'}(t)$ involves flow-time dependent coefficients:
\begin{equation}\label{eq:Resu:deil}
  \begin{aligned}
    \tilde{\mathcal{O}}'(t)
    = \xi(t)\tilde{\mathcal{O}}(t)\,,\qquad
    \xi(t) = \zeta'(t)T\zeta^{-1}(t)\,,
  \end{aligned}
\end{equation}
where $\zeta(t)$ and $\zeta'(t)$ are the matching matrices to the
\msbar\ scheme for $\tilde{\mathcal{O}}$ and $\tilde{\mathcal{O}'}$,
respectively, and $T$ relates the unflowed operators as
\begin{equation}\label{eq:Resu:isbn}
  \begin{aligned}
      \mathcal{O}' =T\mathcal{O}\,.
  \end{aligned}
\end{equation}
In certain cases, the \sftx\ of the unphysical operators is known. For
example, for $\tilde{\mathcal{N}}_1[t;\Lambda]$, it can be obtained by
combining the results for the \sftx\ of $\bar\chi\chi\sim
\tilde{\mathcal{O}}_{\text{\abbrev{S}}}$~\cite{Borgulat:2023xml} and
$\bar\chi\slashed{\mathcal{D}}\chi$~\cite{Harlander:2018zpi}.  At this point,
we want to focus on a particular dimension-\five\ operator which is related to
the kinetic operator appearing in the context of the heavy-quark
expansion~\cite{Shifman:1984wx,Bigi:1995jr,Dassinger:2006md}:
\begin{equation}
  \mathcal{O}^\bare_{\dsquared}[\Lambda]=
  \mathcal{O}^{\prime\bare}_{1}[\Lambda]\equiv\bar{\bsym{\psi}}
  \left(\Lambda D^2+\overleftarrow{D}^2\Lambda\right)\bsym{\psi}\,.
\end{equation}
It is related to the original basis through
\begin{equation}\label{eq:RelO1Ok}
  \mathcal{O}^\bare_{\dsquared}[\Lambda]=
  \mathcal{O}^\bare_\text{\abbrev{CM}}[\Lambda]
  +2\mathcal{O}^\bare_3[\Lambda]+\mathcal{N}^\bare_2[\Lambda]\,.
\end{equation}
One can therefore define the new basis
\begin{equation}\label{eq:Resu:high}
  \begin{aligned}
    \mathcal{O}^{\prime\bare}[\Lambda]=\left(\mathcal{O}^\bare_{\dsquared}[\Lambda],
    \mathcal{O}^\bare_2[\Lambda],\ldots,\mathcal{O}^\bare_{14}[\Lambda]\right)^T\,,
  \end{aligned}
\end{equation}
and analogously for the renormalized and the flowed operators
$\mathcal{O}'[\Lambda]$ and $\tilde{\mathcal{O}}'[t;\Lambda]$.  We are
interested in the matching matrix $\zeta'(t)$ corresponding to this new basis,
i.e.
\begin{equation}\label{eq:sftxprime}
    \tilde{\mathcal{O}}'[t;\Lambda]=\zeta'(t)\mathcal{O}'[\Lambda]\,.
\end{equation}
For physical matrix elements, we may disregard the \eom\ operator
$\mathcal{N}^\bare_2$ on the \rhs\ of \cref{eq:RelO1Ok} and write
\begin{equation}\label{eq:oprimereno}
  T=\mathbbm{1}_{14\times 14}+2e_1e_3^T\,,
\end{equation}
cf.~\cref{eq:Resu:isbn}.  Due to the specific forms of $T$ and the
renormalization matrix of the operators, cf.~\cref{sec:gamma},
\cref{eq:oprimereno} applies to both the bare and the renormalized operators.

Even though \cref{eq:RelO1Ok} also holds for the corresponding flowed
operators, the flowed version of the \eom\ operator,
$\tilde{\mathcal{N}}_2(t)$, does not generally vanish in physical matrix
elements. Consequently, directly transforming the flowed operator basis
requires knowledge of its \sftx, which could be obtained with the same methods
as described above.  However, the explicit calculation can be avoided by
making use of the flow equations of \cref{eq:FlowEq}. For the flowed
$\dsquared$ operator, one finds\footnote{Note that the renormalization scale
$\mu$ is assumed independent of $t$ throughout this paper, unless indicated
otherwise.\label{fn:mut}}
\begin{equation}
  \delta_t\left(\tilde{\mathcal{O}}_\text{\abbrev{S}}[t;\Lambda]\right)
  =\tilde{\mathcal{O}}_{\dsquared}[t;\Lambda]\,,\qquad\textrm{with}\qquad
  \delta_t\equiv\partial_t-\partial_t\ln\zeta_\chi\,,
\end{equation}
where the second term in $\delta_t$ arises from the ringed scheme, with the
conversion factor $\zeta_\chi$ given in
\cref{eq:RingedScheme}. The matching matrix in the new basis therefore follows
directly from our existing results,
\begin{equation}
    \zeta'(t)=\begin{pmatrix}
        \delta_t\zeta_{\text{\abbrev{S}}}^T(t)\\
        \zeta_2^T(t)\\
        \vdots\\
        \zeta_{14}^T(t)
    \end{pmatrix}T^{-1}\,,
\end{equation}
where the vectors $\zeta_i(t)$ denote the \sftx\ coefficients calculated in
this work, see \cref{eq:ZetaVec}.


\subsection{Chromoelectric dipole operator}\label{sec:cedm}

The \cedm\ operator
\begin{equation}\label{eq:Resu:kaki}
  \begin{aligned}
    \mathcal{O}_\text{\abbrev{CE}}^\bare[\Lambda] = -g_\bare\,
      \bar{\bsym{\psi}}\Lambda\sigma_{\mu\nu}\gamma_5T^a\bsym{\psi}\,
      G_{\mu\nu}^a
      = \mathcal{O}_\text{\abbrev{CM}}^\bare[i\Lambda\gamma_5]
  \end{aligned}
\end{equation}
differs from the \cmdm\ operator of \cref{eq:CMDMbare} by an additional $i
\gamma_5$ insertion.  Its \sftx\ can in principle be obtained in a way
analogous to the \cmdm\ operator. The operator basis is obtained from the one
of \cref{sec:basis} by replacing $\bsym{\psi}\to i\gamma_5\bsym{\psi}$ while
leaving $\bar{\bsym{\psi}}$ unchanged, using the \cp-odd counterpart of
$\mathcal{O}_2^\bare$ which involves the dual tensor
$\epsilon_{\alpha\beta\mu\nu}G_{\alpha\beta}$, and discarding all operators
$\sim \mathbbm{1}$. The \nnlo\ \sftx\ therefore does not involve \three-loop
\vevs; only \two-loop integrals are needed. Nevertheless, diagrams where the
flowed operator appears in a closed fermion loop require a consistent
treatment of $\gamma_5$ which typically leads to a significantly larger
calculational effort than the \cmdm\ calculation. Since the operators under
consideration induce a chirality flip, such diagrams can appear only for
operators $\sim \mathrm{Tr}(\Lambda M_\bare^n)$ with $n\geq 1$, which means
that they are absent in the massless case. Aside from
$\mathcal{O}_\text{\abbrev{CE}}[\Lambda]$, the \sftx\ of the flowed
\cedm\ operator in this case only involves
\begin{equation}\label{eq:Resu:imit}
  \begin{aligned}
    \mathcal{O}_\text{\abbrev{P}}[\Lambda]
    = \mathcal{O}_\text{\abbrev{S}}[i\Lambda\gamma_5]
    = \mathcal{O}_{11}[i\Lambda\gamma_5]\,.
  \end{aligned}
\end{equation}
In the massless perturbative calculation, the operator insertion therefore
appears only on open fermion lines. The $\gamma_5$ matrices from the operators
and the corresponding projectors thus belong to the same fermion trace and,
assuming the \ndr\ scheme, can be eliminated. The matching coefficients
consequently coincide in this prescription, and we obtain
\begin{equation}\label{eq:Resu:bode}
  \begin{alignedat}{2}
    \tilde{\mathcal{O}}_\text{\abbrev{CE}}[t;\Lambda] &=
    \tilde{\mathcal{O}}_\text{\abbrev{CM}}[t;i\Lambda\gamma_5] &&=
    (\zeta_\text{\abbrev{CM}}(t))_1
    \mathcal{O}_\text{\abbrev{CM}}[i\Lambda\gamma_5] +
    (\zeta_\text{\abbrev{CM}}(t))_{11}
    \mathcal{O}_\text{\abbrev{S}}[i\Lambda\gamma_5]\,,\\
    \tilde{\mathcal{O}}_\text{\abbrev{P}}[t;\Lambda] &=
    \tilde{\mathcal{O}}_\text{\abbrev{S}}[t;i\Lambda\gamma_5] &&=
    (\zeta_\text{\abbrev{S}}(t))_1
    \mathcal{O}_\text{\abbrev{CM}}[i\Lambda\gamma_5] +
    (\zeta_\text{\abbrev{S}}(t))_{11}
    \mathcal{O}_\text{\abbrev{S}}[i\Lambda\gamma_5]\,,
  \end{alignedat}
\end{equation}
see \cref{eq:ResultCMDM,eq:zetaCM11,eq:ResultBilinear,eq:zetaS11}.  A
comparison with the results obtained in other schemes such as
\abbrev{BMHV}~\cite{tHooft:1972tcz,Breitenlohner:1977hr} requires accounting
for the finite differences between the operator definitions.

The general treatment of the \cedm\ operator will be deferred to future work.



\section{Flowed anomalous dimension}\label{sec:gammatilde}

The flowed anomalous-dimension matrix $\tilde{\gamma}(t)$ governs the
flow-time evolution of the flowed operators, analogously to the
\msbar\ anomalous-dimension
matrix~\cite{Borgulat:2023xml,Harlander:2020duo}. It is defined by (recall
footnote~\ref{fn:mut})
\begin{align}\label{eq:flowAnoDim}
     t\frac{\mathrm{d}}{\mathrm{d}t}\tilde
     {\mathcal{O}}[t;\Lambda]=\tilde{\gamma}(t)\tilde
     {\mathcal{O}}[t;\Lambda]\,, \qquad\text{with}\qquad
     \tilde{\gamma}(t)=
     \left[t\frac{\mathrm{d}}{\mathrm{d}t}\zeta(t)\right]\zeta^{-1}(t)\,.
\end{align}
Let us stress again that, within the context of the finite operator basis
considered in this paper, similar to the matching coefficients, the
$t$-dependence of the flowed anomalous dimension is
exact~\cite{Harlander:2026lte} at each perturbative order. Since the matching
matrix depends not only on the logarithmic terms $\lmut$ but also on explicit
powers of $t$, we arrive at (see \cref{sec:gamma} for more details)
\begin{equation}\label{eq:flowAnoDimFromZeta}
    \tilde{\gamma}(t)=-\left(\beta_\mathrm{s}\frac{\partial}{\partial\apis}\zeta(t)\right)\zeta^{-1}(t)-\zeta(t)\gamma\zeta^{-1}(t)+\gamma_\mathcal{Z}(t)+\left(t\frac{\partial}{\partial t}\zeta(t)\bigg|_{\lmut}\right)\zeta^{-1}(t)\,,
\end{equation}
which is a generalization of Eq.~(4.5) in \citere{Borgulat:2023xml}. $\gamma$
and $\gamma_{\mathcal{Z}}$ are defined in \cref{eq:regAnoDim} and
\cref{eq:GammaZ}, respectively. The notation in the last term is to be
understood such that $\lmut$ is kept fixed, so that the derivative acts only
on the explicit $t$-dependence. We checked that computing $\tilde{\gamma}(t)$
via \cref{eq:flowAnoDim} and \cref{eq:flowAnoDimFromZeta} gives identical
results through $\mathcal{O}(\apis^2)$. Note that the flowed anomalous
dimension follows the block organization of \cref{tab:sftx_structure}, but its
nonzero entries and their perturbative orders need not coincide with those of
$\zeta(t)$.  For the sake of brevity, we set $\mu=\mu_t$ in the results
presented in the following.  Nevertheless, the full logarithmic $t$-dependence
can be reconstructed using
\begin{equation}\label{eq:MuDepFlowAnoDim}
    \mu^2\frac{\textrm{d}}{\textrm{d}\mu^2}\tilde{\gamma}(t)
    =
    \left[\gamma_{\mathcal{Z}}(t),\tilde{\gamma}(t)\right]+t\frac{\mathrm{d}}{\mathrm{d}t}\gamma_\mathcal{Z}(t)\,,
\end{equation}
which follows directly by differentiating \cref{eq:flowAnoDim} and using \cref{eq:GammaZ}.
We again collect the entries relevant for the flow-time evolution of a specific operator into the vectors
\begin{align}\label{eq:AnoDimVec}
    \tilde{\gamma}_i^T(t)
    =
    e_i^T\tilde{\gamma}(t),
\end{align}
and introduce the notation $\tilde{\gamma}_{\text{\abbrev{CM}}}(t)
\equiv\tilde{\gamma}_1(t)$ and $\tilde{\gamma}_{\text{\abbrev{S}}}(t)
\equiv\tilde{\gamma}_{11}(t)$ for the two operators of interest.

For $\tilde{\gamma}_{2},\ldots,\tilde{\gamma}_{10}$, the entries can be
extracted from the flowed anomalous dimensions and matching coefficients given
in \citere{Harlander:2020duo}. Since $\mathcal{O}_{\text{\abbrev{S}}}$ does not appear in the
\sftx\ of $\tilde{\mathcal{O}}_2,\ldots,\tilde{\mathcal{O}}_{10}$, the
inclusion of the $\mathcal{O}(t)$ terms does not modify these
results. Nevertheless, for completeness, we provide the full matrix
$\tilde{\gamma}(t)$ in the ancillary file, see \cref{app:ancillary_file}.

In contrast, the entries $(\tilde{\gamma}_{\text{\abbrev{S}}})_{11},\ldots,
(\tilde{\gamma}_{\text{\abbrev{S}}})_{14}$, which can be extracted from
\citere{Borgulat:2023xml}, are modified by the inclusion of the
$\mathcal{O}(t)$ terms. Note that this is consistent with the fact that the
operator basis in \citere{Borgulat:2023xml} was restricted to mass
dimension~\three. For example, consider the flow-time evolution of the
dimension-\three\ operator $\tilde{\mathcal{O}}_\text{\abbrev{S}}$ through
\cref{eq:flowAnoDim}. In the dimension-\three\ basis, the \rhs\ obviously only
includes operators up to mass dimension \three, while in the
dimension-\five\ basis, it also includes dimension-\five\ operators. The
difference this implies in the evolution of
$\tilde{\mathcal{O}}_\text{\abbrev{S}}$ is formally compensated by a change in
the flowed anomalous dimension.

Through $\order{\apis^2}$, the flowed anomalous dimension of the
\cmdm\ operator is
\newcommand{\otmarker}{o_t}
\begin{align}
    \left(\tilde{\gamma}_{\text{\abbrev{CM}}}\right)_1
    &=
    \apis\bigg[
    -\frac{1}{2}\cca
    +\ccf\left(
        \frac{5}{4}
        -\frac{3}{2}\otmarker
    \right)\bigg]+\apis^2\bigg\{
    -\frac{121}{72}\cca^2
\notag\\
&\quad
    +\cca\ccf\left[
        \frac{479}{144}
        -\frac{11}{12}\ln 2
        -\frac{11}{16}\ln 3
        +\otmarker\left(
            -\frac{29}{12}
            -\frac{13}{4}\ln 2
            +\frac{9}{4}\ln 3
        \right)
    \right]
\notag\\
&\quad
    +\ccf^2\left[
        -\frac{51}{32}
        +\otmarker\left(
            -\frac{3}{8}
            +\frac{25}{2}\ln 2
            -\frac{27}{4}\ln 3
        \right)
    \right]+\frac{67}{144}\cca\nf\ctr
\notag\\
&\quad
    +\ccf\nf\ctr\left[
        -\frac{4}{9}
        +\frac{1}{3}\ln 2
        +\frac{1}{4}\ln 3
        +\frac{1}{3}\otmarker
    \right]
\bigg\}\,,\\
\left(\tilde{\gamma}_{\text{\abbrev{CM}}}\right)_2&=-\apis\,\ctr\left\{
    1
    +\apis\left[
        \frac{3}{8}\cca
        +\ccf\left(
            \frac{5}{8}
            +\frac{5}{4}\otmarker
            -\ln 2
            -\frac{3}{4}\ln 3
        \right)
    \right]
\right\}\,,\\
\left(\tilde{\gamma}_{\text{\abbrev{CM}}}\right)_3&=-3\apis\,\ccf
\left\{
    1+\otmarker
    +\apis\left[
        \cca\left(
            \frac{1025}{288}
            +\otmarker\left(
                \frac{10}{9}
                +\frac{13}{6}\ln 2
                -\frac{3}{2}\ln 3
            \right)
        \right)
\right.\right.
\notag\\
&\left.\left.
        -\ccf\left(
            \frac{71}{48}
            +\otmarker\left(
                -\frac{13}{4}
                +\frac{25}{3}\ln 2
                -\frac{9}{2}\ln 3
            \right)
        \right)
        -\nf\ctr\left(
            \frac{49}{72}
            +\frac{2}{9}\otmarker
        \right)
    \right]
\right\}\,,\\
\left(\tilde{\gamma}_{\text{\abbrev{CM}}}\right)_4&=-\apis^2\,\ccf\left[
    \frac{1}{4}\cca
    +\ccf\left(
        \frac{1}{2}
        +\frac{9}{8}\otmarker
    \right)
\right]\,,\\
\left(\tilde{\gamma}_{\text{\abbrev{CM}}}\right)_5&=6\apis^2\,\ccf\ctr\,,\\
\left(\tilde{\gamma}_{\text{\abbrev{CM}}}\right)_6&=3\apis^2\,\ccf\ctr\,,\\
\left(\tilde{\gamma}_{\text{\abbrev{CM}}}\right)_7&=\frac{\apis\,\ccf\nc}{32\pi^2}
\bigg\{
    21
    -24\otmarker
    +\apis\bigg[
        16\Gamma_7
        +\cca\left(
            -\frac{379}{40}
            -\frac{7}{2}\text{Li}_2\!\left(1\right)+\frac{4849}{15}\ln 2
            \right.
\notag\\
&\quad\quad\quad\left.
            -\frac{4449}{20}\ln 3+\frac{23}{2}\text{Li}_2\!\left(\frac{1}{4}\right)\left.
            +\otmarker\left(
                -\frac{80}{3}
                -52\ln 2
                +36\ln 3
            \right)
        \right)
\right.
\notag\\
&\quad\quad
            +\ccf\left(
            -\frac{226}{15}
            +9\text{Li}_2\!\left(1\right) +\frac{967}{30}\ln 2+\frac{21}{5}\ln 3
            +88\text{Li}_2\!\left(\frac{1}{4}\right)\right.
\notag\\
&\quad\quad\quad\left.
            +\otmarker\left(
                -93
                +506\ln 2
                -279\ln 3
                +36\text{Li}_2\!\left(\frac{1}{4}\right)
            \right)
        \right)
\notag\\
&\quad\quad
        +\nf\ctr\left(
            -\frac{7}{6}
            -4\text{Li}_2\!\left(1\right)
            -42\ln 2
            +21\ln 3
            -4\text{Li}_2\!\left(\frac{1}{4}\right)
            +\frac{16}{3}\otmarker
        \right)
    \bigg]
\bigg\}\,,\\
\left(\tilde{\gamma}_{\text{\abbrev{CM}}}\right)_8&=-\frac{\apis^2\nc\ccf}{2\pi^2}
\left(
    \frac{1}{2}\ctr
    -\Gamma_8
\right)\,,\\
\left(\tilde{\gamma}_{\text{\abbrev{CM}}}\right)_9&=\frac{\apis^2\,\ccf\nc}{2\pi^2}
\left\{
    \Gamma_9
    +\ctr\left(
        \frac{11}{16}
        -\frac{3}{4}\text{Li}_2\!\left(1\right)
        -9\ln 2
        +\frac{9}{2}\ln 3
        +\frac{3}{4}\text{Li}_2\!\left(\frac{1}{4}\right)
    \right)
\right\}\,,\\
\left(\tilde{\gamma}_{\text{\abbrev{CM}}}\right)_{10}&=0\,,\\
\left(\tilde{\gamma}_{\text{\abbrev{CM}}}\right)_{11}&=\frac{3}{2}\frac{\apis\,\ccf}{t}
\left\{
    1
    +\apis\left[
        \cca\left(
            \frac{10}{9}
            +\frac{13}{6}\ln 2
            -\frac{3}{2}\ln 3
        \right)
        -\frac{2}{9}\nf\ctr\right.\right.\notag\\
        &\quad\quad\left.\left.
        +\ccf\left(
            2
            -\frac{25}{3}\ln 2
            +\frac{9}{2}\ln 3
            -\frac{3}{2}\otmarker
        \right)
    \right]
\right\}\,,\\
\left(\tilde{\gamma}_{\text{\abbrev{CM}}}\right)_{12}&=\frac{3}{8}\frac{\apis\,\ccf\nc}{\pi^2 t}
\left\{
    1
    -\otmarker
    -6\ln 2
    +3\ln 3+\apis\left[\nf\ctr\left(
            0.4852
            +\frac{2}{9}\otmarker\right)
\right.\right.
\notag\\
&\quad\quad\left.\left.
        +\cca\left(
            -1.0124
            +\otmarker\left(
                -\frac{10}{9}
                -\frac{13}{6}\ln 2
                +\frac{3}{2}\ln 3
            \right)
        \right)\right.
\right.
\notag\\
&\quad\quad\left.\left.\left.
        +\ccf\left(
            1.6736
            +\otmarker\left(
                -\frac{49}{8}
                +\frac{49}{3}\ln 2
                -\frac{33}{4}\ln 3
            \right)
        \right)
\right.
\right]
\right\}\,,\\
\left(\tilde{\gamma}_{\text{\abbrev{CM}}}\right)_{13}&=\frac{\apis^2\,\ccf\nc\ctr}{2\pi^2 t}
\left(
    1.7054
\right)\,,\\
\left(\tilde{\gamma}_{\text{\abbrev{CM}}}\right)_{14}&=\frac{3}{8}\frac{\apis\,\ccf\nc}{\pi^2 t^2}
\left\{
    \frac{3}{4}
    -4\ln 2
    +2\ln 3+\apis\bigg[
        0.4151\,\cca-0.0868\,\nf\ctr
\right.
\notag\\
&\quad\quad\left.\left.
        +\ccf\left(
            0.1156
            +\otmarker\left(
                -\frac{7}{16}
                +3\ln 2
                -\frac{3}{2}\ln 3
            \right)
        \right)
    \right]
\right\}\,.
\end{align}
Here, $\otmarker\in\{0,1\}$ is a marker indicating the additional
contributions that arise from including the $\mathcal{O}(t)$ terms discussed
in \cref{sec:ordert}. Setting $\otmarker = 0$ omits the $\order{t}$ scalar
matching terms and recovers the dimension-\three\ evolution in the scalar
sector. Setting $\otmarker = 1$ gives the result for the full
dimension-\five\ basis considered here.

The flowed anomalous dimension of the scalar bilinear operator is
\begin{align}
    \left(\tilde{\gamma}_{\text{\abbrev{S}}}\right)_{1}&=t\,\otmarker\bigg\{
    1
    +\frac{1}{4}\apis\,\ccf
    +\apis^2\bigg[
        \cca^2\left(
            \frac{1}{8}\text{Li}_2\!\left(1\right)
            +\frac{47}{24}\ln 2
            -\frac{21}{16}\ln 3
            -\frac{1}{4}\text{Li}_2\!\left(\frac{1}{4}\right)
        \right)
\notag\\
&\quad\quad
        -\frac{1}{18}\ccf\nf\ctr+\cca\ccf\left(
            \frac{61}{48}
            -\frac{23}{48}\text{Li}_2\!\left(1\right)
            -\frac{101}{8}\ln 2
            +\frac{239}{32}\ln 3
            +\frac{23}{24}\text{Li}_2\!\left(\frac{1}{4}\right)
        \right)
\notag\\
&\quad\quad
        +\ccf^2\left(
            -\frac{593}{288}
            +\frac{17}{24}\text{Li}_2\!\left(1\right)
            +\frac{93}{8}\ln 2
            -\frac{47}{8}\ln 3
            -\frac{17}{12}\text{Li}_2\!\left(\frac{1}{4}\right)
        \right)
    \bigg]
\bigg\}\,,\\
    \left(\tilde{\gamma}_{\text{\abbrev{S}}}\right)_{2}&=-\frac{1}{6}\apis\,\ctr\,t\,\otmarker
\left\{
    1
    +\apis\left[
        \cca\left(
            \frac{77}{40}
            -\frac{3}{4}\text{Li}_2\!\left(1\right)
            -\frac{69}{5}\ln 2
            +\frac{51}{10}\ln 3
            +\frac{3}{2}\text{Li}_2\!\left(\frac{1}{4}\right)
        \right)\right.\right.\notag\\
        &\quad\quad\left.\left.+\ccf\left(
            4
            +29\ln 2
            -\frac{93}{4}\ln 3
        \right)
    \right]
\right\}\,,\\
\left(\tilde{\gamma}_{\text{\abbrev{S}}}\right)_{3}&=t\,\otmarker\bigg\{
    2
    +\frac{5}{2}\apis\,\ccf
    +\apis^2\bigg[ -\ccf\nf\ctr\left(
            \frac{41}{24}
            +\frac{1}{2}\text{Li}_2\!\left(1\right)
        \right)
       \notag\\
&\quad\quad
        +\ccf^2\left(
            \frac{547}{144}
            +\frac{29}{48}\text{Li}_2\!\left(1\right)
            -\frac{2195}{72}\ln 2
            +\frac{81}{4}\ln 3
            -\frac{17}{16}\text{Li}_2\!\left(\frac{1}{4}\right)
        \right)
\notag\\
&\quad\quad
     +\cca\ccf\left(
            \frac{2459}{288}
            -\frac{1}{4}\text{Li}_2\!\left(1\right)
            +\frac{2177}{72}\ln 2
            -\frac{829}{32}\ln 3
            -\frac{47}{48}\text{Li}_2\!\left(\frac{1}{4}\right)
        \right)
    \bigg]
\bigg\}\,,\\
    \left(\tilde{\gamma}_{\text{\abbrev{S}}}\right)_{4}&=\frac{3}{4}\apis\,\ccf\,t\,\otmarker
\left\{
    1
    +\apis\left[
        \cca\left(
            \frac{523}{216}
            +\frac{1}{4}\text{Li}_2\!\left(1\right)
            -\frac{719}{54}\ln 2
            +\frac{83}{12}\ln 3
            -\frac{55}{36}\text{Li}_2\!\left(\frac{1}{4}\right)
        \right)
\right.\right.
\notag\\
&\quad\left.\left.
        +\ccf\left(
            -\frac{16}{27}
            -\frac{17}{36}\text{Li}_2\!\left(1\right)
            +\frac{1385}{54}\ln 2
            -\frac{49}{4}\ln 3
            +\frac{3}{4}\text{Li}_2\!\left(\frac{1}{4}\right)
        \right)
        -\frac{4}{9}\nf\ctr
    \right]
\right\}\,,\\
    \left(\tilde{\gamma}_{\text{\abbrev{S}}}\right)_{5}&=0\,,\\
    \left(\tilde{\gamma}_{\text{\abbrev{S}}}\right)_{6}&=\apis^2\,\ccf\ctr\,t\,\otmarker
\left(
    \frac{23}{30}
    -\text{Li}_2\!\left(1\right)
    -\frac{229}{15}\ln 2
    +\frac{87}{10}\ln 3
    +\text{Li}_2\!\left(\frac{1}{4}\right)
\right)\,,\\
    \left(\tilde{\gamma}_{\text{\abbrev{S}}}\right)_{7}&=\frac{\apis\,\ccf\nc\,t\,\otmarker}{4\pi^2}
\left\{
    \frac{17}{8}
    -\frac{47}{2}\ln 2
    +\frac{63}{4}\ln 3
    -3\text{Li}_2\!\left(\frac{1}{4}\right)
\right.
\notag\\
&\quad\left.
    +\apis\bigg[
        3.3451\,\cca
        -2.3092\,\ccf
        +2.7036\,\nf\ctr
        +2\Gamma_7
    \bigg]
\right\}\,,\\
    \left(\tilde{\gamma}_{\text{\abbrev{S}}}\right)_{8}&=-\frac{\apis^2\nc\ccf\,t\,\otmarker}{\pi^2}
\left(
    4.0225\,\ctr
    -\frac{1}{2}\Gamma_8
\right)\,,\\
    \left(\tilde{\gamma}_{\text{\abbrev{S}}}\right)_{9}&=\frac{\apis^2\nc\ccf\,t\,\otmarker}{\pi^2}
\left(
    -0.7575\,\ctr
    +\frac{1}{2}\Gamma_9
\right)\,,\\
    \left(\tilde{\gamma}_{\text{\abbrev{S}}}\right)_{10}&=0\,,\\
    \left(\tilde{\gamma}_{\text{\abbrev{S}}}\right)_{11}&=\apis\,\ccf\left(
    -\frac{3}{4}
    +\frac{3}{2}\otmarker
\right)+\apis^2\,\ccf\bigg\{\ccf\left[
        -\frac{3}{32}
        +\otmarker\left(
            \frac{3}{8}
            -\frac{25}{2}\ln 2
            +\frac{27}{4}\ln 3
        \right)
    \right]
\notag\\
&\quad\quad+\cca\left[
        -\frac{47}{32}
        -\frac{11}{12}\ln 2
        -\frac{11}{16}\ln 3
        +\otmarker\left(
            \frac{91}{24}
            +\frac{13}{4}\ln 2
            -\frac{9}{4}\ln 3
        \right)
    \right]
\notag\\
&\quad\quad
    +\nf\ctr\left[
        \frac{3}{8}
        +\frac{1}{3}\ln 2
        +\frac{1}{4}\ln 3
        -\frac{5}{6}\otmarker
    \right]
\bigg\}\,,\\
    \left(\tilde{\gamma}_{\text{\abbrev{S}}}\right)_{12}&=-\frac{\nc}{4\pi^2}\bigg\{
    1-\otmarker
    +\apis\,\ccf\left[
        \frac{7}{2}
        -\ln 2
        -\frac{3}{4}\ln 3
        +\otmarker\left(
            -\frac{33}{8}
            +8\ln 2
            -\frac{15}{4}\ln 3
        \right)
    \right]
\notag\\
&\quad
    +\apis^2\bigg[
        \frac{c_\chi^{(2)}}{16}\left(
            1-\otmarker
        \right)+\cca\ccf\left[
            \frac{673}{96}
            -\frac{7}{16}\text{Li}_2\!\left(1\right)
            +\frac{1}{12}\ln 2
            +\frac{1}{4}\ln^2 2+\frac{3}{8}\ln 3\right.
\notag\\
&\quad\quad\left.
            -\frac{5}{4}\text{Li}_2\!\left(\frac{1}{4}\right)
            -\frac{3}{2}\text{Li}_3\!\left(1\right)
            +\otmarker\left(
                -4.9968
                +\frac{13}{4}\ln 2
                -\frac{9}{4}\ln 3
            \right)
        \right]
\notag\\
&\quad\quad
        +\ccf^2\left[
            \frac{217}{32}
            +\frac{1}{2}\text{Li}_2\!\left(1\right)
            +\frac{19}{2}\ln 2
            +\frac{1}{4}\ln^2 2
            -\frac{111}{8}\ln 3
            -6\text{Li}_2\!\left(\frac{1}{4}\right)
            +3\text{Li}_3\!\left(1\right)
\right.
\notag\\
&\quad\quad\quad\left.
            +\otmarker\left(
                -1.7239
                -\frac{49}{2}\ln 2
                +\frac{99}{8}\ln 3
            \right)
        \right]
\notag\\
&\quad\quad
        +\ccf\nf\ctr\left[
            -\frac{47}{24}
            +\frac{1}{4}\text{Li}_2\!\left(1\right)
            -\frac{5}{3}\ln 2
            +\frac{3}{2}\ln 3
            +\text{Li}_2\!\left(\frac{1}{4}\right)
            +1.7024\,\otmarker
        \right]
    \bigg]
\bigg\}\,,\\
    \left(\tilde{\gamma}_{\text{\abbrev{S}}}\right)_{13}&=\frac{3\apis^2\,\ccf\nc\ctr}{4\pi^2}
\left(
    1
    -1.0522\,\otmarker
\right)\,,\\
    \left(\tilde{\gamma}_{\text{\abbrev{S}}}\right)_{14}&=\frac{\nc}{8\pi^2 t}\bigg\{
    1
    +\apis\,\ccf\left[
        -\frac{1}{2}
        +\ln 2
        -\frac{3}{4}\ln 3
        +\otmarker\left(
            \frac{13}{8}
            -6\ln 2
            +3\ln 3
        \right)
    \right]
\notag\\
&\quad
    +\apis^2\bigg[
        \frac{c_\chi^{(2)}}{16}
        +\cca\ccf\left(
            -0.5926
            +1.8879\,\otmarker
        \right)
        -\ccf^2\left(
            2.1631
            +0.7697\,\otmarker
        \right)
\notag\\
&\quad\quad
        +\ccf\nf\ctr\left(
            0.6366
            -0.5974\,\otmarker
        \right)
    \bigg]
\bigg\}\,,
\end{align}
which holds again through $\order{\apis^2}$.


\section{Conclusions and outlook}\label{sec:conclusions}

We have presented the conversion of the \cp-even
dimension-\five\ \qcd\ operator basis from the \msbar\ to the gradient-flow
scheme through \nnlo\ in the strong coupling. We have allowed for $\nf$ quark
flavors with arbitrary masses, and a general flavor structure of the
operators. Aside from the corresponding matching matrix, the requirement of
its finiteness has enabled us to determine the subset of previously unknown
entries in the \msbar\ anomalous dimension matrix. A consistent conversion
within this basis also requires the $\order{t}$ coefficients of
dimension-\five\ operators in the matching of the flowed scalar density. In
this paper, we determined these coefficients through \nnlo\ for the first
time.  The results were then used to derive the corresponding flowed anomalous
dimensions to the same perturbative order.  We also showed how the flow
equations allow one to substitute certain flowed operators by others in the
basis.

We have performed a number of checks on our results, including gauge-parameter
independence and consistency with existing results.  In the massless-quark
limit, the corresponding matching coefficients also apply to the \cedm\ and
pseudoscalar operators in the anticommuting-$\gamma_5$ scheme specified in
\cref{sec:cedm}. Extending the \cp-odd matching calculation to nonzero quark
masses is the next step in this program.


\paragraph{Acknowledgments.}
This work was supported by the \textit{German Research Foundation
  (\abbrev{DFG})} through grants~460791904, 513989149, and 396021762
(\abbrev{TRR}~257 ``Particle Physics Phenomenology after the Higgs
Discovery''), and the \textit{National Science Foundation} under grant
\abbrev{PHY}-2209185.  We would like to thank Henry Werthenbach for providing
us with the \three-loop integral reduction to master integrals.


\begin{appendix}


\section{Renormalization constants}
\label{app:rg-functions}

For completeness, we collect the renormalization-group functions required in
this work.  We use dimensional regularization with $D=4-2\epsilon$ and define
\begin{align}
    \apis
    &\equiv
    \frac{\alpha_{\textrm{s}}}{\pi}
    =
    \frac{g^2}{4\pi^2},
    &
    g_\bare^2
    &=
    \left(\frac{\mu^{2} e^{\EulerGamma}}{4\pi}\right)^{\ep}
    Z_\text{s} g^2=\hat\mu^{2\epsilon}Z_\text{s}g^2\,.
\end{align}
Our conventions follow \citeres{Harlander:2020duo,Borgulat:2023xml}. The $D$-dimensional running of the strong coupling is written as
\begin{align}
    \label{eq:betaDef}
    \mu^2\frac{\mathrm{d}}{\mathrm{d}\mu^2}\apis(\mu)
    &=
    \beta_{\mathrm{s}}\bigl(\apis(\mu)\bigr)
    =
    -\epsilon\apis
    -
    \sum_{n=0}^{\infty}
    \beta_{\mathrm{s}}^{(n)}\apis^{n+2}\,.
\end{align}
In the \msbar\ scheme, the coupling-renormalization constant through two-loop order is
\begin{align}
    \label{eq:Zs}
    Z_\text{s}
    &=
    1
    -
    \frac{\beta_{\mathrm{s}}^{(0)}}{\epsilon}\apis
    +
    \apis^2
    \left[
        \frac{\left(\beta_{\mathrm{s}}^{(0)}\right)^2}{\epsilon^2}
        -
        \frac{\beta_{\mathrm{s}}^{(1)}}{2\epsilon}
    \right]
    +
    \mathcal{O}(\apis^3)\,.
\end{align}
The first two coefficients of the beta function are~\cite{Jones:1974pg}
\begin{align}
    \label{eq:beta-coefficients}
    \beta_{\mathrm{s}}^{(0)}
    &=
    \frac{1}{4}
    \left(
        \frac{11}{3}\cca
        -
        \frac{4}{3}\ctr\nf
    \right),
    &
    \beta_{\mathrm{s}}^{(1)}
    &=
    \frac{1}{16}
    \left(
        \frac{34}{3}\cca^2
        -
        4\ccf\ctr\nf
        -
        \frac{20}{3}\cca\ctr\nf
    \right)\,.
\end{align}
Throughout this work, unless stated otherwise, the anomalous dimension $\gamma_\eta$ of a scalar quantity $\eta$ is defined in the \msbar\ scheme by
\begin{align}
    \gamma_\eta(\apis)
    &=
    \beta_{\mathrm{s}}(\apis)
    \frac{\partial}{\partial \apis}\ln Z_\eta
    =
    \sum_{n=0}^{\infty}
    \gamma_\eta^{(n)}\apis^{n+1}.
    \label{eq:gamma-m-definition}
\end{align}
Accordingly, the corresponding renormalization constant is given by
\begin{align}
    Z_\eta
    &=
    1
    -
    \frac{\gamma_\eta^{(0)}}{\epsilon}\apis
    +
    \apis^2
    \left[
        \frac{
            \gamma_\eta^{(0)}
            \left(
                \gamma_\eta^{(0)}
                +
                \beta_{\mathrm{s}}^{(0)}
            \right)
        }{2\epsilon^2}
        -
        \frac{\gamma_\eta^{(1)}}{2\epsilon}
    \right]
    +
    \mathcal{O}(\apis^3)\,.
    \label{eq:ZmDef}
\end{align}
In the context of this work, we require the anomalous dimension of the quark mass, corresponding to $\eta=m$, whose first two perturbative coefficients are
\begin{align}
    \label{eq:GammaMCoffs}
    \gamma_m^{(0)}
    &=
    \frac{3}{4}\ccf,
    &
    \gamma_m^{(1)}
    &=
    \frac{3}{32}\ccf^2
    +
    \frac{97}{96}\cca\ccf
    -
    \frac{5}{24}\ccf\ctr\nf\,.
\end{align}
Furthermore, we require the anomalous dimension of the flowed fermion field $\chi(t,x)$, corresponding to $\eta=\chi$, which was computed in \citeres{Luscher:2013cpa, Harlander:2018zpi} and is given by
\begin{gather}
\begin{aligned}\label{eq:ringedRen}
    \gamma_{\chi}^{(0)}
    &=
    -\frac{3}{4}\ccf\,,& 
    \gamma_{\chi}^{(1)}
    &=
    \left(\frac{1}{2}\ln 2-\frac{223}{96}\right)\cca\ccf
    +
    \left(\frac{3}{32}+\frac{1}{2}\ln 2\right)\ccf^2
    +
    \frac{11}{24}\ccf\ctr\nf\,.
\end{aligned}
\end{gather}
The conversion factor to the ringed scheme, $\zeta_\chi(t)$, is defined via the
\vev\ of the quark kinetic operator,
\begin{equation}\label{eq::gait}
  \begin{aligned}
    \zeta_\chi(t)
    Z_\chi\langle\bar\chi(t)\overleftrightarrow{\slashed{\mathcal{D}}}
    \chi(t)\rangle =
    -\frac{2\nc\nf}{(4\pi t)^2}\,,
  \end{aligned}
\end{equation}
where we adopt the massless-quark limit. The \nnlo\ expression
reads~\cite{Makino:2014taa,Artz:2019bpr}
\begin{gather}\begin{aligned}\label{eq:RingedScheme}
    \zeta_{\chi}(t)
    &=1-a_s\left(\gamma_{\chi}^{(0)}L_{\mu t}+\frac{3}{4}C_F\ln3+C_F\ln2\right)\\
    &\qquad+a_s^2\left\{\frac{\gamma_{\chi}^{(0)}}{2}\left({\gamma_{\chi}^{(0)}-\beta_\textrm{s}^{(0)}}\right)L_{\mu t}^2+\left[\gamma_{\chi}^{(0)}\left(\beta_\textrm{s}^{(0)}-{\gamma_{\chi}^{(0)}}\right)\ln3\right.\right.\\
    &\qquad\hspace{1cm}+\left.\left.\frac{4}{3}\gamma_{\chi}^{(0)}\left(\beta_\textrm{s}^{(0)}-{\gamma_{\chi}^{(0)}}\right)\ln2-{\gamma_{\chi}^{(1)}}\right]L_{\mu t}+\frac{c_{\chi}^{(2)}}{16}\right\}\,,\\
    c_{\chi}^{(2)}&=-23.7947\,\cca\ccf+30.3914\,\ccf^2-3.92255\,\nf\ccf\ctr\,.
\end{aligned}\end{gather}

For a general mass matrix, the anomalous dimension of the vacuum energy density can be decomposed into diagonal and off-diagonal contributions
\begin{equation}
    \mu^2\frac{\mathrm{d}}{\mathrm{d}\mu^2}E_0=\gamma_0^\textrm{d}\,\mathrm{Tr}\left(M^4\right)+\gamma_0^\textrm{nd}\left[\mathrm{Tr}\left(M^2\right)^2-\mathrm{Tr}\left(M^4\right)\right]\,,
    \end{equation}
where $\gamma_0^\textrm{d}$ and $\gamma_0^\textrm{nd}$ are related to the corresponding vacuum renormalization constants ($Z_0^{\mathrm{d}/\mathrm{nd}}$) by
\begin{align}
    \gamma_0^{\mathrm{d}/\mathrm{nd}}(\apis)
    &=
    -\left[4\gamma_m(\apis)+\epsilon\right]Z_0^{\mathrm{d}/\mathrm{nd}}(\apis)
    +
    \beta_{\mathrm{s}}(\apis)
    \frac{\partial}{\partial \apis}Z_0^{\mathrm{d}/\mathrm{nd}}(\apis)\,.
    \label{eq:gamma-zero-definition}
\end{align}
Their perturbative expansion through $\mathcal{O}(\apis^2)$ reads\cite{Spiridonov:1988md, Chetyrkin:1994ex} 
\begin{gather}
\begin{aligned}
    \gamma_0^\textrm{d}&=-\frac{\nc}{(4\pi)^2}\left\{1+\apis\ccf+\apis^2\ccf\left[\cca\left(\frac{109}{32}-\frac{3}{2}\text{Li}_3\!\left(1\right)\right)\right.\right.\\
    &\hspace{3cm}\left.\left.-\ccf\left(\frac{131}{32}-3\text{Li}_3\!\left(1\right)\right)-\ctr\left(3+\frac{5}{8}\nf\right)\right]\right\}\,,\\
    \gamma_0^\textrm{nd}&=\frac{3\apis^2\nc\ccf\ctr}{(4\pi)^2}\,.
\end{aligned}
\end{gather}


  \section{Anomalous dimension in the
    \texorpdfstring{$\boldsymbol{\msbar}$}{MS} scheme}
\label{sec:gamma}
As discussed in \cref{sec:basis}, the mixing of the \cmdm\ operator into the
dimension-five bilinears as well as the identity operators, characterized by
$Z_{1,3},\ldots,Z_{1,10}$ in \cref{eq:RenMatrixStructure}, is, to the best of our
knowledge, not known at $\mathcal{O}(\apis^2)$.  However, our way of
evaluating the matching matrix directly determines the entries of the
renormalization matrix in the $\msbar$ scheme.  For this purpose, we first
define a general ansatz for the entries of interest. Note that the mixing with
identity operators at $\mathcal{O}(\apis^2)$ originates from three-loop
\vev-type diagrams. As a result, the corresponding renormalization matrix
contains poles whose maximal order is one higher than the perturbative
order. Therefore, we define $V$ as the set of indices corresponding to such
identity operators in a given basis.  For the basis described in
\cref{sec:basis}, we have $V=\{7,8,9,10,12,13,14\}$.  Furthermore, we define
\begin{equation}\label{eq:DeltaDef}
    \Delta_i
    =
    \begin{cases}
        1\,, & i\in V\,,\\
        0\,, & \mathrm{otherwise}\,,
    \end{cases}
\end{equation}
which allows us to write the ansatz for the entries of the renormalization
matrix compactly as
\begin{equation}\label{eq:Zansatz}
    Z_{ij}
    =
    \left(\hat\mu^2\right)^{-\epsilon(\Delta_j-\Delta_i)}
    \sum_{n=0}^{\infty}\apis^n
    \sum_{k=0}^{n+(1-\Delta_i)\Delta_j}
    \frac{z_{ij}^{(n,k)}}{\epsilon^k}\,.
\end{equation}
Here, $z_{ij}^{(0,0)}=\delta_{ij}$ and $z_{ij}^{(n,0)}= 0$ for $n\geq 1$,
since we work in the $\msbar$ scheme. The explicit factors of $\hat{\mu}^2$
arise from the difference in the mass dimension between the identity operators
and the field-containing operators.  The coefficients of the poles in this
ansatz are now determined by requiring that the renormalized matching matrix
$\zeta(t)=\zeta^\bare(t)Z^{-1}$ is finite.  A necessary consistency check is
that these coefficients are independent of the flow time $t$.  This requires a
non-trivial cancellation among the pole parts of the \lo, \nlo, and
\nnlo\ contributions. Following this procedure, we obtain the full
renormalization matrix $Z$, or, equivalently, the anomalous dimension matrix
$\gamma$ which determines the $\mu$ dependence of the operators:
\begin{equation}\label{eq:regAnoDim}
    \mu^2\frac{\mathrm{d}}{\mathrm{d}\mu^2}\mathcal{O}[\Lambda]
    =
    \gamma\,\mathcal{O}[\Lambda]\,,
    \qquad
    \mathrm{with}
    \qquad
    \gamma
    =
    \left(\mu^2\frac{\mathrm{d}}{\mathrm{d}\mu^2}Z\right)Z^{-1}.
\end{equation}
It has the same block structure as $Z$, cf.~\cref{eq:RenMatrixStructure},
\begin{equation}\label{eq:regAnoDimStructure}
\begingroup
\renewcommand*{\arraystretch}{1.5}
    \gamma
    =
    \begin{pmatrix}
        \gamma_{1,1} & \gamma_{1\times9} & & \\
        & \gamma_{9\times9} & & \\
        & & \gamma_{3\times3} & \\
        & & & -\gamma_m
    \end{pmatrix}\,.
\endgroup
\end{equation}
$\gamma_{1,1}$ can be obtained from \citere{Gorbahn:2005sa} and $\gamma_{1,2}$
from \citere{Naterop:2024cfx}. They read
\begin{align}
    \gamma_{1,1}&=-\apis\left[
    \frac{5}{4}\ccf
    -
    \frac{1}{2}\cca\right]\notag\\
    &\quad\quad-\apis^2\left[
    -
    \frac{51}{32}\ccf^2
    +
    \frac{1057}{288}\ccf\cca
    -
    \frac{11}{9}\cca^2
    +
    \nf\ctr
    \left(
        \frac{43}{144}\cca
        -
        \frac{41}{72}\ccf
    \right)\right]\,,\\
    \gamma_{1,2}
   &=
   \apis\ctr
   \left[
        1+\apis\left(\frac{5}{4}\cca+\frac{3}{2}\ccf\right)
   \right]\,.
\end{align}
Requiring finiteness of our results, we further find
\begin{gather}
\begin{aligned}\label{eq:regAnoDimResult3}
   \gamma_{1,3}
   &=
   3\apis\ccf
   \left[
        1-\apis\left(\ccf
            -\frac{169}{72}\cca
            +\frac{5}{18}\nf\ctr
        \right)
   \right]\,,\\
   \gamma_{1,4}
   &=\frac{1}{2}\apis^2\ccf\left(\ccf+\frac{1}{2}\cca\right)\,,\\
   \gamma_{1,5}
   &=-6\apis^2\ccf\ctr\,,\\
   \gamma_{1,6}
   &=-\frac{7}{2}\apis^2\ccf\ctr\,.
\end{aligned}
\end{gather}
Our results should reproduce the single-flavor limit considered in
\citere{Naterop:2024cfx}.  Indeed, we have verified that the relation
\begin{equation}
  \sum_{i=3}^{6}\gamma_{1,i}
  =\frac{g}{2m^2}\frac{\partial\dot{m}}{\partial L}\Bigg|_{L=0}
\end{equation}
holds, where $\dot{m}$ (and the notation of the \rhs) is given in Eq.~(A.4) of
\citere{Naterop:2024cfx}. This provides a strong consistency check of our
calculation.

For the mixing with the identity operators, we can summarize the results as
\begin{align}\label{eq:GammasNum}
       \gamma_{1,i}
       &=-\frac{\apis\nc\ccf}{2\pi^2}\left[\delta_{i,7}+\apis(1-\delta_{i,10})
         \Gamma_i\right]\,,\qquad \mathrm{for}\qquad i\in\{7,8,9,10\}\,. 
\end{align}
The $\mathcal{O}(\apis)$ contribution to
$\gamma_{1,7}$
agrees with \citere{Mereghetti:2021nkt}, whereas the $\mathcal{O}(\apis^2)$
contributions, parametrized by the coefficients $\Gamma_i$, are new
results. We obtain
\begin{gather}
\begin{aligned}
    \Gamma_7&=0.5000\,\cca+1.1250\,\ccf-0.4375\,\nf\ctr\,,\\
    \Gamma_8&=3.5000\,\ctr\,,\\
    \Gamma_9&=1.5625\,\ctr\,.
\end{aligned}
\end{gather}
Although we display the numerical results with four digits, the additional
digits obtained in our numerical evaluation allow us to identify the following
analytic expressions:
\begin{gather}\label{eq:GammasAna}
\begin{aligned}
  \Gamma_7&=\left(\frac{1}{2}\,\cca+\frac{9}{8}\ccf-\frac{7}{16}\,
  \nf\ctr\right)+\mathcal{O}\left(10^{-12}\right)\,,\\
  \Gamma_8&=\frac{7}{2}\ctr+\mathcal{O}\left(10^{-14}\right)\,,\\
  \Gamma_9&=\frac{25}{16}\ctr+\mathcal{O}\left(10^{-12}\right)\,.
\end{aligned}
\end{gather}
The deviations from these rational values are smaller than the estimated numerical uncertainties of our calculation. Thus, the numerical results and their precision provide strong evidence for the proposed exact analytic expressions.

For the reader's convenience, we also provide the remaining submatrices of the anomalous dimension matrix.
They are given by~\cite{Spiridonov:1984br,Spiridonov:1988md}

\begingroup
\renewcommand*{\arraystretch}{1.5}
\begin{equation}\label{eq:regAnoDim3}
    \gamma_{9\times9}=
    \begin{pmatrix}
        U_{5\times 5} & V_{5\times4}\\
        0_{4\times5} & -5\gamma_m\mathbbm{1}_{4\times4}
    \end{pmatrix},
\end{equation}
where
\begin{align}
    U_{5\times5} &=
    \begin{pmatrix}
        -\gamma_m-\apis(\beta_\mathrm{s}/\apis)' & 0 & 0 & 0 & -4\apis\gamma_m'\\
        0 & -\gamma_m & 0 & 0 & 0\\
        0 & 0 & -\gamma_m & 0 & 0\\
        0 & 0 & 0 & -\gamma_m & 0\\
        0 & 0 & 0 & 0 & -\gamma_m
    \end{pmatrix},
\end{align}
and
\begin{align}       
    V_{5\times4} &=
    \begin{pmatrix}
        0 & 0 & 4\apis(\gamma_0^{\mathrm d}-\gamma_0^{\mathrm{nd}})' & 4\apis(\gamma_0^{\mathrm{nd}})'\\
        -4(\gamma_0^{\mathrm d}-\gamma_0^{\mathrm{nd}}) & -4\gamma_0^{\mathrm{nd}} & 0 & 0\\
        -4(\gamma_0^{\mathrm d}-\gamma_0^{\mathrm{nd}}) & -4\gamma_0^{\mathrm{nd}} & 0 & 0\\
        0 & -4(\gamma_0^{\mathrm d}-\gamma_0^{\mathrm{nd}}) & 0 & -4\gamma_0^{\mathrm{nd}}\\
        0 & 0 & -4(\gamma_0^{\mathrm d}-\gamma_0^{\mathrm{nd}}) & -4\gamma_0^{\mathrm{nd}}
    \end{pmatrix},
\end{align}
with $f'\equiv\frac{\partial f}{\partial\apis}$. The dimension-three operators renormalize according to
\begin{equation}\label{eq:regAnoDim2}
    \gamma_{3\times3}=
    \begin{pmatrix}
        \gamma_m & -4(\gamma_0^{\mathrm d}-\gamma_0^{\mathrm{nd}}) & -4\gamma_0^{\mathrm{nd}}\\
        0 & -3\gamma_m & 0\\
        0 & 0 & -3\gamma_m
    \end{pmatrix}.
\end{equation}
\endgroup The known anomalous dimensions of the mass, $\gamma_m$, and of the
vacuum energy, $\gamma_0^{\mathrm{d}/\mathrm{nd}}$, were given in
\cref{app:rg-functions}. From the anomalous dimension matrix, one can
reconstruct the coefficients of $Z$ in \cref{eq:Zansatz}.  Expanding the
anomalous dimension matrix as
\begin{align}\label{eq:ExpansionCoff}
    \gamma_{i,j}
    &=
    \sum_{n=0}^{\infty}\apis^{n}\gamma^{(n)}_{i,j}\,,
\end{align}
one obtains the following recursion relation for the entries of the
renormalization matrix:
\begin{equation}
  z_{i,j}^{(n,k+1)}=\frac{1}{n-\Delta_i+\Delta_j}\left[-\sum_{r=0}^{n}
    \sum_l\gamma_{i,l}^{(r)}z_{l,j}^{(n-r,k)}-\sum_{m=0}^{n-1}m
    \beta_\textrm{s}^{(n-m-1)}z_{i,j}^{(m,k)}\right]\,.
\end{equation}
This relation is valid only for $n\neq \Delta_i-\Delta_j$. In the three exceptional cases for which this condition is not satisfied, the corresponding coefficients vanish,
\begin{align}
    z_{i,j}^{(n,k+1)}
    &=
    0
    \qquad\text{for}\qquad
    \begin{cases}
        n=0,\quad i,j\in V\ \text{or}\ i,j\notin V,\\
        n=1,\quad i\in V,\ j\notin V,
    \end{cases}
    \quad k\geq 0\,.
\end{align}
For $n=0$ and $i,j\notin V$, this follows from the absence of \lo\ mixing among field-containing operators. The remaining two cases follow from the multiplicative renormalization of the identity operators.

We want to emphasize that, although we employ the ringed scheme for the fermions, some of the renormalized flowed operators are still $\mu$-dependent. However, the bare ones are renormalization-group-invariant, i.e.\
\begin{equation}\label{eq:RGIops}
    \mu^2\frac{\mathrm{d}}{\mathrm{d}\mu^2}\tilde{\mathcal{O}}^\bare[t;\Lambda]=\mu^2\frac{\mathrm{d}}{\mathrm{d}\mu^2}
    \left[
        \mathcal{Z}^{-1}(t)\tilde {\mathcal{O}}[t;\Lambda]
    \right]
    =
    0\,,
\end{equation}
with $\mathcal{Z}$ defined in \cref{eq:flowedOpsRen} and~\eqref{eq:CursZdef}.
From this, we get the relation
\begin{equation}\label{eq:regAnoDimFromZeta}
    \gamma
    =
    -\zeta^{-1}(t)\,
    \mu^2\frac{\mathrm{d}}{\mathrm{d}\mu^2}\zeta(t)
    +
    \zeta^{-1}(t)
    \gamma_\mathcal{Z}(t)
    \zeta(t)\,,
\end{equation}
with
\begin{equation}\label{eq:GammaZ}
        \gamma_\mathcal{Z}(t)=\mu^2\frac{\mathrm{d}}{\mathrm{d}\mu^2}\ln\mathcal{Z}(t)\,,
\end{equation}
which provides an important check, since the non-trivial $t$-dependence on the
right-hand side must cancel, rendering the left-hand side independent of the
flow time. If, as in the case considered in this paper, the flow-time
dependence of $\mathcal{Z}(t)$ enters only through the ringed scheme, then
$\gamma_{\mathcal{Z}}$ is independent of $t$, and the last term in
\cref{eq:MuDepFlowAnoDim} vanishes.

As mentioned in \cref{sec:sftx}, the logarithmic $t$-dependence can be determined analytically from the anomalous dimension.
To this end, we expand both anomalous dimension matrices, $\gamma$ and $\gamma_{\mathcal{Z}}$, as in \cref{eq:ExpansionCoff}, while keeping the coefficients in matrix form.
The matching matrix $\zeta(t)$ is expanded as
\begin{align}
    \zeta(t)=\sum_{n=0}^\infty\apis^n\sum_{k=0}^{n+1}\lmut^k\zeta_{k}^{(n)}\,.
\end{align}
One then finds the all-order recursive relation
\begin{align}\label{eq:regAnoDimRecursive}
    \zeta_{k+1}^{(n)}
    =
    \frac{1}{k+1}
    \left[
        \sum_{r=0}^{n-2}
        (n-r-1)\beta_{\mathrm{s}}^{(r)}
        \zeta_k^{(n-r-1)}
        -
        \sum_{r=0}^{n}
        \left(
            \zeta_k^{(n-r)}\gamma^{(r)}
            -
            \gamma_{\mathcal{Z}}^{(r)}\zeta_k^{(n-r)}
        \right)
    \right]\,,
\end{align}
which allows us to determine all logarithmic coefficients $\zeta_k^{(n)}$ with
$k\geq 1$ from the non-logarithmic terms, provided the corresponding anomalous
dimensions are known.  Notice that $\zeta(t)$ depends on $t$ both
logarithmically and through explicit powers of $t$, thus
\begin{equation}\label{eq:Reltmu2}
    \mu^2\frac{\mathrm{d}}{\mathrm{d}\mu^2}\zeta(t)
    =
    \beta_{\textrm{s}}\frac{\partial}{\partial\apis}\zeta(t)
    +
    \frac{\partial}{\partial\lmut}\zeta(t)
    =
    \beta_{\textrm{s}}\frac{\partial}{\partial\apis}\zeta(t)
    +
    t\frac{\mathrm{d}}{\mathrm{d}t}\zeta(t)-
     t\frac{\partial}{\partial t}\zeta(t)\Big|_{\lmut}\,,
\end{equation}
where, in the last term, $\lmut$ is kept fixed, such that the derivative acts only on the explicit $t$-dependence.
Inserting this relation into \cref{eq:regAnoDimFromZeta}, we obtain
\cref{eq:flowAnoDimFromZeta}.


\section{Projectors}
\label{sec:Projectors}
In the calculation outlined in \cref{sec:zeta}, it is sufficient to introduce
four distinct flavor classes.  The flavors $p$ and $q$ are selected by
$\Lambda_\star$ and therefore appear explicitly in the operators and
projectors; see \cref{eq:LamStar}.  An additional flavor $r$ is required for
the construction of certain projectors.  For example,
$\mathcal{O}_{10}=\mathrm{Tr}(\Lambda M)\mathrm{Tr}(M^2)^2$ can only be
distinguished from the other identity operators by taking derivatives
\wrt\ three different fermion masses.

The remaining flavor class $s$ does not appear explicitly in the
projectors. Owing to the flavor blindness of \qcd\ and the fact that the
relevant mass scales are set to zero after applying the projectors, the
contribution associated with $s$ is identical to that of any further fermion
flavor added to the theory. Assigning the class $s$ a multiplicity
$n_{\textrm{f}_s}$ therefore accounts for all such additional
flavors. Consequently, setting $n_{\textrm{f}_s}=\nf-3$ recovers the most
general flavor dependence relevant for the present calculation.

In the following, we provide the projectors used in the calculation.
All momenta $k_i$ are considered outgoing, and $m_\bare=0$ denotes $m_1=\ldots=m_{\nf}=0$.
The projectors onto the physical operators are
\begin{align}\label{eq:physProj}
P_1[X]&\equiv\frac{-1}{8\na\ctr g_\bare D(D-1)}\frac{\partial}{\partial k_3^\nu}\Big\langle \bar\psi_p^i(k_1)\psi^j_p(k_2) A^a_\mu(k_3)\Big\vert\sigma_{\mu\nu}T^a_{ji}\,X\Big\vert0\Big\rangle^{\mathrm{1PI,\,amp}}_{k_1,k_2,k_3,m_\bare=0}\,,\\
P_2[X]&\equiv\frac{-1}{4\na D(D-1)}\frac{\partial}{\partial m_{p,\bare}}\frac{\partial}{\partial k_1^\mu}\frac{\partial}{\partial k_2^\mu}\Big\langle A^a_\nu(k_1)A^a_\nu(k_2)\Big\vert X\Big\vert0\Big\rangle^{\mathrm{1PI,\,amp}}_{k_1,k_2,m_\bare=0}\,,\\
P_3[X]&\equiv\frac{-1}{8\nc}\frac{\partial^2}{\partial m_{p,\bare}^2}\Big\langle \bar\psi_p^i(k_1)\psi_p^i(k_2)\Big\vert X\Big\vert0\Big\rangle^{\mathrm{1PI,\,amp}}_{k_1,k_2,m_\bare=0}\,,\\
P_4[X]&\equiv\frac{-1}{4\nc}\frac{\partial}{\partial m_{p,\bare}}\frac{\partial}{\partial m_{q,\bare}}\Big\langle \bar\psi_p^i(k_1)\psi_q^i(k_2)\Big\vert X\Big\vert0\Big\rangle^{\mathrm{1PI,\,amp}}_{k_1,k_2,m_\bare=0}\,,\\
P_5[X]&\equiv\frac{-1}{8\nc}\frac{\partial^2}{\partial m_{r,\bare}^2}\Big\langle \bar\psi_p^i(k_1)\psi_p^i(k_2)\Big\vert X\Big\vert0\Big\rangle^{\mathrm{1PI,\,amp}}_{k_1,k_2,m_\bare=0}\,,\\
P_6[X]&\equiv\frac{-1}{4\nc}\frac{\partial}{\partial m_{p,\bare}}\frac{\partial}{\partial m_{r,\bare}}\Big\langle \bar\psi_r^i(k_1)\psi_r^i(k_2)\Big\vert X\Big\vert0\Big\rangle^{\mathrm{1PI,\,amp}}_{k_1,k_2,m_\bare=0}\,,\\
P_{7}[X]&\equiv\frac{1}{5!}\frac{\partial^5}{\partial m_{p,\bare}^5}\Big\langle0\Big\vert X\Big\vert0\Big\rangle^{\mathrm{1PI,\,amp}}_{m_\bare=0}\,,\\
P_{8}[X]&\equiv\frac{1}{3!}\frac{\partial^3}{\partial m_{p,\bare}^3}\frac{1}{2!}\frac{\partial^2}{\partial m_{r,\bare}^2}\Big\langle0\Big\vert X\Big\vert0\Big\rangle^{\mathrm{1PI,\,amp}}_{m_\bare=0}\,,\\
P_{9}[X]&\equiv\frac{1}{4!}\frac{\partial^4}{\partial m_{r,\bare}^4}\frac{\partial}{\partial m_{p,\bare}}\Big\langle0\Big\vert X\Big\vert0\Big\rangle^{\mathrm{1PI,\,amp}}_{m_\bare=0}\,,\\
P_{10}[X]&\equiv\frac{1}{2}\frac{1}{2!}\frac{\partial^2}{\partial m_{q,\bare}^2}\frac{1}{2!}\frac{\partial^2}{\partial m_{r,\bare}^2}\frac{\partial}{\partial m_{p,\bare}}\Big\langle0\Big\vert X\Big\vert0\Big\rangle^{\mathrm{1PI,\,amp}}_{m_\bare=0}\,,\\
P_{11}[X]&\equiv\frac{-1}{4\nc}\Big\langle \bar\psi^i_p(k_1)\psi^i_q(k_2)\Big\vert X\Big\vert0\Big\rangle^{\mathrm{1PI,\,amp}}_{k_1,k_2,m_\bare=0}\,,\\
P_{12}[X]&\equiv\frac{1}{3!}\frac{\partial^3}{\partial m_{p,\bare}^3}\Big\langle0\Big\vert X\Big\vert0\Big\rangle^{\mathrm{1PI,\,amp}}_{m_\bare=0}\,,\\
P_{13}[X]&\equiv\frac{1}{2!}\frac{\partial^2}{\partial m_{r,\bare}^2}\frac{\partial}{\partial m_{p,\bare}}\Big\langle0\Big\vert X\Big\vert0\Big\rangle^{\mathrm{1PI,\,amp}}_{m_\bare=0}\,,\\
P_{14}[X]&\equiv\frac{\partial}{\partial m_{p,\bare}}\Big\langle0\Big\vert X\Big\vert0\Big\rangle^{\mathrm{1PI,\,amp}}_{m_\bare=0}\,.
\end{align}
The linear combinations used to construct the orthogonal projectors are given
in \cref{eq:OrthProj}.  For this construction, projectors onto the unphysical
operators are required.  We use
\begin{align}\label{eq:NuisProj}
    P_{\mathcal{N}_+}[X]&\equiv\frac{i}{8\nc D}\frac{\partial}{\partial m_{p,\bare}}\frac{\partial}{\partial k_1^\mu}\Big\langle \bar\psi_p^i(k_1)\psi_p^i(k_2)\Big\vert\gamma_\mu X\Big\vert0\Big\rangle^{\mathrm{1PI,\,amp}}_{k_1,k_2,m_\bare=0}\,,\\
    P_{\mathcal{N}_-}[X]&\equiv\frac{i}{4\nc D}\frac{\partial}{\partial m_{q,\bare}}\frac{\partial}{\partial k_1^\mu}\Big\langle \bar\psi_p^i(k_1)\psi_q^i(k_2)\Big\vert\gamma_\mu X\Big\vert0\Big\rangle^{\mathrm{1PI,\,amp}}_{k_1,k_2,m_\bare=0}\,,\\
    P_{\mathcal{N}_1}[X]&\equiv\frac{-i}{4\nc D}\frac{\partial}{\partial m_{p,\bare}}\frac{\partial}{\partial k_2^\mu}\Big\langle \bar\psi_r^i(k_1)\psi_r^i(k_2)\Big\vert\gamma_\mu X\Big\vert0\Big\rangle^{\mathrm{1PI,\,amp}}_{k_1,k_2,m_\bare=0}\,,\\
    P_{\mathcal{N}_2}[X]&\equiv\frac{1}{16\na\ctr g_\bare^2 D}\Big\langle \bar\psi_p^i(k_1)\psi_p^i(k_2)A^a_\mu(k_3)A_\mu^a(k_4)\Big\vert X\Big\vert0\Big\rangle^{\mathrm{1PI,\,amp}}_{k_1,k_2,k_3,k_4,m_\bare=0}\,,\\
    P_{\mathcal{N}_3}[X]&\equiv\frac{i}{4\nc D(D-1)}\frac{\partial}{\partial k_1^{\mu}}\frac{\partial}{\partial k_2^{\nu}}\Big\langle \bar\psi_p^i(k_1)\psi_p^i(k_2)\Big\vert\sigma_{\mu\nu} X\Big\vert0\Big\rangle^{\mathrm{1PI,\,amp}}_{k_1,k_2,m_\bare=0}\,,\\
    P_{\mathcal{N}_4}[X]&\equiv\frac{-i}{4\na\ctr g_\bare D}\frac{\partial}{\partial k_1^{\mu}}\Big\langle \bar\psi_p^i(k_1)\psi^j_p(k_2)A_\mu^a(k_3)\Big\vert T^a_{ji} X\Big\vert0\Big\rangle^{\mathrm{1PI,\,amp}}_{k_1,k_2,k_3,m_\bare=0}\,.
\end{align}


\section{Remaining entries of the matching matrix}
In \cref{sec:Results}, we presented the new results obtained in this work,
namely the \sftx\ of the \cmdm\ operator and the $\mathcal{O}(t)$ contribution
to the scalar bilinear. To determine quantities such as $\tilde{\gamma}(t)$,
however, the full matching matrix is required. The remaining entries have been
calculated previously; see
\cref{tab:sftx_structure,tab:OverviewOfCalcs}. These results were not given
for a general flavor matrix $\Lambda$. Although they can be disentangled from
the existing calculations, for the reader's convenience we provide the
corresponding expressions for a general flavor matrix here. The non-vanishing
coefficients in the \sftx\ of the operator $\tilde{\mathcal{O}}_2(t)$ take the
form (see \cref{tab:sftx_structure})
\begin{equation}\label{eq:SFTXmG2}
    \tilde{\mathcal{O}}_2[t;\Lambda]=\sum_{i\in\{2,6,9,10,13,14\}}\left(\zeta_2\right)_i\mathcal{O}_i[\Lambda]\,,
\end{equation}
where the vector $\zeta_2$ is related to the full matching matrix through Eq.~\eqref{eq:ZetaVec}. The coefficients read
\begin{align}
\left(\zeta_2\right)_2&=1
+\frac{7}{8}\apis\,\cca+\apis^2\bigg\{
    \cca^2\left(
        \frac{227}{180}
        -\frac{87}{80}\ln 2
        +\frac{27}{40}\ln 3
    \right)-\frac{1}{18}\cca\nf\ctr
\notag\\
&\qquad\qquad\qquad
    +\frac{3}{16}\ccf\nf\ctr
    +\lmut\left(
        \frac{3}{32}\cca^2
        +\frac{1}{8}\cca\nf\ctr
        +\frac{1}{4}\ccf\nf\ctr
    \right)
\bigg\}\,,\\
\left(\zeta_2\right)_6&=\apis\,\ccf\bigg\{
    \frac{5}{2}
    +3\lmut
    +\apis\bigg[
        \cca\left(
            \frac{431}{30}
            -\text{Li}_2\!\left(1\right)
            +\frac{4273}{60}\ln 2
            -\frac{2139}{40}\ln 3
            -\frac{1}{4}\text{Li}_2\!\left(\frac{1}{4}\right)
        \right)\notag\\
        &\quad\quad
        +\frac{17}{16}\ccf
        -\nf\ctr\left(
            \frac{15}{4}
            +\text{Li}_2\!\left(1\right)
        \right)
        +\lmut\left(
            \frac{367}{24}\cca
            +\frac{3}{4}\ccf
            -\frac{10}{3}\nf\ctr
        \right)\notag\\
        &\quad\quad
        +\lmut^2\left(
            \frac{11}{4}\cca
            -\nf\ctr
        \right)
    \bigg]
\bigg\}\,,\\
\left(\zeta_2\right)_9&=\frac{\apis\,\ccf\nc}{16\pi^2}
\bigg\{
    5
    -6\text{Li}_2\!\left(1\right)
    -6\lmut
    -6\lmut^2+\apis\bigg[
        -3.3144\,\cca
        -27.5707\,\ccf
\notag\\
&\quad\quad
        +19.5780\,\nf\ctr+\lmut\Biggl[
        \ccf\left(
                -32
                -18\text{Li}_2\!\left(1\right)
                +24\text{Li}_3\!\left(1\right)
            \right)
            \notag\\
&\quad\quad+\nf\ctr\left(
                \frac{20}{3}
                +8\text{Li}_2\!\left(1\right)
            \right)+\cca\left(
                -\frac{351}{20}
                -7\text{Li}_2\!\left(1\right)
                -\frac{4273}{15}\ln 2
                +\frac{2139}{10}\ln 3
                \right.\notag\\
&\quad\quad\left.+\text{Li}_2\!\left(\frac{1}{4}\right)
                -12\text{Li}_3\!\left(1\right)
            \right)\Biggr]
        +\lmut^2\left(
            -\frac{433}{12}\cca
            -\frac{33}{2}\ccf
            +\frac{26}{3}\nf\ctr
        \right)
\notag\\
&\quad\quad
        +\lmut^3\left(
            -\frac{22}{3}\cca
            -6\ccf
            +\frac{8}{3}\nf\ctr
        \right)
    \bigg]
\bigg\}\,,\\
\left(\zeta_2\right)_{10}&=-\frac{\apis^2\,\ccf\nc\ctr}{16\pi^2}
\left(
    34.6667
    +24\lmut
\right)\,,\\
\left(\zeta_2\right)_{13}&=\frac{3}{8}\frac{\apis\,\na\ctr}{\pi^2 t}
\bigg\{
    1
    +\apis\bigg[
        7.4379\,\cca
        +2\,\ccf
        -\frac{10}{9}\,\nf\ctr
\notag\\
&\quad\quad
        +\lmut\left(
            \frac{11}{6}\cca
            +\frac{3}{2}\ccf
            -\frac{2}{3}\nf\ctr
        \right)
    \bigg]
\bigg\}\,,\\
\left(\zeta_2\right)_{14}&=\frac{3\na}{32\pi^2 t^2}
\bigg\{
    1
    +\apis\left[
        \cca\left(
            \frac{13}{9}
            +\frac{11}{6}\ln 2
            -\frac{3}{4}\ln 3
        \right)
        -\frac{2}{9}\nf\ctr
        +\lmut\left(
            \frac{11}{12}\cca
            -\frac{1}{3}\nf\ctr
        \right)
    \right]
\notag\\
&\quad
    +\apis^2\bigg[
        1.7487\,\cca^2
        -1.9728\,\cca\nf\ctr
        +0.3062\,\ccf\nf\ctr
        +0.1210\,\nf^2\ctr^2
\notag\\
&\quad\quad
        +\lmut\left[
            \cca^2\left(
                \frac{725}{216}
                +\frac{121}{36}\ln 2
                -\frac{11}{8}\ln 3
            \right)
            +\cca\nf\ctr\left(
                -\frac{193}{108}
                -\frac{11}{9}\ln 2
                +\frac{1}{2}\ln 3
            \right)
\right.
\notag\\
&\quad\quad\quad\left.
            -\frac{1}{4}\ccf\nf\ctr
            +\frac{4}{27}\nf^2\ctr^2
        \right]
        +\lmut^2\left(
            \frac{121}{144}\cca^2
            -\frac{11}{18}\cca\nf\ctr
            +\frac{1}{9}\nf^2\ctr^2
        \right)
    \bigg]
\bigg\}\,.
\end{align}

The \sftx\ of the scalar bilinear
$\tilde{\mathcal{O}}_\text{\abbrev{S}}[t;\Lambda]$ reads
\begin{equation}\label{eq:SFTXFF}
  \tilde{\mathcal{O}}_\text{\abbrev{S}}[t;\Lambda] =
  \tilde{\mathcal{O}}_{11}[t;\Lambda]
  =\sum_{i\in\{11,12,13,14\}} \left(\zeta_{\text{\abbrev{S}}}\right)_i
  \mathcal{O}_i[\Lambda]+\ldots\,,
\end{equation}
where the ellipsis denotes operators of mass dimension $d>3$. The
contributions from the $d=5$ operators have been presented in
\cref{sec:ordert}, while contributions from $d>5$ are consistently
neglected. The corresponding coefficients read
\begin{align}
\left(\zeta_{\text{\abbrev{S}}}\right)_{11}&=1
-\apis\,\ccf\left(
    \frac{1}{2}
    +\ln 2
    +\frac{3}{4}\ln 3
    +\frac{3}{4}\lmut
\right)+\apis^2\bigg\{
    \frac{c_\chi^{(2)}}{16}    +\ccf\nf\ctr\left[
        \frac{5}{6}
        +\frac{1}{4}\text{Li}_2\!\left(1\right)
    \right]\notag
\\
&\quad\quad+\cca\ccf\left[
        -\frac{197}{48}
        -\frac{7}{16}\text{Li}_2\!\left(1\right)
        -\frac{9}{2}\ln 2
        +\frac{1}{4}\ln^2 2
        +\frac{9}{2}\ln 3
        +\frac{3}{2}\text{Li}_2\!\left(\frac{1}{4}\right)
    \right]\notag
\\
&\quad\quad
    +\ccf^2\left[
        \frac{1}{2}
        +\frac{1}{2}\text{Li}_2\!\left(1\right)
        -\frac{3}{2}\ln 2
        +\frac{1}{4}\ln^2 2
        +\frac{21}{8}\ln 3
        +3\text{Li}_2\!\left(\frac{1}{4}\right)
    \right]\notag\notag
\\
&\quad
    +\lmut\left[
        -\cca\ccf\left(
            \frac{47}{32}
            +\frac{11}{12}\ln 2
            +\frac{11}{16}\ln 3
        \right)
\right.\notag
\\
&\quad\quad\left.
+\ccf^2\left(
            \frac{9}{32}
            +\frac{3}{4}\ln 2
            +\frac{9}{16}\ln 3
        \right)
        +\ccf\nf\ctr\left(
            \frac{3}{8}
            +\frac{1}{3}\ln 2
            +\frac{1}{4}\ln 3
        \right)
    \right]\notag\\
&\quad\quad
    +\lmut^2\left(
        -\frac{11}{32}\cca\ccf
        +\frac{9}{32}\ccf^2
        +\frac{1}{8}\ccf\nf\ctr
    \right)
\bigg\}\,,\label{eq:zetaS11}\\
\left(\zeta_{\text{\abbrev{S}}}\right)_{12}&=-\frac{\nc}{4\pi^2}\bigg\{
    1+\lmut
    +\apis\,\ccf\bigg[
        \frac{7}{2}
        +4\ln 2
        -\frac{21}{4}\ln 3
        -3\text{Li}_2\!\left(\frac{1}{4}\right)
\notag\\
&\quad\quad
        +\lmut\left(
            \frac{11}{4}
            -\ln 2
            -\frac{3}{4}\ln 3
        \right)
        +\frac{3}{4}\lmut^2
    \bigg]
\notag\\
&\quad
    +\apis^2\bigg[
        \frac{c_\chi^{(2)}}{16}
        +1.5137\,\cca\ccf
        +3.5553\,\ccf^2
        -0.8046\,\ccf\nf\ctr
\notag\\
&\quad\quad
        +\lmut\Biggl[
            \frac{c_\chi^{(2)}}{16}+\ccf\nf\ctr\left(
                -\frac{19}{12}
                +\frac{1}{4}\text{Li}_2\!\left(1\right)
                -\frac{4}{3}\ln 2
                +\frac{7}{4}\ln 3
                +\text{Li}_2\!\left(\frac{1}{4}\right)
            \right)
\notag\\
&\quad\quad
            +\cca\ccf\left(
                \frac{133}{24}
                -\frac{7}{16}\text{Li}_2\!\left(1\right)
                -\frac{5}{6}\ln 2
                +\frac{1}{4}\ln^2 2
                -\frac{5}{16}\ln 3
                -\frac{5}{4}\text{Li}_2\!\left(\frac{1}{4}\right)
                -\frac{3}{2}\text{Li}_3\!\left(1\right)
            \right)\notag
\\
&\quad\quad
            +\ccf^2\left(
                \frac{65}{16}
                +\frac{1}{2}\text{Li}_2\!\left(1\right)
                +\frac{13}{2}\ln 2
                +\frac{1}{4}\ln^2 2
                -\frac{159}{16}\ln 3
                -\frac{15}{4}\text{Li}_2\!\left(\frac{1}{4}\right)
                +3\text{Li}_3\!\left(1\right)
            \right)\Biggr]\notag    
\\
&\quad\quad
        +\lmut^2\Biggl[
            \cca\ccf\left(
                \frac{49}{24}
                -\frac{11}{12}\ln 2
                -\frac{11}{16}\ln 3
            \right)
            +\ccf^2\left(
                3
                -\frac{3}{4}\ln 2
                -\frac{9}{16}\ln 3
            \right)
\notag\\
&\quad\quad
            +\ccf\nf\ctr\left(
                -\frac{7}{12}
                +\frac{1}{3}\ln 2
                +\frac{1}{4}\ln 3
            \right)
        \Biggr]
\notag\\
&\quad\quad
        +\lmut^3\left(
            \frac{11}{96}\cca\ccf
            +\frac{21}{32}\ccf^2
            -\frac{1}{24}\ccf\nf\ctr
        \right)
    \bigg]
\bigg\}\,,\label{eq:zetaS12}\\
\left(\zeta_{\text{\abbrev{S}}}\right)_{13}&=\frac{\apis^2\,\ccf\nc\ctr}{4\pi^2}
\left(
    5.3337
    +3\lmut
\right)\,,\label{eq:zetaS13}\\
\left(\zeta_{\text{\abbrev{S}}}\right)_{14}&=-\frac{\nc}{8\pi^2 t}\bigg\{
    1
    +\apis\,\ccf\left(
        1
        +\ln 2
        -\frac{3}{4}\ln 3
        +\frac{3}{4}\lmut
    \right)\notag
\\
&\quad
    +\apis^2\bigg[
        \frac{c_\chi^{(2)}}{16}
        +4.0740\,\cca\ccf
        -0.6718\,\ccf^2
        -0.7422\,\ccf\nf\ctr\notag
\\
&\quad\quad
        +\lmut\left[
            \cca\ccf\left(
                \frac{185}{96}
                +\frac{11}{12}\ln 2
                -\frac{11}{16}\ln 3
            \right)
            +\ccf^2\left(
                \frac{27}{32}
                +\frac{3}{4}\ln 2
                -\frac{9}{16}\ln 3
            \right)
\right.\notag
\\
&\quad\quad\quad\left.
            +\ccf\nf\ctr\left(
                -\frac{13}{24}
                -\frac{1}{3}\ln 2
                +\frac{1}{4}\ln 3
            \right)
        \right]\notag
\\
&\quad\quad
        +\lmut^2\left(
            \frac{11}{32}\cca\ccf
            +\frac{9}{32}\ccf^2
            -\frac{1}{8}\ccf\nf\ctr
        \right)
    \bigg]
\bigg\}\,.\label{eq:zetaS14}
\end{align}


\section{Application to special cases}\label{sec:special}

In order to illustrate how our results for general mass and flavor matrices
$M$ and $\Lambda$ can be applied, it may be useful to consider a few special
cases. In \citeres{Borgulat:2023xml, Harlander:2020duo}, a setup with $\nh$
fermions of degenerate mass and $\nl$ massless quarks is considered. We adopt
this setup in the present section. Ordering the flavors according to their
masses, we write
\begin{equation}
M=\textrm{diag}(\underbrace{m,\ldots,m}_{\nh},\underbrace{0,\ldots,0}_{\nl})\,.
\end{equation}
From this follows $\mathrm{Tr}(M^k)=\nh m^k$. In the following, we consider
several flavor configurations to illustrate how the \sftx\ for a specific
flavor structure can be obtained from our general results. We define the sets
of heavy- and light-flavor indices as $\mathcal{H}=\{1,\ldots,\nh\}$ and
$\mathcal{L}=\{\nh+1,\ldots,\nf\}$, respectively, with $\nf=\nh+\nl$. The
procedure for an arbitrary choice of the flavor matrix follows the same
strategy. First, the operator basis is reduced by inserting the specific
choice of $\Lambda$, identifying operators that vanish and eliminating
linearly dependent structures whenever relations among the operators
arise. The matching matrix in the reduced basis is then obtained by combining
the corresponding entries of the general matching matrix according to the same
linear relations.\\

\textbf{Case 1: $\Lambda_{hh'}$}\\
In the first case, we choose 
\begin{equation}
\left(\Lambda_{hh'}\right)_{ij}=\frac{\delta_{ih}\delta_{jh'}+\delta_{ih'}\delta_{jh}}{1+\delta_{hh'}}\,,\qquad \textrm{with}\qquad h,h'\in\mathcal{H}\,,
\end{equation}
which yields $\mathrm{Tr}(\Lambda_{hh'}M^k)=\delta_{hh'}m^k$. From this, certain linear relations between operators arise, i.e.
\begin{gather}
\begin{aligned}
\mathcal{O}_3[\Lambda_{hh'}]
&=\mathcal{O}_4[\Lambda_{hh'}]
=m^2\mathcal{O}_{\text{\abbrev{S}}}[\Lambda_{hh'}]\,,\\
\mathcal{O}_5[\Lambda_{hh'}]
&=\nh m^2\mathcal{O}_{\text{\abbrev{S}}}[\Lambda_{hh'}]\,,\\
\mathcal{O}_8[\Lambda_{hh'}]
&=\mathcal{O}_9[\Lambda_{hh'}]
=\nh\mathcal{O}_7[\Lambda_{hh'}]
=\delta_{hh'}\,\nh m^5\mathbbm{1}\,,\\
\mathcal{O}_{10}[\Lambda_{hh'}]
&=\nh^2\mathcal{O}_7[\Lambda_{hh'}]
=\delta_{hh'}\,\nh^2 m^5\mathbbm{1}\,,\\
\mathcal{O}_{13}[\Lambda_{hh'}]
&=\nh\mathcal{O}_{12}[\Lambda_{hh'}]
=\delta_{hh'}\,\nh m^3\mathbbm{1}\,,\\
\mathcal{O}_{14}[\Lambda_{hh'}]
&=\delta_{hh'}\,m\mathbbm{1}\,.
\end{aligned}
\end{gather}
Thus, all operators reduce to the structures $\mathcal{O}_1$, $\mathcal{O}_2$, $\mathcal{O}_6$, $\mathcal{O}_{\text{\abbrev{S}}}$, and $\mathbbm{1}$, and the \sftx s read
\begin{gather}\label{eq:SFTXhh}
\begin{aligned}
\tilde{\mathcal O}_i[t;\Lambda_{hh'}]
&=
\left(\zeta_{i}\right)_1\mathcal{O}_{\text{\abbrev{CM}}}[\Lambda_{hh'}]+
\Bigg\{
m^2\Big[
\left(\zeta_{i}\right)_3
+
\left(\zeta_{i}\right)_4
+
n_h\left(\zeta_{i}\right)_5
\Big]
+
\left(\zeta_{i}\right)_{11}
\Bigg\}
\mathcal{O}_{\text{\abbrev{S}}}[\Lambda_{hh'}]
\\[1mm]
&+
\bigg\{
\left(\zeta_{i}\right)_2
\mathcal{O}_{2}[\Lambda_{hh'}]
+
\left(\zeta_{i}\right)_6
\mathcal{O}_6[\Lambda_{hh'}]
\bigg\}
\\[1mm]
&+
\delta_{hh'}
\bigg\{
m^5
\Big[
\left(\zeta_{i}\right)_7
+
n_h\Big(
\left(\zeta_{i}\right)_8
+
\left(\zeta_{i}\right)_9
\Big)
+
n_h^2\left(\zeta_{i}\right)_{10}
\Big]
\\
&\hspace{2.4cm}
+
m^3
\Big[
\left(\zeta_{i}\right)_{12}
+
n_h\left(\zeta_{i}\right)_{13}
\Big]
+
m\left(\zeta_{i}\right)_{14}
\bigg\}
\mathbbm{1}
+\ldots\,.
\end{aligned}
\end{gather}
\textbf{Case 2: $\Lambda_{ll'}$}\\
We now select two quarks from the massless sector with the flavor matrix
\begin{equation}
\left(\Lambda_{ll'}\right)_{ij}=\frac{\delta_{il}\delta_{jl'}+\delta_{il'}\delta_{jl}}{1+\delta_{ll'}}\,,\qquad \textrm{with}\qquad l,l'\in\mathcal{L}\,.
\end{equation}
In this case, the results simplify further because $\mathrm{Tr}(\Lambda_{ll'}M^k)=0$, such that
\begin{equation}
    \mathcal{O}_j[\Lambda_{ll'}]=0\,,\qquad\forall j\in\{1,\ldots,14\}\setminus\{1,5,11\}\,,
\end{equation}
while
\begin{equation}
    \mathcal{O}_5[\Lambda_{ll'}]=m^2\nh \mathcal{O}_{\text{\abbrev{S}}}[\Lambda_{ll'}]\,.
\end{equation}
The corresponding \sftx s therefore reduce to
\begin{equation}
    \tilde{\mathcal{O}}_i[t;\Lambda_{ll'}]=\left(\zeta_{i}\right)_1\mathcal{O}_{\text{\abbrev{CM}}}[\Lambda_{ll'}]+\bigg[m^2 \nh \left(\zeta_{i}\right)_5+\left(\zeta_{i}\right)_{11}\bigg]\mathcal{O}_{\text{\abbrev{S}}}[\Lambda_{ll'}]\,.
\end{equation}

\textbf{Case 3: $\Lambda_{hl}$}\\
As a final example, we consider an off-diagonal mixed-flavor structure that couples one massless and one massive quark to the \cmdm,
\begin{equation}
\left(\Lambda_{hl}\right)_{ij}=\delta_{il}\delta_{jh}+\delta_{ih}\delta_{jl}\,,\qquad \textrm{with}\qquad l\in\mathcal{L},\,h\in\mathcal{H}\,.
\end{equation}
In this case, simplifications similar to those in Case 2 occur:
\begin{equation}
    \mathcal{O}_j[\Lambda_{hl}]=0\,,\qquad\forall j\in\{1,\ldots,14\}\setminus\{1,3,5,11\}\,,
\end{equation}
with
\begin{equation}
    \mathcal{O}_3[\Lambda_{hl}]=\frac{1}{2}m^2 \mathcal{O}_{\text{\abbrev{S}}}[\Lambda_{hl}]\,,\qquad
    \mathcal{O}_5[\Lambda_{hl}]=m^2\nh \mathcal{O}_{\text{\abbrev{S}}}[\Lambda_{hl}]\,.
\end{equation}
Thus, the \sftx s reduce to
\begin{equation}
    \tilde{\mathcal{O}}_i[t;\Lambda_{hl}]=\left(\zeta_{i}\right)_1\mathcal{O}_{\text{\abbrev{CM}}}[\Lambda_{hl}]+\bigg[m^2 \left(\frac{1}{2}\left(\zeta_i\right)_3+\nh \left(\zeta_{i}\right)_5\right)+\left(\zeta_{i}\right)_{11}\bigg]\mathcal{O}_{\text{\abbrev{S}}}[\Lambda_{hl}]\,.
\end{equation}
This procedure generalizes straightforwardly to arbitrary flavor matrices. Another important example is the flavor-singlet case, where one may either set $\Lambda_{\mathrm{S}}=\mathbbm{1}_{\nf\times\nf}$ and apply the procedure described above directly, or exploit the linearity in the flavor matrix and use
\begin{equation}
    \mathcal{O}_i[\Lambda_{\mathrm{S}}]
    =
    \sum_{h=1}^{\nh}\mathcal{O}_i[\Lambda_{hh}]
    +
    \sum_{l=\nh+1}^{\nf}\mathcal{O}_i[\Lambda_{ll}]\,.
\end{equation}
Both approaches lead to identical results.


\section{Ancillary file}\label{app:ancillary_file}
For the reader's convenience, we provide the full $14\times 14$ matching
matrix $\zeta(t)$ and the flowed anomalous dimension matrix
$\tilde{\gamma}(t)$ in machine-readable form, both as a Mathematica and as a
Python file. Since it is most useful in applications, we also provide the
inverse matching matrix $\zeta^{-1}(t)$, which expresses the unflowed
operators in terms of the flowed ones,
$\mathcal{O}=\zeta^{-1}(t)\,\tilde{\mathcal{O}}(t)$. The results are given in
the ringed scheme; their translation to the \msbar\ scheme is
straightforward. They are expressed in terms of the variables listed in
\cref{tab:AncFile}. The replacement rules \texttt{RepcChi}, \texttt{RepQCD} and
\texttt{RepGamma} insert $c_{\chi}^{(2)}$ according to
\cref{eq:RingedScheme}, the values of the color factors of \qcd\ and the $\Gamma_i$ according to
\cref{eq:GammasAna}, respectively. Coefficients that are known only
numerically are given to six significant digits; all others are given
analytically. As in Section~\ref{sec:gammatilde}, $\zeta^{-1}$ and $\tilde{\gamma}$ contain
the marker \texttt{ot}, which tags all contributions that stem from the
$\mathcal{O}(t)$ entries $(\zeta_{\text{\abbrev{S}}})_j$, $j=1,\ldots,9$. These entries
contribute at $\mathcal{O}(t^0)$ only upon inversion. Setting
$\texttt{ot}=1$ gives the complete result, while $\texttt{ot}=0$ omits these
contributions.

\textbf{Mathematica file:}
The matrices $\zeta$, $\zeta^{-1}$, and $\tilde{\gamma}$ are stored as
Mathematica arrays named \texttt{zeta}, \texttt{zetaInv}, and
\texttt{tildeGamma}. The command \texttt{zeta[[i]]} returns the vector of
coefficients in the \sftx\ of the $i$th operator, as defined in
\cref{eq:ZetaVec}; the results presented in \cref{sec:Results} correspond to
\texttt{i=1} and \texttt{i=11}. All standard Mathematica operations apply.
For example, \texttt{zeta[[i,j]]} returns the entry $\zeta_{ij}$, and
\texttt{Coefficient[zeta[[i,j]], api, n]} its coefficient of $\apis^n$. The
same holds for \texttt{zetaInv} and \texttt{tildeGamma}.

\textbf{Python file:}
The Python file contains the same objects under the same names, with the
matrices given as SymPy matrices and the replacement rules as dictionaries.
The function \texttt{coefs(M, n, i, j)} returns the coefficient of $\apis^n$
in the entry $M_{ij}$ of a matrix \texttt{M}, and
\texttt{entry(M, n, a\_s, i, j)} returns $M_{ij}$ through order $\apis^n$ at
$\apis=\texttt{a\_s}$, where \texttt{a\_s} may be a number or a symbol. As in
this paper, the indices $i$ and $j$ start at 1. Both functions take two
optional arguments. With \texttt{numeric=True}, all analytically known
constants are evaluated numerically, and the result is rounded to six
significant digits, the precision of the numerically known coefficients.
With \texttt{subs}, substitutions are made first, for example
\texttt{RepcChi} and \texttt{RepGamma}, or numerical values for $\nf$ and
the color factors.

\begin{table}[h!]
\centering
\caption{Notation in the ancillary files.}
\label{tab:AncFile}
\begin{tabular}{l|*{13}{c}}
\hline
\textbf{symbol}  & \texttt{nc} & \texttt{cf} & \texttt{ca} & \texttt{tr}
                 & \texttt{api} & \texttt{Lmut} & \texttt{nf} & \texttt{cChi}
                 & \texttt{ot} & \texttt{gam7} & \texttt{gam8} & \texttt{gam9} & \texttt{t}\\
\textbf{meaning} & $\nc$ & $\ccf$ & $\cca$ & $\ctr$
                 & $\apis$ & $\lmut$ & $\nf$ & $c_\chi^{(2)}$
                 & $\otmarker$ & $\Gamma_7$ & $\Gamma_8$ & $\Gamma_9$ & $t$\\
\hline
\end{tabular}
\end{table}


\end{appendix}


\bibliographystyle{utphys}
\bibliography{non-inspire,literature}


\end{document}